\documentclass[12pt]{iopart}

\usepackage{latexsym,amsthm,amssymb}
\usepackage{cite}

\newcommand{\parens}[1]{\left(#1\right)}
\newcommand{\bracks}[1]{\left[#1\right]}

\newcommand{\stareq}{\mathop{=}\limits^{\star}}
\newcommand{\ud}{\;\mathrm{d}}

\newcommand{\finsarg}{(t,\mathbf{x})}
\newcommand{\sinsarg}{(t,\mathbf{x'})}
\newcommand{\retarg}{(t',\mathbf{x'})}

\newcommand{\uarg}{(t-\frac{ |\mathbf{x}|}{c})}
\newcommand{\varg}{(t+\frac{ |\mathbf{x}|}{c})}
\newcommand{\targ}{(t)}

\newcommand{\xarg}{(\mathbf{x})}
\newcommand{\xparg}{(\mathbf{x'})}

\newcommand{\gpoisint}[1]{\Delta^{-1}\Bigg[\xpb #1 \Bigg]}
\newcommand{\retint}[1]{\Box^{-1}_{\mathrm{R}}\Bigg[ #1  \Bigg]}
\newcommand{\gretint}[1]{\Box^{-1}_{\mathrm{R}}\Bigg[\xpb #1  \Bigg]}

\newcommand{\gnearint}[1]{-\frac{1}{4\pi}\int_{|\mathbf{x'}|<\mathcal{R}} \xpb \frac{#1}{|\mathbf{x}-\mathbf{x'}|}\ud^3\mathbf{x'}}

\newcommand{\gfarint}[1]{-\frac{1}{4\pi}\int_{\mathcal{R}<|\mathbf{x'}|} \xpb \frac{#1}{|\mathbf{x}-\mathbf{x'}|}\ud^3\mathbf{x'}}

\newcommand{\x}{|\mathbf{x}|}
\newcommand{\xp}{|\mathbf{x'}|}

\newcommand{\A}{\mathrm{A}}
\newcommand{\fp}{\mathop{\mathrm{FP}}\limits_{B=0}}
\newcommand{\fpa}{\mathop{\mathrm{FP}}\limits_{B=0}\mathrm{A}}

\newcommand{\resa}{\mathop{\mathrm{Residue}}\limits_{B=0}\mathrm{A}}
\newcommand{\ellfpa}{\mathop{\mathrm{FP}}\limits_{j=\ell}\mathrm{A}}

\newcommand{\xb}{\parens{\frac{\x}{r_0}}^B}
\newcommand{\xpb}{\parens{\frac{\xp}{r_0}}^B}
\newcommand{\re}{\mathrm{Re}}
\newcommand{\im}{\mathrm{Im}}
\newcommand{\lnx}{\parens{\ln{\frac{\x}{r_0}}}}
\newcommand{\lnxp}{\parens{\ln{\frac{\xp}{r_0}}}}

\newcommand{\tfrac}[2]{#1/#2}
\newcommand{\tbfrac}[2]{\bracks{#1/#2}}

\newcommand{\txb}{\parens{\x/r_0}^B}
\newcommand{\txpb}{\parens{\xp/r_0}^B}
\newcommand{\tlnx}{\parens{\ln{\parens{\x/r_0}}}}

\theoremstyle{definition}\newtheorem{theorem}{Theorem}[section]

\theoremstyle{remark}\newtheorem*{defa1}{Definition A.1}
\theoremstyle{remark}\newtheorem*{defa2}{Definition A.2}
\theoremstyle{remark}\newtheorem*{lemma1}{Lemma A.1}
\theoremstyle{remark}\newtheorem*{lemma2}{Lemma A.2}

\theoremstyle{remark}\newtheorem*{defa3}{Definition A.3}
\theoremstyle{remark}\newtheorem*{defa4}{Definition A.4}

\theoremstyle{remark}\newtheorem*{defkirch}{Kirchhoff's no-incoming radiation condition}
\theoremstyle{remark}\newtheorem*{formurod}{Rodrigues' formula}

\eqnobysec

\begin{document}


\title[Alternative method for matching ...]{Alternative method for matching post-Newtonian expansion to post-Minkowskian field}

\author{Abbas Mirahmadi}

\address{Department of Physics, University of Tehran, North Kargar Ave., Tehran, Iran}
\ead{abbasmirahmadi@ut.ac.ir}
\vspace{10pt}
\begin{indented}
\item[]February 2021
\end{indented}

\begin{abstract}
In 2002, Poujade and Blanchet succeeded in matching the post-Newtonian solution to the Einstein field equation to the post-Minkowskian field up to any arbitrary order as well as reproducing, in a different way, the results of the 1998 paper by Blanchet in which he showed how to match the post-Minkowskian series to the post-Newtonian expansion. Comparing these two papers, it might be asked whether it is possible to match the post-Newtonian field to the post-Minkowskian one by means of a method similar to the one used in the 1998 paper. The answer is affirmative, and in the present paper we provide this alternative method. Furthermore, detailed proofs of several properties and results stated in previous papers are given.\\

\end{abstract}


\section{Introduction}\label{sec:1}

\subsection{Historical background and motivation}\label{subsec:1.1}

Due to the nonlinearities of the Einstein field equation, employing approximate solution methods  for the analytical study of the gravitational waves is unavoidable. Einstein himself solved the problem of the linear approximation to the gravitational radiation generated by a localized time-dependent source and, by approximating 
$|\mathbf{x}-\mathbf{x'}|$
appearing in the retarded argument and the denominator of the integrand as constant for large distances and applying the conservation equation, found his famous quadrupole formula
\cite{E1916, E1918}.
More systematically, in the cases of self-gravitating sources with internal speeds very small compared with the speed of light, the post-Minkowskian and post-Newtonian expansions are used. By substituting these two asymptotic expansions into the Einstein field equation, we find the post-Minkowskian and post-Newtonian equations governing the gravitational field. Although the post-Minkowskian expansion, if one combines it with a multipole expansion
\cite{B1959, T1980},
can provide approximate solutions up to a certain finite order, there is no guarantee that higher-order approximations are obtainable
\cite{BD1986}.
Moreover, there are several works, such as
\cite{K1980a, K1980b},
showing that the post-Newtonian expansion, whose domain of validity is the near-zone
$\lambda \ll \x$
(where
$\lambda$
is a typical wavelength of the emitted gravitational wave), breaks down beyond a specific order due to the divergence of integrals, which itself is owing to the behavior of the lower-order approximations in the far zone.

The inefficiency of the usual methods of taking the retarded and Poisson integrals does not mean that the solutions to the post-Minkowskian and post-Newtonian equations cannot be obtained up to any order. As suggested by Fock, the problem of describing the gravitational field everywhere in
$\mathbb{R}^3$
needs to be split into two subproblems, one concerning the near zone and the other outside the source, and then, a matching procedure must be employed in order to determine the unknown terms of the general solutions to the subproblems
\cite{F1964}.
Based on the Fock's idea, Blanchet and Damour established their own algorithm in which they employed the post-Minkowskian expansion to describe the gravitational field outside the source, a region where
$T^{\mu\nu}\finsarg=0$,
considered the post-Newtonian expansion as the approximate solution to the Einstein field equation in the near zone and assumed that the domains of validity of both expansions under discussion overlap (the mathematical statement of this last assumption is called the 
\textit{matching equation})
\cite{B2014}.
Thanks to the method introduced by Riesz
\cite{R1939},
which depends deeply on analytic continuation, Blanchet and Damour could provide the most general solution to each order of the post-Minkowskian approximation in 1986
\cite{BD1986}.
Then, in the following years, they matched the post-Minkowskian series to the post-Newtonian field order by order
\cite{BD1989, B1995, B1996}.

Instead of proceeding with the order-by-order matching procedure in which a coordinate transformation between the fields is looked for, in 1998 Blanchet matched the resummation of the post-Minkowskian expansion (containing all the coefficients from
$n=1$
to infinity) to the post-Newtonian field
\cite{B1998}.
Roughly speaking, the method used was to convert the near-zone integrals brought about in the course of computations into the far-zone integrals by means of a process of analytic continuation and then to apply the matching equation. Later, in 2002, Poujade and Blanchet matched the resummation of the post-Newtonian expansion to the post-Minkowskian field with a method different from the one employed in the 1998 paper
\cite{PB2002}.
They substituted the near-zone expansion of the post-Minkowskian solution and the far-zone expansion of the post-Newtonian field into the matching equation, and then, after the identical terms cancelling out, they determined the unknown post-Minkowskian and post-Newtonian moments of the general solutions by comparing the remaining unknown and known terms. In this paper, we intend to reproduce the 2002 paper results of matching the post-Newtonian expansion to the post-Minkowskian expansion by virtue of a method similar to the one used in
\cite{B1998}.
As we will see later, in contrast to
\cite{B1998},
here we need to transform the far-zone integrals into the near-zone integrals, and this makes the computations more sophisticated. Before proceeding further, we devote the next subsection to stating the results\footnote{
The detailed proofs of these results can be found in
\cite{THESIS}.}
of the previous papers that we need to achieve our aim.


\subsection{Review of the past results}\label{subsec:1.2}

In this paper, we use the Landau-Lifshitz formulation of the Einstein field equation in which the main variables are the components of
$h^{\mu\nu}= \sqrt{-g} g^{\mu\nu} - \eta^{\mu\nu}$,
where
$g=\mathrm{det}[g_{\mu\nu}]$,
$[g^{\mu\nu}]=[g_{\mu\nu}]^{-1}$
and
$\eta^{\mu\nu}=\mathrm{diag}\left(-1, 1, 1, 1\right)$.
We have
\begin{equation}\label{1.2.1}
\Box h^{\mu\nu}\finsarg = \frac{16\pi G}{c^4}\tau ^{\mu\nu}\finsarg,
\end{equation}
\begin{equation}\label{1.2.2}
\partial_\mu h^{\mu\nu}\finsarg = 0.
\end{equation}
The first of the above equations is called the {\it relaxed Einstein field equation} and
$\tau^{\mu\nu}$
appearing in it the {\it effective energy-momentum pseudotensor}.
$\Box=\partial_\alpha \partial^\alpha = \eta^{\alpha\beta} \partial_\alpha \partial_\beta$
and
$\tau^{\mu\nu}$
is given by
\begin{equation}\label{1.2.3}
\tau^{\mu\nu} = \parens{-g} T^{\mu\nu}+\frac{c^4}{16\pi G}\Lambda^{\mu\nu},
\end{equation}                       
where,
defining
$[\mathfrak{g}^{\mu\nu}]=[\eta^{\mu\nu}+h^{\mu\nu}]$
and
$[\mathfrak{g}_{\mu\nu}]=[\eta^{\mu\nu}+h^{\mu\nu}]^{-1}$,
$\Lambda^{\mu\nu}$
reads
\begin{eqnarray}\label{1.2.4}
\Lambda^{\mu\nu} &= \partial_\alpha h^{\mu\beta} \partial_\beta h^{\nu\alpha}- h^{\alpha\beta}\partial_{\alpha\beta} h^{\mu\nu}+\frac{1}{2}\mathfrak{g}^{\mu\nu}\mathfrak{g}_{\lambda\alpha}\partial_\rho h^{\lambda\beta}\partial_\beta h^{\alpha\rho} \nonumber\\
& - \mathfrak{g}^{\mu\lambda}\mathfrak{g}_{\alpha\beta}\partial_\rho h^{\nu\beta}\partial_\lambda h^{\alpha\rho} - \mathfrak{g}^{\nu\lambda}\mathfrak{g}_{\alpha\beta}\partial_\rho h^{\mu\beta}\partial_\lambda h^{\alpha\rho}+ \mathfrak{g}_{\lambda\alpha}\mathfrak{g}^{\beta\rho}\partial_\beta h^{\mu\lambda}\partial_\rho h^{\nu\alpha}\nonumber\\  
 &+ \frac{1}{8} \left( 2\mathfrak{g}^{\mu\lambda}\mathfrak{g}^{\nu\alpha} - \mathfrak{g}^{\mu\nu}\mathfrak{g}^{\lambda\alpha}\right) \left( 2\mathfrak{g}_{\beta\rho}\mathfrak{g}_{\sigma\tau} - \mathfrak{g}_{\rho\sigma}\mathfrak{g}_{\beta\tau} \right)\partial_\lambda h^{\beta\tau}\partial_\alpha h^{\rho\sigma}.
\end{eqnarray}
Moreover,
(\ref{1.2.1}),
together with (\ref{1.2.2})
called the
\textit{harmonic gauge condition}, imply that
$\tau^{\mu\nu}$
is conserved, i.e.,
\begin{equation}\label{1.2.5}
\partial_\mu \tau^{\mu\nu}\finsarg = 0.
\end{equation}

We restrict our attention to the gravitational waves sources that can be treated as perfect fluids with internal speeds very small in comparison to the speed of light. We also choose a coordinate system in which the origin of the spatial coordinates is located within the source and assume that the material energy-momentum tensor of the source
$T^{\mu\nu} \finsarg$
is compactly supported (i.e., there is a positive constant
$d$
such that
$T^{\mu\nu} \finsarg = 0$
for
$\x>d$)
and belongs to
$C^\infty \left(\mathbb{R}^4\right)$. Furthermore, we consider the following conditions for
$h^{\mu\nu} \finsarg$
\cite{BD1986}:
\begin{equation}\label{1.2.6}
 h^{\mu\nu}\finsarg \in C^\infty \left(\mathbb{R}^4\right),
\end{equation}
\begin{equation}\label{1.2.7}
\partial _t h^{\mu\nu}\finsarg = 0 \; \textrm{when}\; t \le -\mathcal{T},
\end{equation}
\begin{equation}\label{1.2.8}
\mathop{\mathop{\lim}\limits_{\x\to \infty}}\limits_{t=\mathrm{const}} h^{\mu\nu}\finsarg =0 \; \textrm{when}\; t \le -\mathcal{T},
\end{equation}
where by
$-\mathcal{T}$
we mean an instant in the past.

We denote the post-Minkowskian and post-Newtonian expansions of a general function by
$\mathcal{M}(f)\finsarg=\sum_{n=n_0}^{\infty} G^n f_{\left(n\right)}\finsarg$
and
$\bar{f}\finsarg=\sum_{m=m_0}^{\infty} \parens{\tfrac{1}{c}}^m \bar{f}_{\left(m\right)}\finsarg$
respectively, where
$n_0$
and
$m_0$
depend on the function. Having introduced the notation and taking the domains of validity of the two approximations into account, by substituting the post-Minkowskian expansions of the functions appearing in
(\ref{1.2.1}), (\ref{1.2.2}), (\ref{1.2.5}), (\ref{1.2.6}), (\ref{1.2.7}) and (\ref{1.2.8}),
we find
\begin{eqnarray}
\Box h^{\mu\nu}_{\parens{n}}\finsarg =\Lambda^{\mu\nu}_{\parens{n}}\finsarg \hspace{27.5mm} &\textrm{for}\;n\ge 1, \label{1.2.9}\\
\partial_\mu h^{\mu\nu}_{\parens{n}}\finsarg = 0  &\textrm{for}\;n\ge1, \label{1.2.10}\\
\partial_\mu \Lambda^{\mu\nu}_{\parens{n}}\finsarg= 0  &\textrm{for}\;n\ge1, \label{1.2.11}\\
h^{\mu\nu}_{\parens{n}}\finsarg \; \textrm{is smooth at}\; |\mathbf{x}|>d  &\textrm{for}\;n\ge 1, \label{1.2.12}\\
\partial _t h^{\mu\nu}_{\parens{n}}\finsarg = 0\; \textrm{when}\; t \le -\mathcal{T} & \textrm{for}\;n\ge 1, \label{1.2.13}\\
\mathop{\mathop{\lim}\limits_{\x\to \infty}}\limits_{t=\mathrm{const}} h^{\mu\nu}_{\parens{n}}\finsarg = 0\; \textrm{when}\; t \le -\mathcal{T}  &\textrm{for}\;n\ge 1, \label{1.2.14}
\end{eqnarray}
while substitution of the post-Newtonian expansions results in
\begin{eqnarray}
\Delta \bar{ h}^{\mu\nu}_{\parens{n}}\finsarg = 16\pi G\bar{\tau} ^{\mu\nu}_{\parens{n-4}}\finsarg + \partial^2_t   \bar{ h}^{\mu\nu}_{\parens{n-2}}\finsarg \hspace{10mm} &\textrm{for}\;n\ge2, \label{1.2.15}\\
\partial_t \bar{h}^{0\nu}_{\parens{n-1}}\finsarg + \partial_i \bar{h}^{i\nu}_{\parens{n}}\finsarg= 0 &\textrm{for}\;n\ge 2, \label{1.2.16}\\
\partial_t \bar{\tau}^{0\nu}_{\parens{n-5}}\finsarg + \partial_i \bar{\tau}^{i\nu}_{\parens{n-4}}\finsarg = 0 &\textrm{for}\;n\ge 2, \label{1.2.17}\\
\bar{ h}^{\mu\nu}_{\parens{n}}\finsarg \; \textrm{is smooth at} \; |\mathbf{x}| < \mathcal{R} &\textrm{for}\;n\ge 2, \label{1.2.18}\\
\partial _t \bar{ h}^{\mu\nu}_{\parens{n}}\finsarg = 0\; \textrm{when}\; t \le -\mathcal{T} &\textrm{for}\;n\ge 2, \label{1.2.19}
\end{eqnarray}
where
$\mathcal{R}$
is the radius at which we take the boundary of the near zone to be
($d<\mathcal{R}\ll \lambda$).
In both cases, the general solution at each order is written as the sum of the general solution to the corresponding homogeneous equation (which is subject to some restrictions  mainly brought about by the harmonic gauge condition) and a particular solution to the inhomogeneous equation derived from the relaxed Einstein field equation
\cite{BD1986, PB2002}.
That particular solution is given with the use of the following theorem
\cite{BD1986, PB2002}:
\begin{theorem}\label{th:1.2.1}
A particular solution to the equation
$\mathbf{L}f\finsarg=g\finsarg$,
where
$\mathbf{L}$
is either
$\Delta$
or
$\Box$,
is
$\fpa\mathbf{L}^{-1}\left[\txpb g (t'',\mathbf{x'})\right]$,
where
$\mathbf{L}^{-1}$
is either
$\Delta^{-1}$
(with
$t''=t$)
or
$\Box^{-1}_{\mathrm{R}}$
(with
$t''=t'=t-\tfrac{|\mathbf{x}-\mathbf{x'}|}{c}$).
In this particular solution,
$B$
is a complex number and
$r_0$
an arbitrary constant, and by
$\A$
and
$\fp$
we mean \textit{the analytic continuation of ...} and \textit{the coefficient of the zeroth power of}
$B$
\textit{in the Laurent expansion of ... about}
$B=0$,
respectively. Evidently, for this to be a particular solution,
$\mathbf{L}^{-1}\left[\txpb g (t'',\mathbf{x'})\right]$
needs to be analytic in some original domain and analytically continuable to some (punctured) neighborhood of
$B=0$.
\end{theorem}

In terms of the post-Minkowskian approximation, the only results that we need are as follows. It can be proven that the structure of
$\mathcal{M}\parens{h^{\mu\nu}}\finsarg$
and
$\Lambda^{\mu\nu}\parens{\mathcal{M}\parens{h}}\finsarg$,
the (untruncated) post-Minkowskian expansions of
$h^{\mu\nu}\finsarg$
and
$\Lambda^{\mu\nu}\finsarg$,
read
\cite{BD1986}
\begin{eqnarray}\label{1.2.20}
\fl\mathcal{M}\parens{h^{\mu\nu}}\finsarg &= \mathcal{M}\parens{h^{\mu\nu}_{\mathrm{AS}}}\xarg + \mathcal{M}\parens{h^{\mu\nu}_{\mathrm{PZ}}}\finsarg\nonumber\\
&=\sum_{\ell=0}^{\infty}\sum_{a=-\infty}^{-1}\hat{n}^L \x^k\hat{C}^{\mu\nu}_{L,k}+\sum_{q=0}^{\infty}\sum_{a=-\infty}^{\infty}\sum_{p=0}^{\infty}\hat{n}^Q\x^a\lnx^p\hat{F}^{\mu\nu}_{Q,a,p}\targ \nonumber \\  
&+ R^{\mu\nu}\finsarg,
\end{eqnarray}
\begin{eqnarray}\label{1.2.21}
\fl\Lambda^{\mu\nu}(\mathcal{M}\parens{h})\finsarg &= \Lambda^{\mu\nu}_{\mathrm{AS}}\parens{\mathcal{M}\parens{h}}\xarg + \Lambda^{\mu\nu}_{\mathrm{PZ}}\parens{\mathcal{M}\parens{h}}\finsarg\nonumber\\
&=\sum_{\ell=0}^{\infty}\sum_{a=-\infty}^{-4}\hat{n}^L \x^k\hat{C}'^{\;\mu\nu}_{L,k}+\sum_{q=0}^{\infty}\sum_{a=-\infty}^{\infty}\sum_{p=0}^{\infty}\hat{n}^Q\x^a\lnx^p\hat{F}'^{\;\mu\nu}_{Q,a,p}\targ\nonumber \\  
&+ R'^{\;\mu\nu}\finsarg,
\end{eqnarray}
where in each case the term with the subscript AS denotes the first term on the RHS of the second equality and the one with the subscript PZ the remaining terms.
$\hat{n}^L$
is the symmetric-trace-free part of
$n^L=n^{I_\ell}=n^{i_1}\cdots n^{i_\ell}$,
the constant
$r_0$
is the same as in theorem
\ref{th:1.2.1}
given earlier, and
$\hat{C}^{\mu\nu}_{L,k}$
and
$\hat{C}'^{\;\mu\nu}_{L,k}$
are constant, while
$\hat{F}^{\mu\nu}_{Q,a,p}\targ$,
$\hat{F}'^{\;\mu\nu}_{Q,a,p}\targ$,
$ R^{\mu\nu}\finsarg$
and
$ R'^{\;\mu\nu}\finsarg$
are 
\textit{past-zero}.\footnote{
We denote the multi-index
$i_1...i_{\ell-m}$
by
$L-m$,
and by
$\hat{T}^Q$
we mean the symmetric-trace-free part of
$T^Q$.
In addition, past-zero functions are the ones that equate to zero at
$t\le -\mathcal{T}$.
Last but not least, the subscripts AS and PZ stand for 
\textit{always-stationary}
and
\textit{past-zero}
respectively.}
Furthermore,
$ R^{\mu\nu}\finsarg$
and
$ R'^{\;\mu\nu}\finsarg$
are
$\textrm{O}\parens{\x^N}$
(called the {\it big-O})
as
$\x \to 0$
where
$N$
tends to infinity, and hence,
$\x^m R^{\mu\nu}\finsarg$
and
$\x^m  R'^{\;\mu\nu}\finsarg$
with any
$m$
are well-behaved in any punctured neighborhood of
$\x=0$.
The set of equations governing
$\mathcal{M}\parens{h^{\mu\nu}}\finsarg$
and
$\Lambda^{\mu\nu}\parens{\mathcal{M}\parens{h}}\finsarg$
are
\begin{equation}\label{1.2.22}
\Box \mathcal{M}\parens{h^{\mu\nu}}\finsarg=\Lambda^{\mu\nu}(\mathcal{M}\parens{h})\finsarg,
\end{equation}
\begin{equation}\label{1.2.23}
 \partial_\mu \mathcal{M}\parens{h^{\mu\nu}}\finsarg=0,
\end{equation}
\begin{equation}\label{1.2.24}
 \partial_\mu \Lambda^{\mu\nu}(\mathcal{M}\parens{h})\finsarg=0.
\end{equation}

With regard to the post-Newtonian approximation, it can be shown that, for any arbitrary
$n$,
$\bar{\Lambda}^{\mu\nu}_{\parens{n}}\finsarg$
is smooth in the near zone while outside the near zone its structure is of the form
\begin{equation}\label{1.2.25}
\bar{\Lambda}^{\mu\nu}_{\parens{n}}\finsarg= \sum_{q=0}^{\infty}\sum_{a=-\infty}^{a_{\mathrm{max}}\parens{n}}\sum_{p=0}^{p_{\mathrm{max}}\parens{n}}\hat{n}^Q\x^a\lnx^p\hat{E}^{\mu\nu}_{\parens{n}Q,a,p}\targ,
\end{equation}
where
$a_{\mathrm{max}}\parens{n}>0$
and
$\hat{E}^{\mu\nu}_{\parens{n}Q,a,p}\targ$
is a 
\textit{past-stationary}\footnote{
Past-stationarity means being stationary at
$t\le - \mathcal{T}$.}
function.
These structural properties of
$\bar{\Lambda}^{\mu\nu}_{\parens{n}}\finsarg$
in the near zone and outside of it guarantee that the general solution to
$n$th-order
problem is given by
\cite{PB2002}
\begin{eqnarray}\label{1.2.26}
\bar{h}^{\mu\nu}_{\parens{n}}\finsarg & \stareq \fpa\gpoisint{\parens{16\pi G \bar{\tau}^{\mu\nu}_{\parens{n-4}}\sinsarg+\partial^2_t \bar{h}^{\mu\nu}_{\parens{n-2}}\sinsarg}}\nonumber\\
&+\sum_{\ell=0}^{\infty}\hat{n}^L\x^\ell \hat{B}^{\mu\nu}_{\parens{n}L}\targ,
\end{eqnarray}
where by the sign
$\stareq$
we mean that the LHS of this sign equals its RHS provided that
$\int_{\mathbb{R }^3}$
appearing on the RHS is written as the sum of
$\int_{|\mathbf{x'}|<\mathcal{R}}$
and
$\int_{\mathcal{R}<|\mathbf{x'}|}$;
otherwise, it cannot be used in computations
\cite{THESIS}.
Similar to
$\bar{\Lambda}^{\mu\nu}_{\parens{n}}\finsarg$,
$\bar{h}^{\mu\nu}_{\parens{n}}\finsarg$
is smooth in the near zone. Furthermore, its structure outside the near zone, provided that
$\hat{B}^{\mu\nu}_{\parens{n}L}\targ$'s
with
$\ell>\ell_{\mathrm{max}}\parens{n}$
are taken to be zero\footnote{
This assumption is technically inevitable as in the post-Minkwskian solution
\cite{M2008}.},
can be written as
\begin{equation}\label{1.2.27}
\bar{h}^{\mu\nu}_{\parens{n}}\finsarg= \sum_{q=0}^{\infty}\sum_{a=-\infty}^{a'_{\mathrm{max}}\parens{n}}\sum_{p=0}^{p'_{\mathrm{max}}\parens{n}}\hat{n}^Q\x^a\lnx^p\hat{G}^{\mu\nu}_{\parens{n}Q,a,p}\targ,
\end{equation}
where
$\hat{G}^{\mu\nu}_{\parens{n}Q,a,p}\targ$
is past-stationary. In contrast to
\cite{PB2002}
in which this structure has been assumed, it has been obtained in
\cite{THESIS}.
Moreover, as it can be seen in the above equation, no remainder term appears in the structure. By summing the post-Newtonian coefficients
$\bar{h}^{\mu\nu}_{\parens{n}}\finsarg$
(multiplied by
$\tfrac{1}{c^n}$)
over all values of
$n$
and using (3.9) in
\cite{PB2002},
one can readily reach
\begin{eqnarray}\label{1.2.28}
\fl\bar{h}^{\mu\nu}\finsarg \hspace{-10mm}&\stareq \frac{16\pi G}{c^4}\sum_{k=0}^{\infty}\frac{1}{c^{2k}}\partial^{2k}_t\fpa\Bigg[-\frac{1}{4\pi}\int_{\mathbb{R}^3}\xpb\frac{{|\mathbf{x}-\mathbf{x'}|}^{2k-1}}{\parens{2k}!}\bar{\tau}^{\mu\nu}(t,\mathbf{x'})\ud^3\mathbf{x'}\Bigg]\nonumber \\  
\hspace{-10mm}& +\sum_{k=0}^{\infty}\sum_{\ell=0}^{\infty}\frac{1}{c^{2k}}\frac{\parens{2\ell+1}!!}{\parens{2k}!!\parens{2\ell+2k+1}!!}\x^{2k}\hat{x}^{L}\partial^{2k}_t\hat{B}^{\mu\nu}_{L}\targ.
\end{eqnarray}
It is also obvious that, since all the post-Newtonian coefficints
$\bar{h}^{\mu\nu}_{\parens{n}}\finsarg$
and
$\bar{\Lambda}^{\mu\nu}_{\parens{n}}\finsarg$
are smooth in the near zone, so do the (untruncated) post-Newtonian expansions 
$\bar{h}^{\mu\nu}\finsarg$
and
$\bar{\Lambda}^{\mu\nu}\finsarg$.
Additionally, considering the structures given in
(\ref{1.2.25}) and (\ref{1.2.27}),
the structures of
$\bar{h}^{\mu\nu}\finsarg$
and
$\bar{\Lambda}^{\mu\nu}\finsarg$
outside the near zone can be expressed as
\begin{equation}\label{1.2.29}
\bar{h}^{\mu\nu}\finsarg= \sum_{q=0}^{\infty}\sum_{a=-\infty}^{\infty}\sum_{p=0}^{\infty}\hat{n}^Q\x^a\lnx^p\hat{G}^{\mu\nu}_{Q,a,p}\targ,
\end{equation}
\begin{equation}\label{1.2.30}
\bar{\Lambda}^{\mu\nu}\finsarg= \sum_{q=0}^{\infty}\sum_{a=-\infty}^{\infty}\sum_{p=0}^{\infty}\hat{n}^Q\x^a\lnx^p\hat{E}^{\mu\nu}_{Q,a,p}\targ.
\end{equation}
Finally, the set of equations governing them are as follows:
\begin{equation}\label{1.2.31}
\Box \bar{h}^{\mu\nu}\finsarg=\frac{16\pi G}{c^4}\bar{\tau}^{\mu\nu}\finsarg,
\end{equation}
\begin{equation}\label{1.2.32}
 \partial_\mu \bar{h}^{\mu\nu}\finsarg=0,
\end{equation}
\begin{equation}\label{1.2.33}
 \partial_\mu \bar{\tau}^{\mu\nu}\finsarg=0.
\end{equation}

Last but not least, let us provide the matching equation. It is clear that we can write
\cite{B1998}
\begin{equation}\label{1.2.34}
\fl\overline{\mathcal{M}\parens{h^{\mu\nu}}}\finsarg=\mathop{\mathop{\lim}\limits_{\x\to 0}}\limits_{t=\mathrm{const}}\mathcal{M}\parens{h^{\mu\nu}}\finsarg=  \sum_{q=0}^{\infty}\sum_{a=-\infty}^{\infty}\sum_{p=0}^{\infty}\hat{n}^Q\x^a\lnx^p\hat{F}^{\mu\nu}_{Q,a,p}\targ,
\end{equation}
\begin{equation}\label{1.2.35}
\fl\mathcal{M}\parens{\bar{h}^{\mu\nu}}\finsarg=\mathop{\mathop{\lim}\limits_{\x\to \infty}}\limits_{t=\mathrm{const}}\bar{h}^{\mu\nu}\finsarg= \sum_{q=0}^{\infty}\sum_{a=-\infty}^{\infty}\sum_{p=0}^{\infty}\hat{n}^Q\x^a\lnx^p\hat{G}^{\mu\nu}_{Q,a,p}\targ.
\end{equation}
If
(\ref{1.2.34}) and (\ref{1.2.35})
had been written in terms of the \textit{rescaled} variables,
it would have been obvious that
$\overline{\mathcal{M}\parens{h^{\mu\nu}}}\finsarg$
and
$\mathcal{M}\parens{\bar{h}^{\mu\nu}}\finsarg$
are in fact the post-Minkowskian and post-Newtonian expansions of
$h^{\mu\nu}\finsarg$ within the common region of validity of these expansions. Therefore, recalling the assumption of existence of the matching region discussed before, we have
\cite{B1998}
\begin{equation}\label{4.1.3}
\overline{\mathcal{M}\parens{h^{\mu\nu}}}\finsarg=\mathcal{M}\parens{\bar{h}^{\mu\nu}}\finsarg,
\end{equation}
which is in full agreement with the similarity of the structures of
$\overline{\mathcal{M}\parens{h^{\mu\nu}}}\finsarg$
and
$\mathcal{M}\parens{\bar{h}^{\mu\nu}}\finsarg$.
(\ref{4.1.3})
is the matching equation and we use it to match
$\mathcal{M}\parens{h^{\mu\nu}}\finsarg$
and
$\bar{h}^{\mu\nu}\finsarg$
together.

Having reviewed the past results, we are now in a position to match the post-Newtonian series to the post-Minkowskian field. This is what we will do in section 
\ref{sec:2}.
In section 
\ref{sec:3},
it will be shown in great detail that
$\bar{h}^{\mu\nu}\finsarg$
obtained in section
\ref{sec:2}
indeed fulfills the harmonic gauge condition. In section
\ref{sec:4},
we will determine the moments
$\hat{B}^{\mu\nu}_{L}\targ$,
and in section
\ref{sec:5},
provide the detailed computation of the closed form\footnote{
We will later state what this means.}
of
$\bar{h}^{\mu\nu}\finsarg$.
Finally, in section
\ref{sec:6},
a brief summary of the work will be given. At the end of this paper, two appendices have also been included. Appendix A contains a collection of useful definitions, lemmas and formulae, and appendix B
has been devoted to proving that
$\bar{h}^{\mu\nu}\finsarg$
is independent of the constant
$r_0$.


\section{Determination of $\bar{h}^{\mu\nu}\finsarg$}\label{sec:2}

We start our investigation with the trivial equality
\begin{eqnarray}\label{4.3.1}
\fl \Box \Bigg[\xb\bar{h}^{\mu\nu}\finsarg\Bigg]&=\xb\Box\bar{h}^{\mu\nu}\finsarg\nonumber\\
&+\xb\bigg[2B\x^{-1}\partial_{\x}\bar{h}^{\mu\nu}\finsarg+B\parens{B+1}\x^{-2}\bar{h}^{\mu\nu}\finsarg\bigg].
\end{eqnarray}
As we saw in subsection
\ref{subsec:1.2},
the structure of
$\bar{h}^{\mu\nu}\finsarg$
outside the near zone is of the form
$\sum_{q=0}^{\infty}\sum_{a=-\infty}^{\infty}\sum_{p=0}^{\infty}\hat{n}^Q\x^a\tlnx^p\hat{G}^{\mu\nu}_{Q,a,p}\targ$.
Since
$\bar{h}^{\mu\nu}\finsarg$
is smooth inside the near zone, it is reasonable to assume that there exists a radius of convergence
$r < \mathcal{R}$
within which, by using the Taylor expansion for functions of three variables (see appendix A), we can write
\begin{eqnarray}\label{4.3.2-fist part}
\fl\bar{h}^{\mu\nu}\finsarg = \sum_{j=0}^{\infty}\frac{1}{j!}x^J\left[\partial_J\bar{h}^{\mu\nu}\finsarg{\bigg|}_{{\mathbf{x}}=\mathbf{0}}\right] =\sum_{j=0}^{\infty}\frac{1}{j!}\x^j n^J\left[\partial_J\bar{h}^{\mu\nu}\finsarg{\bigg|}_{{\mathbf{x}}=\mathbf{0}}\right],
\end{eqnarray}
and by means of
(\ref{a.2.4}),
we reach
\begin{eqnarray}\label{4.3.2}
\fl\bar{h}^{\mu\nu}\finsarg\nonumber\\
\fl=\sum_{j=0}^{\infty}\sum_{k=0}^{\left[\frac{j}{2}\right]}\frac{1}{j!}\frac{\parens{2j-4k+1}!!}{\parens{2j-2k+1}!!}\x^j\delta^{\{i_1 i_2}\cdots\delta^{i_{2k-1}i_{2k}}\hat{n}^{i_{2k+1}...i_j\}}\left[\partial_J\bar{h}^{\mu\nu}\finsarg{\bigg|}_{{\mathbf{x}}=\mathbf{0}}\right] \nonumber\\
\fl= \sum_{j=0}^{\infty}\sum_{k=0}^{\left[\frac{j}{2}\right]}\frac{1}{j!}\frac{\parens{2j-4k+1}!!}{\parens{2j-2k+1}!!}\frac{j!}{2^k k!\parens{j-2k}!}\x^j\delta^{i_1 i_2}\cdots\delta^{i_{2k-1}i_{2k}}\hat{n}^{i_{2k+1}...i_j}\left[\partial_J\bar{h}^{\mu\nu}\finsarg{\bigg|}_{{\mathbf{x}}=\mathbf{0}}\right] \nonumber\\
\fl=\sum_{j=0}^{\infty}\sum_{k=0}^{\left[\frac{j}{2}\right]}\frac{1}{2^k k!\parens{j-2k}!}\frac{\parens{2\parens{j-2k}+1}!!}{\parens{2\parens{j-2k}+2k+1}!!}\x^j\hat{n}^{J-2k}\left[\partial_{J-2k}\Delta^k\bar{h}^{\mu\nu}\finsarg{\bigg|}_{{\mathbf{x}}=\mathbf{0}}\right]  \nonumber\\
\fl= \sum_{\ell=0}^{\infty}\sum_{k=0}^{\infty}\frac{1}{2^k k! \ell !}\frac{\parens{2\ell+1}!!}{\parens{2\ell+2k+1}!!}\x^{\ell+2k}\hat{n}^L\left[\partial_{L}\Delta^k\bar{h}^{\mu\nu}\finsarg{\bigg|}_{{\mathbf{x}}=\mathbf{0}}\right] \nonumber\\
\fl=\sum_{q=0}^{\infty}\sum_{k=0}^{\infty}\hat{n}^Q\x^{q+2k}\bar{h}^{\mu\nu}_{Q,k}\targ, 
\end{eqnarray}
where to obtain the second equality, we have used the fact that, since all the indices of
$J$
are dummy and
$\partial_J$
is a symmetric tensor, we can replace
$\delta^{\{i_1 i_2}\cdots\delta^{i_{2k-1}i_{2k}}\hat{n}^{i_{2k+1}...i_j\}}$
by only one of its terms times the number of its terms. Due to
$\hat{n}^{J-2k}$
and the Kronecker delta being totally symmetric tensors, the number of the terms of
$\delta^{\{i_1 i_2}\cdots\delta^{i_{2k-1}i_{2k}}\hat{n}^{i_{2k+1}...i_j\}}$
is equal to the number of ways in which one can select
$2k$
objects from
$j$
distinguishable ones and put them in
$k$
indistinguishable boxes such that each contains two objects, i.e.,
$\tfrac{{j\choose2}{j-2\choose2}\cdots{j-2k+2\choose2}}{k!}=\tfrac{j!}{\bracks{2^k k!\parens{j-2k}!}}$
where
$k!$
has appeared in the denominator because of the indistinguishability of the boxes. Hence, we have replaced
$\delta^{\{i_1 i_2}\cdots\delta^{i_{2k-1}i_{2k}}\hat{n}^{i_{2k+1}...i_j\}}$
by
$\tbfrac{j!}{2^k k!\parens{j-2k}!}\delta^{i_1 i_2}\cdots\delta^{i_{2k-1}i_{2k}}\hat{n}^{i_{2k+1}...i_j}$.
Moreover, to write the fourth equality, or in other words, to relabel the summations, it has been noted that, since
$j$
takes all nonnegative integer values, so does
$k$
because its maximum value is
$\left[\tfrac{j}{2}\right]$,
and further,
so does
$j-2k$
due to the inequality
$j\ge 2k$.

For any arbitrary values of
$a$,
$q$,
$p$
and
$k$,
one is able to connect
$\hat{n}^Q\x^a\tlnx^p\hat{G}^{\mu\nu}_{Q,a,p}\targ$
to
$\hat{n}^Q\x^{q+2k}\bar{h}^{\mu\nu}_{Q,k}\targ$
(the indices of
$Q$
are not summed over) smoothly. (Their functional dependence on
$\theta$
and
$\varphi$
are the same. Therefore, one only needs to connect
$\hat{G}^{\mu\nu}_{Q,a,p}\targ$
to
$\bar{h}^{\mu\nu}_{Q,k}\targ$
and
$\x^a\tlnx^p$
to
$\x^{q+2k}$
in a smooth manner. It is trivially obvious that a number of mappings are available for doing so.) For
$a\le0$,
we connect each
$\hat{n}^Q\x^a\tlnx^p\hat{G}^{\mu\nu}_{Q,a,p}\targ$
to
$A_{a,p}\hat{n}^Q\x^q\bar{h}^{\mu\nu}_{Q,0}\targ$
where
$\sum_{a=-\infty}^{0}\sum_{p=0}^{\infty}A_{a,p}=1$,
and for
$a>0$,
to
$B_p\hat{n}^Q\x^{q+2a}\bar{h}^{\mu\nu}_{Q,a}\targ$
where
$\sum_{p=0}^{\infty}B_p=1$.
We denote the functions constructed in this way by
$f^{\mu\nu}_{q,a,p}\finsarg$.
It is clear that in the regions
$\x<r$
and
$\x>\mathcal{R}$
we have
$\sum_{q=0}^{\infty}\sum_{a=-\infty}^{\infty}\sum_{p=0}^{\infty}f^{\mu\nu}_{q,a,p}\finsarg=\bar{h}^{\mu\nu}\finsarg$.
We demand this equation also hold in the region
$r<\x<\mathcal{R}$.
Now note that an equality similar to 
(\ref{4.3.1})
can also be written for
$f^{\mu\nu}_{q,a,p}\finsarg$,
that is
\begin{eqnarray}\label{4.3.3}
\fl\Box \Bigg[\xb f^{\mu\nu}_{q,a,p}\finsarg\Bigg]&=\xb\Box f^{\mu\nu}_{q,a,p}\finsarg\nonumber \\  
&+\xb\bigg[2B\x^{-1}\partial_{\x}f^{\mu\nu}_{q,a,p}\finsarg+B\parens{B+1}\x^{-2}f^{\mu\nu}_{q,a,p}\finsarg\bigg].\nonumber\\
\end{eqnarray}
Equating the real and imaginary parts separately, we get
\begin{eqnarray}\label{4.3.4}
\fl\Box \Bigg[\re\Bigg[\xb\Bigg] f^{\mu\nu}_{q,a,p}\finsarg\Bigg]&=\re\Bigg[\xb\Bigg]\Box f^{\mu\nu}_{q,a,p}\finsarg\nonumber \\
&+2\;\re\Bigg[B\xb\Bigg] \x^{-1}\partial_{\x}f^{\mu\nu}_{q,a,p}\finsarg\nonumber \\  
&+\re\Bigg[B\parens{B+1}\xb\Bigg]\x^{-2}f^{\mu\nu}_{q,a,p}\finsarg,
\end{eqnarray}
\begin{eqnarray}\label{4.3.5}
\fl\Box \Bigg[\im\Bigg[\xb\Bigg] f^{\mu\nu}_{q,a,p}\finsarg\Bigg]&=\im\Bigg[\xb\Bigg]\Box f^{\mu\nu}_{q,a,p}\finsarg\nonumber \\
&+2\;\im\Bigg[B\xb\Bigg] \x^{-1}\partial_{\x}f^{\mu\nu}_{q,a,p}\finsarg\nonumber \\  
&+\im\Bigg[B\parens{B+1}\xb\Bigg]\x^{-2}f^{\mu\nu}_{q,a,p}\finsarg.
\end{eqnarray}
Outside the near zone,
$ f^{\mu\nu}_{q,a,p}\finsarg$
is equal to
$\hat{n}^Q\x^a \tlnx^p\hat{G}^{\mu\nu}_{Q,a,p}\targ$,
and hence, due to the past-stationarity of
$\hat{G}^{\mu\nu}_{Q,a,p}\targ$,
both real and imaginary parts of
$\txb f^{\mu\nu}_{q,a,p}\finsarg$
fulfill the no-incoming radiation condition (see appendix A) provided that
$\re(B)+a<0$.
Thus, if the retarded integrals converge, we can write
\begin{eqnarray}\label{4.3.6}
\fl\re\Bigg[\xb\Bigg]f^{\mu\nu}_{q,a,p}\finsarg&=\retint{\re\Bigg[\xpb\Bigg]\Box'f^{\mu\nu}_{q,a,p}\retarg}\nonumber \\  
&+\retint{2\;\re\Bigg[B\xpb\Bigg]\xp^{-1}\parens{\partial_{\xp}f^{\mu\nu}_{q,a,p}\retarg}_{t'}}\nonumber \\  
&+\retint{\re\Bigg[B\parens{B+1}\xpb\Bigg]\xp^{-2}f^{\mu\nu}_{q,a,p}\retarg},
\end{eqnarray}
\begin{eqnarray}\label{4.3.7}
\fl\im\Bigg[\xb\Bigg]f^{\mu\nu}_{q,a,p}\finsarg&=\retint{\im\Bigg[\xpb\Bigg]\Box'f^{\mu\nu}_{q,a,p}\retarg}\nonumber \\  
&+\retint{2\;\im\Bigg[B\xpb\Bigg]\xp^{-1}\parens{\partial_{\xp}f^{\mu\nu}_{q,a,p}\retarg}_{t'}}\nonumber \\  
&+\retint{\im\Bigg[B\parens{B+1}\xpb\Bigg]\xp^{-2}f^{\mu\nu}_{q,a,p}\retarg},
\end{eqnarray}
where
$\parens{\partial_{\xp}f^{\mu\nu}_{q,a,p}\retarg}_{t'}$
denotes the partial derivative of
$f^{\mu\nu}_{q,a,p}\retarg$
with respect to
$\xp$,
ignoring the contribution from the variable
$t'$.
Each of the retarded integrals appearing in the above equations can be rewritten as the sum of two integrals, one over the region
$\xp<\mathcal{R}$
and the other over
$\xp>\mathcal{R}$.
Considering lemma A.1, the near zone integrals are convergent providing
$-q-3<\re(B)+q-2$
if
$a\le 0$,
and
$-q-3<\re(B)+q+2a-2$
if
$a>0$.
(The near-zone integrals whose integrands include
$\partial_{t'}^2 f^{\mu\nu}_{q,a,p}\retarg$
converge if
$-q-3<\re(B)+q$
[for
$a\le 0$],
and
$-q-3<\re(B)+q+2a$
[for
$a>0$].
These are clearly weaker conditions, however, for the convergence of all the near-zone integrals, the stronger conditions stated earlier must be satisfied.) On the other hand, taking lemma A.2 into account, the far-zone integrals converge if
$\re(B)+a-2<q-2$.
(Since
$ f^{\mu\nu}_{q,a,p}\finsarg$
is past-stationary and hence
$\partial^2_{t} f^{\mu\nu}_{q,a,p}\finsarg$
past-zero, for the far-zone integrals in which
$\partial_{t'}^2 f^{\mu\nu}_{q,a,p}\retarg$
appear to be convergent, the stronger condition
$\re(B)+a<q-2$
is not needed.) In light of these considerations, for
$a\le 0$
if
$-2q-1<\re(B)<q-a$,
and for
$a>0$
if
$-2q-2a-1<\re(B)<q-a$,
the retarded integrals are convergent. Therefore, irrespective of the value of
$a$,
there exists a vertical strip in the complex plane (whose width depends on the value of
$a$)
in which both
(\ref{4.3.6}) and (\ref{4.3.7})
hold. Consequently, in this region we can write
\begin{eqnarray}\label{4.3.8}
\xb f^{\mu\nu}_{q,a,p}\finsarg&=\retint{\xpb\Box'f^{\mu\nu}_{q,a,p}\retarg}\nonumber \\  
&+\retint{2 B\xpb\xp^{-1}\parens{\partial_{\xp}f^{\mu\nu}_{q,a,p}\retarg}_{t'}}\nonumber \\  
&+\retint{B\parens{B+1}\xpb\xp^{-2}f^{\mu\nu}_{q,a,p}\retarg}.
\end{eqnarray}
It is straightforward to show that all the complex functions appearing in the above equation are also analytic in the aforementioned region. Hence, as a result of the identity theorem, the equality between the analytic continuations of each side of
(\ref{4.3.8})
must hold in their common region of definition. Thus, considering that
$\txb f^{\mu\nu}_{q,a,p}\finsarg$
is entire, we have
\begin{eqnarray}\label{4.3.9}
\xb f^{\mu\nu}_{q,a,p}\finsarg&=\A\retint{\xpb\Box'f^{\mu\nu}_{q,a,p}\retarg}\nonumber \\  
&+\A\retint{2 B\xpb\xp^{-1}\parens{\partial_{\xp}f^{\mu\nu}_{q,a,p}\retarg}_{t'}}\nonumber \\  
&+\A\retint{B\parens{B+1}\xpb\xp^{-2}f^{\mu\nu}_{q,a,p}\retarg},
\end{eqnarray}
which, after writing each retarded integral as the sum of the near-zone and far-zone integrals and summing over all values of
$q$,
$a$
and
$p$,
reads
\begin{eqnarray}\label{4.3.10}
\fl\xb\bar{h}^{\mu\nu}\finsarg\nonumber\\
\fl=\sum_{q=0}^{\infty}\sum_{a=-\infty}^{\infty}\sum_{p=0}^{\infty}\A\Bigg[\gnearint{\Box'f^{\mu\nu}_{q,a,p}\retarg}\Bigg]\nonumber \\
\fl+\sum_{q=0}^{\infty}\sum_{a=-\infty}^{\infty}\sum_{p=0}^{\infty}\A\Bigg[\gfarint{\Box'f^{\mu\nu}_{q,a,p}\retarg}\Bigg] \nonumber\\
\fl+\sum_{q=0}^{\infty}\sum_{a=-\infty}^{\infty}\sum_{p=0}^{\infty}\A\Bigg[\gnearint{2 B\xp^{-1}\parens{\partial_{\xp}f^{\mu\nu}_{q,a,p}\retarg}_{t'}}\Bigg] \nonumber\\
\fl+\sum_{q=0}^{\infty}\sum_{a=-\infty}^{\infty}\sum_{p=0}^{\infty}\A\Bigg[\gfarint{2 B\xp^{-1}\parens{\partial_{\xp}f^{\mu\nu}_{q,a,p}\retarg}_{t'}}\Bigg] \nonumber\\
\fl+ \sum_{q=0}^{\infty}\sum_{a=-\infty}^{\infty}\sum_{p=0}^{\infty}\A\Bigg[\gnearint{B\parens{B+1}\xp^{-2}f^{\mu\nu}_{q,a,p}\retarg}\Bigg] \nonumber\\
\fl+\sum_{q=0}^{\infty}\sum_{a=-\infty}^{\infty}\sum_{p=0}^{\infty}\A\Bigg[\gfarint{B\parens{B+1}\xp^{-2}f^{\mu\nu}_{q,a,p}\retarg}\Bigg].
\end{eqnarray}
Since there exists a common region in which all the integrals in each term on the RHS of  the above equation are analytic (depending on whether the region of integration is, this common region is either some right or left half-plane), using
$\sum_{q=0}^{\infty}\sum_{a=-\infty}^{\infty}\sum_{p=0}^{\infty}f^{\mu\nu}_{q,a,p}\finsarg=\bar{h}^{\mu\nu}\finsarg$,
one can write
\begin{eqnarray}\label{4.3.11}
\fl\xb\bar{h}^{\mu\nu}\finsarg&=\A\Bigg[\gnearint{\Box'\bar{h}^{\mu\nu}\retarg}\Bigg]\nonumber \\
&+\A\Bigg[\gfarint{\Box'\bar{h}^{\mu\nu}\retarg}\Bigg] \nonumber\\
&+\A\Bigg[\gnearint{2 B\xp^{-1}\parens{\partial_{\xp}\bar{h}^{\mu\nu}\retarg}_{t'}}\Bigg] \nonumber\\
&+\A\Bigg[\gfarint{2 B\xp^{-1}\parens{\partial_{\xp}\bar{h}^{\mu\nu}\retarg}_{t'}}\Bigg] \nonumber\\
&+\A\Bigg[\gnearint{B\parens{B+1}\xp^{-2} \bar{h}^{\mu\nu}\retarg}\Bigg] \nonumber\\
&+\A\Bigg[\gfarint{B\parens{B+1}\xp^{-2}\bar{h}^{\mu\nu}\retarg}\Bigg].
\end{eqnarray}
As
$\bar{h}^{\mu\nu}\finsarg$
is smooth in the near zone, and its structure outside of that region is of the form
$\sum_{q=0}^{\infty}\sum_{a=-\infty}^{\infty}\sum_{p=0}^{\infty}\hat{n}^Q\x^a\tlnx^p\hat{G}^{\mu\nu}_{Q,a,p}\targ$,
it can be readily shown that all the analytic continuations appearing on the RHS of 
(\ref{4.3.11})
are defined in some punctured neighborhood of
$B=0$.
Therefore, each of them possesses a Laurent expansion about
$B=0$.
Further, it is obvious that the LHS of this equation has a Taylor expansion about
$B=0$.
Thus, since the coefficients of
$B^n$
on both sides of the equation must be equal for each
$n$,
taking the coefficients of
$B^0$
into account, we reach
\begin{eqnarray}\label{4.3.12}
\fl\bar{h}^{\mu\nu}\finsarg \hspace{-10mm}&=\fpa\Bigg[\gnearint{\Box'\bar{h}^{\mu\nu}\retarg}\Bigg]\nonumber \\  
\hspace{-10mm}&+\fpa\Bigg[\gfarint{\Box'\bar{h}^{\mu\nu}\retarg}\Bigg]\nonumber \\  
\hspace{-10mm}&+\fpa\Bigg[\gnearint{2 B\xp^{-1}\parens{\partial_{\xp}\bar{h}^{\mu\nu}\retarg}_{t'}}\Bigg]\nonumber\\
\hspace{-10mm}&+\fpa\Bigg[\gfarint{2 B\xp^{-1}\parens{\partial_{\xp}\bar{h}^{\mu\nu}\retarg}_{t'}}\Bigg]\nonumber \\  
\hspace{-10mm}&+\fpa\Bigg[\gnearint{B\parens{B+1}\xp^{-2} \bar{h}^{\mu\nu}\retarg}\Bigg] \nonumber \\  
\hspace{-10mm}&+\fpa\Bigg[\gfarint{B\parens{B+1}\xp^{-2}\bar{h}^{\mu\nu}\retarg}\Bigg].
\end{eqnarray}
Taking
$B$
and
$B\parens{B+1}$
out of the integrals appearing respectively in the third and fifth terms on the RHS of
(\ref{4.3.12}),
the integrals left after doing that are analytic at
$B=0$
due to
$\bar{h}^{\mu\nu}\finsarg$
being smooth in the near zone. Thus, owing to the aforementioned complex coefficients, the third and fifth terms on the RHS of
(\ref{4.3.12})
vanish. This, together with
(\ref{1.2.31}), yield
\begin{eqnarray}\label{4.3.13}
\fl\bar{h}^{\mu\nu}\finsarg \hspace{-10mm} &\stareq \frac{16\pi G}{c^4}\fpa\gretint{\bar{\tau}^{\mu\nu}\retarg}\nonumber \\  
\hspace{-10mm}&+\fpa\Bigg[\gfarint{2 B\xp^{-1}\parens{\partial_{\xp}\bar{h}^{\mu\nu}\retarg}_{t'}}\Bigg]\nonumber \\   
\hspace{-10mm}&+\fpa\Bigg[\gfarint{B\parens{B+1}\xp^{-2}\bar{h}^{\mu\nu}\retarg}\Bigg].
\end{eqnarray}
Since everywhere outside the near zone we have
$\bar{h}^{\mu\nu}\finsarg=\mathcal{M}\parens{\bar{h}^{\mu\nu}}\finsarg$,
one can rewrite
(\ref{4.3.13})
as
\begin{eqnarray}\label{4.3.14}
\fl\bar{h}^{\mu\nu}\finsarg \hspace{-10mm}&\stareq \frac{16\pi G}{c^4}\fpa\gretint{\bar{\tau}^{\mu\nu}\retarg}\nonumber\\  
\hspace{-10mm}&+\fpa\Bigg[\gfarint{2 B\xp^{-1}\parens{\partial_{\xp}\mathcal{M}\parens{\bar{h}^{\mu\nu}}\retarg}_{t'}}\Bigg]\nonumber\\   
\hspace{-10mm}&+\fpa\Bigg[\gfarint{B\parens{B+1}\xp^{-2}\mathcal{M}\parens{\bar{h}^{\mu\nu}}\retarg}\Bigg].
\end{eqnarray}
If
$\x<\mathcal{R}$,
substituting the structure given in
(\ref{1.2.35})
for
$\mathcal{M}\parens{\bar{h}^{\mu\nu}}\finsarg$,
taking the derivatives and using the 3-dimensional Taylor expansion, we get
\begin{eqnarray}\label{4.3.15}
\fl\bar{h}^{\mu\nu}\finsarg \hspace{-10mm} &\stareq \frac{16\pi G}{c^4}\fpa\gretint{\bar{\tau}^{\mu\nu}\retarg}\nonumber \\  
\hspace{-10mm}&-\frac{1}{4\pi}\sum_{q=0}^{\infty}\sum_{a=-\infty}^{\infty}\sum_{p=0}^{\infty}\fp\parens{2aB+B\parens{B+1}}\cdot\A\int_{\mathcal{R}<\xp}\xpb\hat{n}'^{I_q}\xp^{a-2}\nonumber \\  
\hspace{-10mm}&\times \lnxp^p\left[\sum_{j=0}^{\infty}\frac{\parens{-1}^j}{j!}x^{I'_j}\partial'_{I'_j}\parens{\frac{\hat{G}^{\mu\nu}_{I_q,a,p}(t-\frac{\xp}{c})}{|\mathbf{x'}|}}\right]\ud^3\mathbf{x'}\nonumber \\   
\hspace{-10mm}&-\frac{1}{4\pi}\sum_{q=0}^{\infty}\sum_{a=-\infty}^{\infty}\sum_{p=1}^{\infty}\fp\parens{2pB}\cdot\A\int_{\mathcal{R}<\xp}\xpb\hat{n}'^{I_q}\xp^{a-2}\nonumber \\  
\hspace{-10mm}&\times \lnxp^{p-1}\left[\sum_{j=0}^{\infty}\frac{\parens{-1}^j}{j!}x^{I'_j}\partial'_{I'_j}\parens{\frac{\hat{G}^{\mu\nu}_{I_q,a,p}(t-\frac{\xp}{c})}{|\mathbf{x'}|}}\right]\ud^3\mathbf{x'}.
\end{eqnarray}
Now note that with the use of
(\ref{a.2.5})
and
(\ref{a.1.1}),
we have
\begin{eqnarray}\label{4.3.16}
\fl\sum_{j=0}^{\infty}\frac{\parens{-1}^j}{j!}x^{I'_j}\partial'_{I'_j}\parens{\frac{\hat{G}^{\mu\nu}_{I_q,a,p}(t-\frac{\xp}{c})}{|\mathbf{x'}|}} \nonumber \\ 
\fl =\sum_{j=0}^{\infty}\frac{\parens{-1}^j}{j!}x^{I'_j}\sum_{k=0}^{\left[\frac{j}{2}\right]}\frac{\parens{2j-4k+1}!!}{\parens{2j-2k+1}!!}\delta_{\{i'_1 i'_2} \cdots \delta_{i'_{2k-1}i'_{2k}}\hat{\partial}'_{i'_{2k+1}...i'_j\}}\Delta'^k\parens{\frac{\hat{G}^{\mu\nu}_{I_q,a,p}(t-\frac{\xp}{c})}{|\mathbf{x'}|}} \nonumber\\
\fl =\sum_{j=0}^{\infty} \sum_{k=0}^{\left[\frac{j}{2}\right]}\frac{\parens{-1}^j}{j!}\frac{\parens{2j-4k+1}!!}{\parens{2j-2k+1}!!}x^{I'_j}\delta_{\{i'_1 i'_2} \cdots \delta_{i'_{2k-1}i'_{2k}}\hat{\partial}'_{i'_{2k+1}...i'_j\}}\parens{\frac{{}^{\parens{2k}}\hat{G}^{\mu\nu}_{I_q,a,p}(t-\frac{\xp}{c})}{c^{2k}|\mathbf{x'}|}} \nonumber\\
\fl =\sum_{j=0}^{\infty} \sum_{k=0}^{\left[\frac{j}{2}\right]}\frac{\parens{-1}^j}{j!}\frac{\parens{2j-4k+1}!!}{\parens{2j-2k+1}!!}\frac{j!}{2^k k!\parens{j-2k}!}x^{I'_j}\delta_{i'_1 i'_2} \cdots \delta_{i'_{2k-1}i'_{2k}}\nonumber\\
\fl \times \hat{\partial}'_{i'_{2k+1}...i'_j}\parens{\frac{{}^{\parens{2k}}\hat{G}^{\mu\nu}_{I_q,a,p}(t-\frac{\xp}{c})}{c^{2k}|\mathbf{x'}|}} \nonumber\\
\fl =\sum_{j=0}^{\infty} \sum_{k=0}^{\left[\frac{j}{2}\right]}\frac{\parens{-1}^j}{2^k k!\parens{j-2k}!}\frac{\parens{2j-4k+1}!!}{\parens{2j-2k+1}!!}\x^{2k}x^{i'_{2k+1}...i'_j}\hat{\partial}'_{i'_{2k+1}...i'_j}\parens{\frac{{}^{\parens{2k}}\hat{G}^{\mu\nu}_{I_q,a,p}(t-\frac{\xp}{c})}{c^{2k}|\mathbf{x'}|}} \nonumber\\
\fl =\sum_{j=0}^{\infty} \sum_{k=0}^{\left[\frac{j}{2}\right]}\frac{\parens{-1}^{j-2k}}{2^k k!\parens{j-2k}!}\frac{\parens{2\parens{j-2k}+1}!!}{\parens{2\parens{j-2k}+2k+1}!!}\x^{2k}x^{I'_{J-2k}}\hat{\partial}'_{I'_{J-2k}}\parens{\frac{{}^{\parens{2k}}\hat{G}^{\mu\nu}_{I_q,a,p}(t-\frac{\xp}{c})}{c^{2k}|\mathbf{x'}|}} \nonumber\\
\fl =\sum_{\ell=0}^{\infty} \sum_{k=0}^{\infty}\frac{\parens{-1}^{\ell}}{\ell !}\frac{\parens{2\ell+1}!!}{\parens{2k}!!\parens{2\ell+2k+1}!!}\x^{2k}x^{I'_\ell}\hat{\partial}'_{I'_\ell}\parens{\frac{{}^{\parens{2k}}\hat{G}^{\mu\nu}_{I_q,a,p}(t-\frac{\xp}{c})}{c^{2k}|\mathbf{x'}|}},
\end{eqnarray}
where the third and sixth equalities have been obtained by arguments similar to those following
(\ref{4.3.2}).
By means of
(\ref{a.2.9}),
(\ref{4.3.16})
takes the form
\begin{eqnarray}\label{4.3.17}
\fl\sum_{j=0}^{\infty}\frac{\parens{-1}^j}{j!}x^{I'_j}\partial'_{I'_j}\parens{\frac{\hat{G}^{\mu\nu}_{I_q,a,p}(t-\frac{\xp}{c})}{|\mathbf{x'}|}}&=\sum_{\ell=0}^{\infty}\sum_{k=0}^{\infty}\sum_{i=0}^{\ell}\frac{1}{\ell !}\frac{\parens{\ell+i}!}{2^i i!\parens{\ell-i}!}\frac{\parens{2\ell+1}!!}{\parens{2k}!!\parens{2\ell+2k+1}!!}\nonumber \\  
&\times\frac{\x^{2k}\hat{x}^{I'_\ell}}{c^{2k+\ell-i}}\bracks{\hat{n}'_{I'_\ell}\frac{{}^{\parens{2k+\ell-i}}\hat{G}^{\mu\nu}_{I_q,a,p}(t-\frac{\xp}{c})}{\xp^{i+1}}}.
\end{eqnarray}
Combining
(\ref{4.3.15}) and (\ref{4.3.17})
and using
(\ref{a.1.3}),
we reach
\begin{eqnarray}\label{4.3.18}
\fl\bar{h}^{\mu\nu}\finsarg \hspace{-10mm} &\stareq \frac{16\pi G}{c^4}\fpa\gretint{\bar{\tau}^{\mu\nu}\retarg}\nonumber \\  
\hspace{-10mm}&-\frac{1}{4\pi}\sum_{\ell=0}^{\infty}\sum_{k=0}^{\infty}\sum_{i=0}^{\ell}\sum_{q=0}^{\infty}\sum_{a=-\infty}^{\infty}\sum_{p=0}^{\infty}\frac{1}{\ell !}\frac{\parens{\ell+i}!}{2^i i!\parens{\ell-i}!}\frac{\parens{2\ell+1}!!}{\parens{2k}!!\parens{2\ell+2k+1}!!}\frac{\x^{2k}\hat{x}^{I'_\ell}}{c^{2k+\ell-i}} \nonumber \\  
\hspace{-10mm}&\times \int \hat{n}'^{I_q}\hat{n}'_{I'_\ell}\ud\Omega'\fp\parens{2aB+B\parens{B+1}}\frac{\partial^p}{\partial B^p}\Bigg[\frac{1}{r_0^B}\A\int_{\mathcal{R}}^{\infty}\xp^{B+a-i-1}\nonumber \\   
\hspace{-10mm}&\times {}^{\parens{2k+\ell-i}}\hat{G}^{\mu\nu}_{I_q,a,p}(t-\frac{\xp}{c})\ud\xp\Bigg]\nonumber \\ 
\hspace{-10mm}&-\frac{1}{4\pi}\sum_{\ell=0}^{\infty}\sum_{k=0}^{\infty}\sum_{i=0}^{\ell}\sum_{q=0}^{\infty}\sum_{a=-\infty}^{\infty}\sum_{p=1}^{\infty}\frac{1}{\ell !}\frac{\parens{\ell+i}!}{2^i i!\parens{\ell-i}!}\frac{\parens{2\ell+1}!!}{\parens{2k}!!\parens{2\ell+2k+1}!!}\frac{\x^{2k}\hat{x}^{I'_\ell}}{c^{2k+\ell-i}} \nonumber \\  
\hspace{-10mm}&\times \int \hat{n}'^{I_q}\hat{n}'_{I'_\ell}\ud\Omega'\fp\parens{2pB}\frac{\partial^{p-1}}{\partial B^{p-1}}\Bigg[\frac{1}{r_0^B}\A\int_{\mathcal{R}}^{\infty}\xp^{B+a-i-1}\nonumber \\   
\hspace{-10mm}&\times {}^{\parens{2k+\ell-i}}\hat{G}^{\mu\nu}_{I_q,a,p}(t-\frac{\xp}{c})\ud\xp\Bigg],
\end{eqnarray}
where we have also used the identity theorem. Assuming that the (1-variable) Taylor expansion of
$\bar{h}^{\mu\nu}\finsarg$
and hence that of
$\hat{G}^{\mu\nu}_{I_q,a,p}\targ$
have infinite radii of convergence about any arbitrary point
$t$,
we can write
\begin{equation}\label{onlyAC-first part}
\hat{G}^{\mu\nu}_{I_q,a,p}(t-\frac{\xp}{c})=\sum_{j=0}^\infty\frac{\parens{-1}^j}{c^j j!} \;{}^{\parens{j}}\hat{G}^{\mu\nu}_{I_q,a,p}\targ\xp^j,
\end{equation}
and noting that
$\partial_t \hat{G}^{\mu\nu}_{I_q,a,p}(t-\tfrac{\xp}{c})= {}^{\parens{1}}\hat{G}^{\mu\nu}_{I_q,a,p}(t-\tfrac{\xp}{c})$,
we find
\begin{eqnarray}\label{onlyAC-second part}
\A \int_\mathcal{R}^\infty\xp^{B+a-i-1}\;{}^{\parens{2k+\ell-i}}\hat{G}^{\mu\nu}_{I_q,a,p}(t-\frac{\xp}{c})\ud\xp\nonumber \\  
=-\sum_{j=0}^\infty\frac{\parens{-1}^j}{c^j j!}\frac{\mathcal{R}^{B+a-i+j}}{B+a-i+j}\; {}^{\parens{2k+\ell-i+j}}\hat{G}^{\mu\nu}_{I_q,a,p}\targ\nonumber \\  
=-\A \int^\mathcal{R}_0\xp^{B+a-i-1}\;{}^{\parens{2k+\ell-i}}\hat{G}^{\mu\nu}_{I_q,a,p}(t-\frac{\xp}{c})\ud\xp.
\end{eqnarray}
As a result, one is allowed to make the replacement
$\A\int_{\mathcal{R}}^{\infty} \to -\A\int^{\mathcal{R}}_0$
in
(\ref{4.3.18}),
and therefore, after reversing the order of computations resulting in that equation, reaches
\begin{eqnarray}\label{4.3.31}
\fl\bar{h}^{\mu\nu}\finsarg \hspace{-10mm}&\stareq\frac{16\pi G}{c^4}\fpa\gretint{\bar{\tau}^{\mu\nu}\retarg}\nonumber \\  
\hspace{-10mm}&+\frac{1}{4\pi}\sum_{\ell=0}^{\infty}\sum_{k=0}^{\infty}\sum_{i=0}^{\ell}\frac{1}{\ell !}\frac{\parens{\ell+i}!}{2^i i!\parens{\ell-i}!}\frac{\parens{2\ell+1}!!}{\parens{2k}!!\parens{2\ell+2k+1}!!}\frac{\x^{2k}\hat{x}^{L}}{c^{2k+\ell-i}} \nonumber\\  
\hspace{-10mm}&\times\fpa\int_{\xp<\mathcal{R}}\xpb\frac{\hat{x}'_{L}}{\xp^{\ell+i+1}}\partial_t^{2k+\ell-i}\bigg[2B\xp^{-1}\parens{\partial_{\xp}\mathcal{M}\parens{\bar{h}^{\mu\nu}}(u',\mathbf{x'})}_{u'}\nonumber\\  
\hspace{-10mm}&+B\parens{B+1}\xp^{-2}\mathcal{M}\parens{\bar{h}^{\mu\nu}}(u',\mathbf{x'})\bigg]\ud^3\mathbf{x'},
\end{eqnarray}
where
$u'=t-\tfrac{\xp}{c}$.
In the case where
$\mathcal{R}<\x<\infty$,
(\ref{4.3.14})
can be rewritten as
\begin{eqnarray}\label{4.3.32}
\fl\bar{h}^{\mu\nu}\finsarg \hspace{-10mm}&\stareq \frac{16\pi G}{c^4}\fpa\gretint{\bar{\tau}^{\mu\nu}\retarg}\nonumber \\  
\hspace{-10mm}&+\fpa\Bigg[-\frac{1}{4\pi}\int_{\mathcal{R}<|\mathbf{x'}|<\x}\xpb \frac{2 B\xp^{-1}\parens{\partial_{\xp}\mathcal{M}\parens{\bar{h}^{\mu\nu}}\retarg}_{t'}}{|\mathbf{x}-\mathbf{x'}|}\ud^3\mathbf{x'}\Bigg]\nonumber \\  
\hspace{-10mm}&+\fpa\Bigg[-\frac{1}{4\pi}\int_{\mathcal{R}<|\mathbf{x'}|<\x}\xpb \frac{B\parens{B+1}\xp^{-2}\mathcal{M}\parens{\bar{h}^{\mu\nu}}\retarg}{|\mathbf{x}-\mathbf{x'}|}\ud^3\mathbf{x'}\Bigg]\nonumber \\  
\hspace{-10mm}&+\fpa\Bigg[-\frac{1}{4\pi}\int_{\x<\xp}\xpb \frac{2 B\xp^{-1}\parens{\partial_{\xp}\mathcal{M}\parens{\bar{h}^{\mu\nu}}\retarg}_{t'}}{|\mathbf{x}-\mathbf{x'}|}\ud^3\mathbf{x'}\Bigg]\nonumber \\  
\hspace{-10mm}&+\fpa\Bigg[-\frac{1}{4\pi}\int_{\x<\xp}\xpb \frac{B\parens{B+1}\xp^{-2}\mathcal{M}\parens{\bar{h}^{\mu\nu}}\retarg}{|\mathbf{x}-\mathbf{x'}|}\ud^3\mathbf{x'}\Bigg].
\end{eqnarray}
Factorizing
$B$
and
$B\parens{B+1}$
out of the integrals in the second and third lines respectively, we are left with the integrals which can be straightforwardly shown to be analytic at
$B=0$,
and the second and third terms on the RHS of
(\ref{4.3.32})
are thereby, due to the complex coefficients factorized earlier, zero. Further, since the integrals appearing in the fourth and fifth terms on the RHS are over the region
$\xp>\x$,
by proceeding along the same lines as in the case where
$\x<\mathcal{R}$,
we can obtain
\begin{eqnarray}\label{4.3.33}
\fl\bar{h}^{\mu\nu}\finsarg \hspace{-10mm}&\stareq \frac{16\pi G}{c^4}\fpa\gretint{\bar{\tau}^{\mu\nu}\retarg}\nonumber \\  
\hspace{-10mm}&-\frac{1}{4\pi}\sum_{\ell=0}^{\infty}\sum_{k=0}^{\infty}\sum_{i=0}^{\ell}\sum_{q=0}^{\infty}\sum_{a=-\infty}^{\infty}\sum_{p=0}^{\infty}\frac{1}{\ell !}\frac{\parens{\ell+i}!}{2^i i!\parens{\ell-i}!}\frac{\parens{2\ell+1}!!}{\parens{2k}!!\parens{2\ell+2k+1}!!}\frac{\x^{2k}\hat{x}^{I'_\ell}}{c^{2k+\ell-i}} \nonumber \\  
\hspace{-10mm}&\times \int \hat{n}'^{I_q}\hat{n}'_{I'_\ell}\ud\Omega'\fp\parens{2aB+B\parens{B+1}}\frac{\partial^p}{\partial B^p}\Bigg[\frac{1}{r_0^B}\A\int_{\x}^{\infty}\xp^{B+a-i-1}\nonumber \\   
\hspace{-10mm}&\times {}^{\parens{2k+\ell-i}}\hat{G}^{\mu\nu}_{I_q,a,p}(t-\frac{\xp}{c})\ud\xp\Bigg]\nonumber \\ 
\hspace{-10mm}&-\frac{1}{4\pi}\sum_{\ell=0}^{\infty}\sum_{k=0}^{\infty}\sum_{i=0}^{\ell}\sum_{q=0}^{\infty}\sum_{a=-\infty}^{\infty}\sum_{p=1}^{\infty}\frac{1}{\ell !}\frac{\parens{\ell+i}!}{2^i i!\parens{\ell-i}!}\frac{\parens{2\ell+1}!!}{\parens{2k}!!\parens{2\ell+2k+1}!!}\frac{\x^{2k}\hat{x}^{I'_\ell}}{c^{2k+\ell-i}} \nonumber \\  
\hspace{-10mm}&\times \int \hat{n}'^{I_q}\hat{n}'_{I'_\ell}\ud\Omega'\fp\parens{2pB}\frac{\partial^{p-1}}{\partial B^{p-1}}\Bigg[\frac{1}{r_0^B}\A\int_{\x}^{\infty}\xp^{B+a-i-1}\nonumber \\   
\hspace{-10mm}&\times {}^{\parens{2k+\ell-i}}\hat{G}^{\mu\nu}_{I_q,a,p}(t-\frac{\xp}{c})\ud\xp\Bigg].
\end{eqnarray}
One can easily show that the
$\int_{\mathcal{R}}^{\x}$
counterpart of the radial integral appearing in the second and third terms on the RHS is entire. Therefore, due to the coefficients
$2aB+B\parens{B+1}$
and
$2pB$,
we are entitled to write the RHS of the above equation as in
(\ref{4.3.18}),
and this means that even in the case under study one can obtain the same equation as
(\ref{4.3.31}),
thereby extending the domain of validity of that equation to be the whole space (except for
$\x \to \infty$).

Now it is time to use the matching equation. Replacing
$\mathcal{M}\parens{\bar{h}^{\mu\nu}}$
by
$\overline{\mathcal{M}\parens{h^{\mu\nu}}}$
in 
(\ref{4.3.31}),
we get
\begin{eqnarray}\label{4.3.36}
\fl\bar{h}^{\mu\nu}\finsarg \hspace{-10mm}&\stareq \frac{16\pi G}{c^4}\fpa\gretint{\bar{\tau}^{\mu\nu}\retarg}\nonumber \\  
\hspace{-10mm}&+\frac{1}{4\pi}\sum_{\ell=0}^{\infty}\sum_{k=0}^{\infty}\sum_{i=0}^{\ell}\frac{1}{\ell !}\frac{\parens{\ell+i}!}{2^i i!\parens{\ell-i}!}\frac{\parens{2\ell+1}!!}{\parens{2k}!!\parens{2\ell+2k+1}!!}\frac{\x^{2k}\hat{x}^{L}}{c^{2k+\ell-i}} \nonumber\\  
\hspace{-10mm}&\times\fpa\int_{\xp<\mathcal{R}}\xpb\frac{\hat{x}'_{L}}{\xp^{\ell+i+1}}\partial_t^{2k+\ell-i}\bigg[2B\xp^{-1}\parens{\partial_{\xp}\overline{\mathcal{M}\parens{h^{\mu\nu}}}(u',\mathbf{x'})}_{u'}\nonumber \\  
\hspace{-10mm}&+B\parens{B+1}\xp^{-2}\overline{\mathcal{M}\parens{h^{\mu\nu}}}(u',\mathbf{x'})\bigg]\ud^3\mathbf{x'},
\end{eqnarray}
and since we are able to demand that, as
$\x \to 0$,
the function
$R'^{\;\mu\nu}\finsarg$
be so that the integrals
$\int_{\xp<\mathcal{R}}\txpb\tbfrac{\hat{x}'_{L}}{\xp^{\ell+i+1}}\partial_t^{2k+\ell-i}\bracks{\xp^{-1}\parens{\partial_{\xp}R'^{\;\mu\nu}(u',\mathbf{x'})}_{u'}}\ud^3\mathbf{x'}$
and
$\int_{\xp<\mathcal{R}}\txpb\tbfrac{\hat{x}'_{L}}{\xp^{\ell+i+1}}\partial_t^{2k+\ell-i}\bracks{\xp^{-2}R'^{\;\mu\nu}(u',\mathbf{x'})}\ud^3\mathbf{x'}$
are analytic in some neighborhood of
$B=0$
(including
$B=0$),
we can write
\begin{eqnarray}\label{4.3.37}
\fl\bar{h}^{\mu\nu}\finsarg\hspace{-10mm}&\stareq \frac{16\pi G}{c^4}\fpa\gretint{\bar{\tau}^{\mu\nu}\retarg}\nonumber \\  
\hspace{-10mm}&+\frac{1}{4\pi}\sum_{\ell=0}^{\infty}\sum_{k=0}^{\infty}\sum_{i=0}^{\ell}\frac{1}{\ell !}\frac{\parens{\ell+i}!}{2^i i!\parens{\ell-i}!}\frac{\parens{2\ell+1}!!}{\parens{2k}!!\parens{2\ell+2k+1}!!}\frac{\x^{2k}\hat{x}^{L}}{c^{2k+\ell-i}} \nonumber\\  
\hspace{-10mm}&\times\fpa\int_{\xp<\mathcal{R}}\xpb\frac{\hat{x}'_{L}}{\xp^{\ell+i+1}}\partial_t^{2k+\ell-i}\bigg[2B\xp^{-1}\parens{\partial_{\xp}\mathcal{M}\parens{h^{\mu\nu}}(u',\mathbf{x'})}_{u'}\nonumber \\  
\hspace{-10mm}&+B\parens{B+1}\xp^{-2}\mathcal{M}\parens{h^{\mu\nu}}(u',\mathbf{x'})\bigg]\ud^3\mathbf{x'}.
\end{eqnarray}
The integrals
$\int_{\mathcal{R}<\xp}\txpb\tbfrac{\hat{x}'_{L}}{\xp^{\ell+i+1}}\partial_t^{2k+\ell-i}\bracks{\xp^{-1}\partial_{\xp}\mathcal{M}\parens{h^{\mu\nu}_{\mathrm{AS}}}\xparg}\ud^3\mathbf{x'}$
and
$\int_{\mathcal{R}<\xp}\txpb\tbfrac{\hat{x}'_{L}}{\xp^{\ell+i+1}}\partial_t^{2k+\ell-i}\bracks{\xp^{-2}\mathcal{M}\parens{h^{\mu\nu}_{\mathrm{AS}}}\xparg}\ud^3\mathbf{x'}$
are identically zero if
$k>0$
or
$ i < \ell$. In the case where
$k=0$
and
$ i = \ell$,
due to the maximal power of
$\x$
in the structure of
$\mathcal{M}\parens{h^{\mu\nu}_{\mathrm{AS}}}\xarg$
being
$-1$,
these integrals are analytic in some neighborhood of
$B=0$
(including
$B=0$).
Moreover, the integrals
$\int_{\mathcal{R}<\xp}\txpb\tbfrac{\hat{x}'_{L}}{\xp^{\ell+i+1}}\partial_t^{2k+\ell-i}\bracks{\xp^{-1}\parens{\partial_{\xp}\mathcal{M}\parens{h^{\mu\nu}_{\mathrm{PZ}}}(u',\mathbf{x'})}_{u'}}\ud^3\mathbf{x'}$
and
$\int_{\mathcal{R}<\xp}\txpb\tbfrac{\hat{x}'_{L}}{\xp^{\ell+i+1}}\partial_t^{2k+\ell-i}\bracks{\xp^{-2}\mathcal{M}\parens{h^{\mu\nu}_{\mathrm{PZ}}}(u',\mathbf{x'})}\ud^3\mathbf{x'}$
are analytic in some neighborhood of
$B=0$
(including
$B=0$)
due to
$\mathcal{M}\parens{h^{\mu\nu}_{\mathrm{PZ}}}\finsarg$
being past-zero.
Therefore, owing to the coefficients
$B$
and
$B\parens{B+1}$,
we have
\begin{eqnarray}\label{4.3.38}
\fl\fpa\int_{\mathcal{R}<\xp}\xpb\frac{\hat{x}'_{L}}{\xp^{\ell+i+1}}\partial_t^{2k+\ell-i}
&\bigg[2B\xp^{-1}\parens{\partial_{\xp}\mathcal{M}\parens{h^{\mu\nu}}(u',\mathbf{x'})}_{u'} \nonumber \\ 
&+ B\parens{B+1}\xp^{-2}\mathcal{M}\parens{h^{\mu\nu}}(u',\mathbf{x'})\bigg]\ud^3\mathbf{x'}=0.\nonumber\\
\end{eqnarray}
Adding this vanishing expression to its counterpart on the RHS of
(\ref{4.3.37}),
it becomes
\begin{eqnarray}\label{4.3.39}
\fl\bar{h}^{\mu\nu}\finsarg \hspace{-10mm}&\stareq \frac{16\pi G}{c^4}\fpa\gretint{\bar{\tau}^{\mu\nu}\retarg}\nonumber \\  
\hspace{-10mm}&+\frac{1}{4\pi}\sum_{\ell=0}^{\infty}\sum_{k=0}^{\infty}\sum_{i=0}^{\ell}\frac{1}{\ell !}\frac{\parens{\ell+i}!}{2^i i!\parens{\ell-i}!}\frac{\parens{2\ell+1}!!}{\parens{2k}!!\parens{2\ell+2k+1}!!}\frac{\x^{2k}\hat{x}^{L}}{c^{2k+\ell-i}} \nonumber\\  
\hspace{-10mm}&\times\fpa\int_{\xp<\mathcal{R}}\xpb\frac{\hat{x}'_{L}}{\xp^{\ell+i+1}}\partial_t^{2k+\ell-i}\bigg[2B\xp^{-1}\parens{\partial_{\xp}\mathcal{M}\parens{h^{\mu\nu}}(u',\mathbf{x'})}_{u'}\nonumber \\  
\hspace{-10mm}&+B\parens{B+1}\xp^{-2}\mathcal{M}\parens{h^{\mu\nu}}(u',\mathbf{x'})\bigg]\ud^3\mathbf{x'}\nonumber\\
\hspace{-10mm}&+\frac{1}{4\pi}\sum_{\ell=0}^{\infty}\sum_{k=0}^{\infty}\sum_{i=0}^{\ell}\frac{1}{\ell !}\frac{\parens{\ell+i}!}{2^i i!\parens{\ell-i}!}\frac{\parens{2\ell+1}!!}{\parens{2k}!!\parens{2\ell+2k+1}!!}\frac{\x^{2k}\hat{x}^{L}}{c^{2k+\ell-i}} \nonumber\\  
\hspace{-10mm}&\times\fpa\int_{\mathcal{R}<\xp}\xpb\frac{\hat{x}'_{L}}{\xp^{\ell+i+1}}\partial_t^{2k+\ell-i}\bigg[2B\xp^{-1}\parens{\partial_{\xp}\mathcal{M}\parens{h^{\mu\nu}}(u',\mathbf{x'})}_{u'}\nonumber \\  
\hspace{-10mm}&+B\parens{B+1}\xp^{-2}\mathcal{M}\parens{h^{\mu\nu}}(u',\mathbf{x'})\bigg]\ud^3\mathbf{x'}.
\end{eqnarray}
We will discuss the reason behind what we did later. Now consider the following equation:
\begin{eqnarray}\label{4.3.40-first step}
\Box \Bigg[\xb \mathcal{M}\parens{h^{\mu\nu}}\finsarg\Bigg] \nonumber\\
=\xb \Box\mathcal{M}\parens{h^{\mu\nu}}\finsarg+ \xb\bigg[2B\x^{-1}\partial_{\x}\mathcal{M}\parens{h^{\mu\nu}}\finsarg\nonumber \\  
+B\parens{B+1}\x^{-2}\mathcal{M}\parens{h^{\mu\nu}}\finsarg\bigg],
\end{eqnarray}
which, after using
(\ref{1.2.22}),
can be rearranged as
\begin{eqnarray}\label{4.3.40-second step}
\xb\bigg[2B\x^{-1}\partial_{\x}\mathcal{M}\parens{h^{\mu\nu}}\finsarg+B\parens{B+1}\x^{-2}\mathcal{M}\parens{h^{\mu\nu}}\finsarg\bigg] \nonumber \\ 
=\Delta\Bigg[\xb\mathcal{M}\parens{h^{\mu\nu}}\finsarg\Bigg]-\frac{1}{c^2}\xb\partial_t^2 \mathcal{M}\parens{h^{\mu\nu}}\finsarg \nonumber\\
 -\xb\Lambda^{\mu\nu}\parens{\mathcal{M}\parens{h}}\finsarg,
\end{eqnarray}
and, defining
$u=t-\tfrac{\x}{c}$,
it is apparent that we can write
\begin{eqnarray}\label{4.3.40}
\xb\bigg[2B\x^{-1}\parens{\partial_{\x}\mathcal{M}\parens{h^{\mu\nu}}(u,\mathbf{x})}_u + B\parens{B+1}\x^{-2}\mathcal{M}\parens{h^{\mu\nu}}(u,\mathbf{x})\bigg] \nonumber \\ 
=\Bigg(\Delta\Bigg[\xb\mathcal{M}\parens{h^{\mu\nu}}(u,\mathbf{x})\Bigg]\Bigg)_u - \frac{1}{c^2}\xb \partial_u^2\mathcal{M}\parens{h^{\mu\nu}}(u,\mathbf{x}) \nonumber\\
 -\xb\Lambda^{\mu\nu}\parens{\mathcal{M}\parens{h}}(u,\mathbf{x}).
\end{eqnarray}
With the use of the above equation,
(\ref{4.3.39})
can be rewritten as
\begin{eqnarray}\label{4.3.41}
\fl\bar{h}^{\mu\nu}\finsarg \hspace{-10mm}&\stareq \frac{16\pi G}{c^4}\fpa\gretint{\bar{\tau}^{\mu\nu}\retarg}\nonumber \\  
\hspace{-10mm} &-\frac{1}{4\pi}\sum_{\ell=0}^{\infty}\sum_{k=0}^{\infty}\sum_{i=0}^{\ell}\frac{1}{\ell !}\frac{\parens{\ell+i}!}{2^i i!\parens{\ell-i}!}\frac{\parens{2\ell+1}!!}{\parens{2k}!!\parens{2\ell+2k+1}!!}\frac{\x^{2k}\hat{x}^{L}}{c^{2k+\ell-i}} \nonumber\\  
 \hspace{-10mm}&\times\Bigg[\fpa\int_{\xp<\mathcal{R}}\xpb\frac{\hat{x}'_{L}}{\xp^{\ell+i+1}}\partial_t^{2k+\ell-i}\Lambda^{\mu\nu}\parens{\mathcal{M}\parens{h}}(u',\mathbf{x'})\ud^3\mathbf{x'}\nonumber \\  
 \hspace{-10mm}&+\fpa\int_{\mathcal{R}<\xp}\xpb\frac{\hat{x}'_{L}}{\xp^{\ell+i+1}}\partial_t^{2k+\ell-i}\Lambda^{\mu\nu}\parens{\mathcal{M}\parens{h}}(u',\mathbf{x'})\ud^3\mathbf{x'}\Bigg]\nonumber\\
\hspace{-10mm} &+\frac{1}{4\pi}\sum_{\ell=0}^{\infty}\sum_{k=0}^{\infty}\sum_{i=0}^{\ell}\frac{1}{\ell !}\frac{\parens{\ell+i}!}{2^i i!\parens{\ell-i}!}\frac{\parens{2\ell+1}!!}{\parens{2k}!!\parens{2\ell+2k+1}!!}\frac{\x^{2k}\hat{x}^{L}}{c^{2k+\ell-i}} \nonumber\\  
\hspace{-10mm} &\times\partial_t^{2k+\ell-i}\Bigg[\fpa\int_{\xp<\mathcal{R}}\frac{\hat{x}'_{L}}{\xp^{\ell+i+1}}\Bigg\{\Bigg(\Delta'\Bigg[\xpb\mathcal{M}\parens{h^{\mu\nu}}(u',\mathbf{x'})\Bigg]\Bigg)_{u'}\nonumber \\  
\hspace{-10mm} & - \frac{1}{c^2} \xpb\partial_{u'}^2\mathcal{M}\parens{h^{\mu\nu}}(u',\mathbf{x'})\Bigg\}\ud^3\mathbf{x'}\nonumber\\
\hspace{-10mm} &+\fpa\int_{\mathcal{R}<\xp}\frac{\hat{x}'_{L}}{\xp^{\ell+i+1}}\Bigg\{\Bigg(\Delta'\Bigg[\xpb\mathcal{M}\parens{h^{\mu\nu}}(u',\mathbf{x'})\Bigg]\Bigg)_{u'}\nonumber \\  
\hspace{-10mm} & - \frac{1}{c^2} \xpb\partial_{u'}^2\mathcal{M}\parens{h^{\mu\nu}}(u',\mathbf{x'})\Bigg\}\ud^3\mathbf{x'}\Bigg].
\end{eqnarray}
It is not difficult to show that
\begin{eqnarray}\label{4.3.44}
\fl\Bigg(\Delta'\Bigg[\xpb\mathcal{M}\parens{h^{\mu\nu}}(u',\mathbf{x'})\Bigg]\Bigg)_{u'} - \frac{1}{c^2}\xpb \partial_{u'}^2\mathcal{M}\parens{h^{\mu\nu}}(u',\mathbf{x'})\nonumber\\ 
\fl=\Delta'\Bigg[\xpb\mathcal{M}\parens{h^{\mu\nu}}(u',\mathbf{x'})\Bigg] + \frac{2}{c}\partial'^j\Bigg[\xpb \xp^{-1} x'_j \partial_{u'}\mathcal{M}\parens{h^{\mu\nu}}(u',\mathbf{x'})\Bigg]\nonumber\\
\fl - \frac{2}{c}\xpb \xp^{-1}\partial_{u'}\mathcal{M}\parens{h^{\mu\nu}}(u',\mathbf{x'}),
\end{eqnarray}
and hence, introducing
\begin{eqnarray}\label{D}
D_i=\cases{\{\mathbf{x'}\in \mathbb{R }^3 \left|\right. \xp<\mathcal{R}\}, & $i=1$,\\
\{\mathbf{x'}\in \mathbb{R }^3 \left|\right. \mathcal{R}<\xp\}, & $i=2$,\\}
\end{eqnarray}
by means of the Gauss' and Green's theorems we can write
\begin{eqnarray}\label{4.3.45}
\A\int_{D_i}\frac{\hat{x}'_{L}}{\xp^{\ell+i+1}}\Bigg\{\Bigg(\Delta'\Bigg[\xpb\mathcal{M}\parens{h^{\mu\nu}}(u',\mathbf{x'})\Bigg]\Bigg)_{u'} \nonumber\\
- \frac{1}{c^2}\xpb \partial_{u'}^2\mathcal{M}\parens{h^{\mu\nu}}(u',\mathbf{x'}) \Bigg\}\ud^3\mathbf{x'} \nonumber \\ 
=\int_{D_i}\xpb\Bigg[\Delta' \left(\frac{\hat{x}'_{L}}{\xp^{\ell+i+1}}\right)\Bigg]\mathcal{M}\parens{h^{\mu\nu}}(u',\mathbf{x'})\ud^3\mathbf{x'} \nonumber\\
 - \frac{2}{c}\int_{D_i}\xpb\xp^{-1} x'^j \Bigg[\partial'_j\left(\frac{\hat{x}'_{L}}{\xp^{\ell+i+1}}\right)\Bigg]\partial_{u'}\mathcal{M}\parens{h^{\mu\nu}}(u',\mathbf{x'})\ud^3\mathbf{x'}\nonumber\\
  - \frac{2}{c}\int_{D_i}\xpb\frac{\hat{x}'_{L}}{\xp^{\ell+i+2}}\partial_{u'}\mathcal{M}\parens{h^{\mu\nu}}(u',\mathbf{x'})\ud^3\mathbf{x'}\nonumber\\
+\int_{\partial D_i} \frac{\hat{x}'_{L}}{\xp^{\ell+i+1}}\partial'^k\Bigg[\xpb \mathcal{M}\parens{h^{\mu\nu}}(u',\mathbf{x'})\Bigg]\ud\sigma'_{\parens{i}k}\nonumber\\
 -\int_{\partial D_i} \Bigg[\partial'^k\left(\frac{\hat{x}'_{L}}{\xp^{\ell+i+1}}\right)\Bigg]\xpb \mathcal{M}\parens{h^{\mu\nu}}(u',\mathbf{x'})\ud\sigma'_{\parens{i}k}\nonumber\\
 +\frac{2}{c}\int_{\partial D_i} \frac{x'^k \hat{x}'_{L}}{\xp^{\ell+i+2}}\xpb\partial_{u'}\mathcal{M}\parens{h^{\mu\nu}}(u',\mathbf{x'})\ud\sigma'_{\parens{i}k},
\end{eqnarray}
providing we restrict ourselves to the original domain of analyticity of the LHS, namely in either some right or left half-plane depending on whether
$ i $
is. For
$i=1$
we have
\begin{eqnarray}\label{4.3.46}
\A\int_{\xp<\mathcal{R}}\frac{\hat{x}'_{L}}{\xp^{\ell+i+1}} \Bigg\{\Bigg(\Delta'\Bigg[\xpb\mathcal{M}\parens{h^{\mu\nu}}(u',\mathbf{x'})\Bigg]\Bigg)_{u'} \nonumber\\
- \frac{1}{c^2}\xpb \partial_{u'}^2\mathcal{M}\parens{h^{\mu\nu}}(u',\mathbf{x'}) \Bigg\}\ud^3\mathbf{x'} \nonumber \\ 
=\int_{\xp<\mathcal{R}} \xpb\Bigg[\Delta' \left(\frac{\hat{x}'_{L}}{\xp^{\ell+i+1}}\right)\Bigg] \mathcal{M}\parens{h^{\mu\nu}} (u',\mathbf{x'})\ud^3\mathbf{x'}\nonumber\\
 -\frac{2}{c}\int_{\xp<\mathcal{R}} \xpb\xp^{-1} x'^j \Bigg[\partial'_j\left(\frac{\hat{x}'_{L}}{\xp^{\ell+i+1}}\right)\Bigg]\partial_{u'}\mathcal{M}\parens{h^{\mu\nu}}(u',\mathbf{x'})\ud^3\mathbf{x'}\nonumber\\
- \frac{2}{c}\int_{\xp<\mathcal{R}} \xpb\frac{\hat{x}'_{L}}{\xp^{\ell+i+2}}\partial_{u'}\mathcal{M}\parens{h^{\mu\nu}}(u',\mathbf{x'})\ud^3\mathbf{x'}\nonumber\\
+\int_{\xp=\mathcal{R}} \frac{\hat{x}'_{L}}{\xp^{\ell+i+1}}\partial'^k\Bigg[\xpb \mathcal{M}\parens{h^{\mu\nu}}(u',\mathbf{x'})\Bigg]\ud\sigma'_{\parens{1}k}\nonumber\\
 -\int_{\xp=\mathcal{R}} \Bigg[\partial'^k\left(\frac{\hat{x}'_{L}}{\xp^{\ell+i+1}}\right)\Bigg]\xpb \mathcal{M}\parens{h^{\mu\nu}}(u',\mathbf{x'})\ud\sigma'_{\parens{1}k}\nonumber\\
 +\frac{2}{c}\int_{\xp=\mathcal{R}} \frac{x'^k \hat{x}'_{L}}{\xp^{\ell+i+2}}\xpb\partial_{u'}\mathcal{M}\parens{h^{\mu\nu}}(u',\mathbf{x'})\ud\sigma'_{\parens{1}k}.
\end{eqnarray}
As a result of the identity theorem, one can equate the LHS of the above equation
and the analytic continuation of its RHS wherever they are both defined. This, together with the fact that the surface integrals are entire, yield
\begin{eqnarray}\label{4.3.47}
\A\int_{\xp<\mathcal{R}}\frac{\hat{x}'_{L}}{\xp^{\ell+i+1}}\Bigg\{\Bigg(\Delta'\Bigg[\xpb\mathcal{M}\parens{h^{\mu\nu}}(u',\mathbf{x'})\Bigg]\Bigg)_{u'}\nonumber\\
- \frac{1}{c^2}\xpb \partial_{u'}^2\mathcal{M}\parens{h^{\mu\nu}}(u',\mathbf{x'}) \Bigg\}\ud^3\mathbf{x'} \nonumber \\ 
=\A\int_{\xp<\mathcal{R}}\xpb\Bigg[\Delta'\left(\frac{\hat{x}'_{L}}{\xp^{\ell+i+1}}\right)\Bigg]\mathcal{M}\parens{h^{\mu\nu}}(u',\mathbf{x'})\ud^3\mathbf{x'} \nonumber\\
 -\frac{2}{c} \A\int_{\xp<\mathcal{R}} \xpb\xp^{-1} x'^j \Bigg[\partial'_j\left(\frac{\hat{x}'_{L}}{\xp^{\ell+i+1}}\right)\Bigg]\partial_{u'}\mathcal{M}\parens{h^{\mu\nu}}(u',\mathbf{x'})\ud^3\mathbf{x'}\nonumber\\
 - \frac{2}{c}\A\int_{\xp<\mathcal{R}} \xpb\frac{\hat{x}'_{L}}{\xp^{\ell+i+2}}\partial_{u'}\mathcal{M}\parens{h^{\mu\nu}}(u',\mathbf{x'})\ud^3\mathbf{x'}\nonumber\\
+\int_{\xp=\mathcal{R}} \frac{\hat{x}'_{L}}{\xp^{\ell+i+1}}\partial'^k\Bigg[\xpb \mathcal{M}\parens{h^{\mu\nu}}(u',\mathbf{x'})\Bigg]\ud\sigma'_{\parens{1}k}\nonumber\\
 -\int_{\xp=\mathcal{R}} \Bigg[\partial'^k\left(\frac{\hat{x}'_{L}}{\xp^{\ell+i+1}}\right)\Bigg]\xpb \mathcal{M}\parens{h^{\mu\nu}}(u',\mathbf{x'})\ud\sigma'_{\parens{1}k}\nonumber\\
 +\frac{2}{c}\int_{\xp=\mathcal{R}} \frac{x'^k \hat{x}'_{L}}{\xp^{\ell+i+2}}\xpb\partial_{u'}\mathcal{M}\parens{h^{\mu\nu}}(u',\mathbf{x'})\ud\sigma'_{\parens{1}k}.
\end{eqnarray}
For
$i=2$
we get
\begin{eqnarray}\label{4.3.48}
\A\int_{\mathcal{R}<\xp}\frac{\hat{x}'_{L}}{\xp^{\ell+i+1}}\Bigg\{\Bigg(\Delta'\Bigg[\xpb\mathcal{M}\parens{h^{\mu\nu}}(u',\mathbf{x'})\Bigg]\Bigg)_{u'}\nonumber\\
- \frac{1}{c^2}\xpb \partial_{u'}^2\mathcal{M}\parens{h^{\mu\nu}}(u',\mathbf{x'}) \Bigg\}\ud^3\mathbf{x'} \nonumber \\ 
=\int_{\mathcal{R}<\xp} \xpb\Bigg[\Delta'\left(\frac{\hat{x}'_{L}}{\xp^{\ell+i+1}}\right)\Bigg]\mathcal{M}\parens{h^{\mu\nu}} (u',\mathbf{x'})\ud^3\mathbf{x'} \nonumber\\
 -\frac{2}{c}\int_{\mathcal{R}<\xp} \xpb\xp^{-1} x'^j \Bigg[\partial'_j\left(\frac{\hat{x}'_{L}}{\xp^{\ell+i+1}}\right)\Bigg]\partial_{u'}\mathcal{M}\parens{h^{\mu\nu}}(u',\mathbf{x'})\ud^3\mathbf{x'}\nonumber\\
 - \frac{2}{c}\int_{\mathcal{R}<\xp}\xpb\frac{\hat{x}'_{L}}{\xp^{\ell+i+2}}\partial_{u'}\mathcal{M}\parens{h^{\mu\nu}}(u',\mathbf{x'})\ud^3\mathbf{x'}\nonumber\\
+\int_{\xp=\mathcal{R}} \frac{\hat{x}'_{L}}{\xp^{\ell+i+1}}\partial'^k\Bigg[\xpb \mathcal{M}\parens{h^{\mu\nu}}(u',\mathbf{x'})\Bigg]\ud\sigma'_{\parens{2}k}\nonumber\\
-\int_{\xp=\mathcal{R}} \Bigg[\partial'^k\left(\frac{\hat{x}'_{L}}{\xp^{\ell+i+1}}\right)\Bigg]\xpb \mathcal{M}\parens{h^{\mu\nu}}(u',\mathbf{x'})\ud\sigma'_{\parens{2}k}\nonumber\\
 +\frac{2}{c}\int_{\xp=\mathcal{R}} \frac{x'^k \hat{x}'_{L}}{\xp^{\ell+i+2}}\xpb\partial_{u'}\mathcal{M}\parens{h^{\mu\nu}}(u',\mathbf{x'})\ud\sigma'_{\parens{2}k}\nonumber\\
+\int_{\xp\to\infty} \frac{\hat{x}'_{L}}{\xp^{\ell+i+1}}\partial'^k\Bigg[\xpb\mathcal{M}\parens{h^{\mu\nu}}(u',\mathbf{x'})\Bigg]\ud\sigma'_{\parens{2}k}\nonumber\\
 -\int_{\xp\to\infty} \Bigg[\partial'^k\left(\frac{\hat{x}'_{L}}{\xp^{\ell+i+1}}\right)\Bigg]\xpb \mathcal{M}\parens{h^{\mu\nu}}(u',\mathbf{x'})\ud\sigma'_{\parens{2}k}\nonumber\\
+\frac{2}{c} \int_{\xp\to\infty} \frac{x'^k \hat{x}'_{L}}{\xp^{\ell+i+2}}\xpb\partial_{u'}\mathcal{M}\parens{h^{\mu\nu}}(u',\mathbf{x'})\ud\sigma'_{\parens{2}k}.
\end{eqnarray}
Taking
$\re(B)$
to be a sufficiently large negative number, the last three surface integrals on the RHS of the above equation vanish. Again, as a consequence of the identity theorem, the sum of the analytic continuations of the remaining terms on the RHS of
(\ref{4.3.48})
and the LHS of it must be equal wherever they are all defined. Thus, considering that the surface integrals are entire functions, we reach
\begin{eqnarray}\label{4.3.49}
\A\int_{\mathcal{R}<\xp}\frac{\hat{x}'_{L}}{\xp^{\ell+i+1}}\Bigg\{\Bigg(\Delta'\Bigg[\xpb\mathcal{M}\parens{h^{\mu\nu}}(u',\mathbf{x'})\Bigg]\Bigg)_{u'}\nonumber\\
- \frac{1}{c^2}\xpb \partial_{u'}^2\mathcal{M}\parens{h^{\mu\nu}}(u',\mathbf{x'}) \Bigg\}\ud^3\mathbf{x'} \nonumber \\ 
=\A\int_{\mathcal{R}<\xp} \xpb\Bigg[\Delta'\left(\frac{\hat{x}'_{L}}{\xp^{\ell+i+1}}\right)\Bigg]\mathcal{M}\parens{h^{\mu\nu}}(u',\mathbf{x'})\ud^3\mathbf{x'} \nonumber\\
- \frac{2}{c}\A\int_{\mathcal{R}<\xp} \xpb\xp^{-1} x'^j \Bigg[\partial'_j\left(\frac{\hat{x}'_{L}}{\xp^{\ell+i+1}}\right)\Bigg]\partial_{u'}\mathcal{M}\parens{h^{\mu\nu}}(u',\mathbf{x'})\ud^3\mathbf{x'}\nonumber\\
- \frac{2}{c}\A\int_{\mathcal{R}<\xp} \xpb\frac{\hat{x}'_{L}}{\xp^{\ell+i+2}}\partial_{u'}\mathcal{M}\parens{h^{\mu\nu}}(u',\mathbf{x'})\ud^3\mathbf{x'}\nonumber\\
+\int_{\xp=\mathcal{R}} \frac{\hat{x}'_{L}}{\xp^{\ell+i+1}}\partial'^k\Bigg[\xpb \mathcal{M}\parens{h^{\mu\nu}}(u',\mathbf{x'})\Bigg]\ud\sigma'_{\parens{2}k}\nonumber\\
 -\int_{\xp=\mathcal{R}} \Bigg[\partial'^k\left(\frac{\hat{x}'_{L}}{\xp^{\ell+i+1}}\right)\Bigg]\xpb \mathcal{M}\parens{h^{\mu\nu}}(u',\mathbf{x'})\ud\sigma'_{\parens{2}k}\nonumber\\
 +\frac{2}{c}\int_{\xp=\mathcal{R}}\frac{x'^k \hat{x}'_{L}}{\xp^{\ell+i+2}}\xpb\partial_{u'}\mathcal{M}\parens{h^{\mu\nu}}(u',\mathbf{x'})\ud\sigma'_{\parens{2}k}.
\end{eqnarray}
Now, taking
(\ref{4.3.47}) and (\ref{4.3.49}) into account,
we find
\begin{eqnarray}\label{4.3.50}
\A\int_{D_i}\frac{\hat{x}'_{L}}{\xp^{\ell+i+1}}\Bigg\{\Bigg(\Delta'\Bigg[\xpb\mathcal{M}\parens{h^{\mu\nu}}(u',\mathbf{x'})\Bigg]\Bigg)_{u'} \nonumber\\
- \frac{1}{c^2}\xpb \partial_{u'}^2\mathcal{M}\parens{h^{\mu\nu}}(u',\mathbf{x'}) \Bigg\}\ud^3\mathbf{x'} \nonumber \\ 
=\A\int_{D_i} \xpb\Bigg[\Delta'\left(\frac{\hat{x}'_{L}}{\xp^{\ell+i+1}}\right)\Bigg]\mathcal{M}\parens{h^{\mu\nu}}(u',\mathbf{x'})\ud^3\mathbf{x'} \nonumber\\
 - \frac{2}{c}\A\int_{D_i} \xpb\xp^{-1} x'^j \Bigg[\partial'_j\left(\frac{\hat{x}'_{L}}{\xp^{\ell+i+1}}\right)\Bigg]\partial_{u'}\mathcal{M}\parens{h^{\mu\nu}}(u',\mathbf{x'})\ud^3\mathbf{x'}\nonumber\\
 - \frac{2}{c}\A\int_{D_i} \xpb\frac{\hat{x}'_{L}}{\xp^{\ell+i+2}}\partial_{u'}\mathcal{M}\parens{h^{\mu\nu}}(u',\mathbf{x'})\ud^3\mathbf{x'}\nonumber\\
 +\int_{\xp=\mathcal{R}} \frac{\hat{x}'_{L}}{\xp^{\ell+i+1}}\partial'^k\Bigg[\xpb\mathcal{M}\parens{h^{\mu\nu}}(u',\mathbf{x'})\Bigg]\ud\sigma'_{\parens{i}k}\nonumber\\
 -\int_{\xp=\mathcal{R}} \Bigg[\partial'^k\left(\frac{\hat{x}'_{L}}{\xp^{\ell+i+1}}\right)\Bigg]\xpb \mathcal{M}\parens{h^{\mu\nu}}(u',\mathbf{x'})\ud\sigma'_{\parens{i}k}\nonumber\\
 +\frac{2}{c}\int_{\xp=\mathcal{R}} \frac{x'^k \hat{x}'_{L}}{\xp^{\ell+i+2}}\xpb\partial_{u'}\mathcal{M}\parens{h^{\mu\nu}}(u',\mathbf{x'})\ud\sigma'_{\parens{i}k}.
\end{eqnarray}
Irrespective of whether
$i$
is, it can be easily shown that all the terms appearing in the above equation are analytic in some punctured neighborhood of
$B=0$.
Therefore, each of them has a Laurent expansion around
$B=0$
(the last three terms on the RHS are analytic at
$B=0$
and each of them thereby possesses a Taylor expansion about this point). Since the coefficients of
$B^n$
on both sides of
(\ref{4.3.50})
must be equal for each
$n$,
considering the coefficients of
$B^0$,
we end with
\begin{eqnarray}\label{4.3.51}
\fpa\int_{D_i}\frac{\hat{x}'_{L}}{\xp^{\ell+i+1}}\Bigg\{\Bigg(\Delta'\Bigg[\xpb\mathcal{M}\parens{h^{\mu\nu}}(u',\mathbf{x'})\Bigg]\Bigg)_{u'}\nonumber\\
- \frac{1}{c^2}\xpb \partial_{u'}^2\mathcal{M}\parens{h^{\mu\nu}}(u',\mathbf{x'}) \Bigg\}\ud^3\mathbf{x'} \nonumber \\ 
=\fpa\int_{D_i} \xpb\Bigg[\Delta' \left(\frac{\hat{x}'_{L}}{\xp^{\ell+i+1}}\right)\Bigg]\mathcal{M}\parens{h^{\mu\nu}}(u',\mathbf{x'})\ud^3\mathbf{x'} \nonumber\\
 - \frac{2}{c}\fpa\int_{D_i} \xpb\xp^{-1} x'^j \Bigg[\partial'_j\left(\frac{\hat{x}'_{L}}{\xp^{\ell+i+1}}\right)\Bigg]\partial_{u'}\mathcal{M}\parens{h^{\mu\nu}}(u',\mathbf{x'})\ud^3\mathbf{x'}\nonumber\\
 - \frac{2}{c}\fpa\int_{D_i} \xpb\frac{\hat{x}'_{L}}{\xp^{\ell+i+2}}\partial_{u'}\mathcal{M}\parens{h^{\mu\nu}}(u',\mathbf{x'})\ud^3\mathbf{x'}\nonumber\\
 +\int_{\xp=\mathcal{R}} \frac{\hat{x}'_{L}}{\xp^{\ell+i+1}}\partial'^k\mathcal{M}\parens{h^{\mu\nu}}(u',\mathbf{x'})\ud\sigma'_{\parens{i}k}\nonumber\\
 -\int_{\xp=\mathcal{R}} \Bigg[\partial'^k\left(\frac{\hat{x}'_{L}}{\xp^{\ell+i+1}}\right)\Bigg]\mathcal{M}\parens{h^{\mu\nu}}(u',\mathbf{x'})\ud\sigma'_{\parens{i}k}\nonumber\\
 +\frac{2}{c}\int_{\xp=\mathcal{R}} \frac{x'^k \hat{x}'_{L}}{\xp^{\ell+i+2}}\partial_{u'}\mathcal{M}\parens{h^{\mu\nu}}(u',\mathbf{x'})\ud\sigma'_{\parens{i}k}.
\end{eqnarray}
Thus, after summing
(\ref{4.3.51})
over
$i=1$
and
$i=2$,
applying 
$\partial_{u'}\mathcal{M}\parens{h^{\mu\nu}}(u',\mathbf{x'})=\partial_{t}\mathcal{M}\parens{h^{\mu\nu}}(u',\mathbf{x'})$
and combining the result with
(\ref{4.3.41}), we get
\begin{eqnarray}\label{4.3.52}
\fl \bar{h}^{\mu\nu}\finsarg \hspace{-10mm}&\stareq \frac{16\pi G}{c^4}\fpa\gretint{\bar{\tau}^{\mu\nu}\retarg}\nonumber \\  
\hspace{-10mm}& -\frac{1}{4\pi}\sum_{\ell=0}^{\infty}\sum_{k=0}^{\infty}\sum_{i=0}^{\ell}\frac{1}{\ell !}\frac{\parens{\ell+i}!}{2^i i!\parens{\ell-i}!}\frac{\parens{2\ell+1}!!}{\parens{2k}!!\parens{2\ell+2k+1}!!}\frac{\x^{2k}\hat{x}^{L}}{c^{2k+\ell-i}} \nonumber\\  
\hspace{-10mm}& \times\fpa\int_{\mathbb{R}^3}\xpb\frac{\hat{x}'_{L}}{\xp^{\ell+i+1}}\partial_t^{2k+\ell-i}\Lambda^{\mu\nu}\parens{\mathcal{M}\parens{h}}(u',\mathbf{x'})\ud^3\mathbf{x'}\nonumber \\  
\hspace{-10mm}& +\frac{1}{4\pi}\sum_{\ell=0}^{\infty}\sum_{k=0}^{\infty}\sum_{i=0}^{\ell}\frac{1}{\ell !}\frac{\parens{\ell+i}!}{2^i i!\parens{\ell-i}!}\frac{\parens{2\ell+1}!!}{\parens{2k}!!\parens{2\ell+2k+1}!!}\frac{\x^{2k}\hat{x}^{L}}{c^{2k+\ell-i}} \nonumber\\  
\hspace{-10mm}& \times\fpa\int_{\mathbb{R}^3}\xpb\Bigg[\Delta'\left(\frac{\hat{x}'_{L}}{\xp^{\ell+i+1}}\right)\Bigg]\partial_t^{2k+\ell-i}\mathcal{M}\parens{h^{\mu\nu}}(u',\mathbf{x'})\ud^3\mathbf{x'}\nonumber \\  
\hspace{-10mm}& -\frac{1}{4\pi}\sum_{\ell=0}^{\infty}\sum_{k=0}^{\infty}\sum_{i=0}^{\ell}\frac{2}{\ell !}\frac{\parens{\ell+i}!}{2^i i!\parens{\ell-i}!}\frac{\parens{2\ell+1}!!}{\parens{2k}!!\parens{2\ell+2k+1}!!}\frac{\x^{2k}\hat{x}^{L}}{c^{2k+\ell-i+1}} \nonumber\\  
\hspace{-10mm}& \times\fpa\int_{\mathbb{R}^3}\xpb\xp^{-1} x'^j \Bigg[\partial'_j\left(\frac{\hat{x}'_{L}}{\xp^{\ell+i+1}}\right)\Bigg]\partial_t^{2k+\ell-i+1}\mathcal{M}\parens{h^{\mu\nu}}(u',\mathbf{x'})\ud^3\mathbf{x'}\nonumber \\  
\hspace{-10mm}& -\frac{1}{4\pi}\sum_{\ell=0}^{\infty}\sum_{k=0}^{\infty}\sum_{i=0}^{\ell}\frac{2}{\ell !}\frac{\parens{\ell+i}!}{2^i i!\parens{\ell-i}!}\frac{\parens{2\ell+1}!!}{\parens{2k}!!\parens{2\ell+2k+1}!!}\frac{\x^{2k}\hat{x}^{L}}{c^{2k+\ell-i+1}} \nonumber\\  
\hspace{-10mm}& \times\fpa\int_{\mathbb{R}^3}\xpb\frac{\hat{x}'_{L}}{\xp^{\ell+i+2}}\partial_t^{2k+\ell-i+1}\mathcal{M}\parens{h^{\mu\nu}}(u',\mathbf{x'})\ud^3\mathbf{x'}.
\end{eqnarray}
where we have written
$\int_{\mathbb{R }^3}$
instead of the sum of
$\int_{|\mathbf{x'}|<\mathcal{R}}$
and
$\int_{\mathcal{R}<|\mathbf{x'}|}$
for the sake of brevity, and the surface integrals have not appeared due to
$\vec{\ud\sigma'_{\parens{1}}}$
being the opposite of
$\vec{\ud\sigma'_{\parens{2}}}$
at each point on the surface
$\xp=\mathcal{R}$.
Using
(\ref{a.2.6}),
the third term on the RHS of the above equation takes the form
\begin{eqnarray}\label{4.3.53}
\fl \bracks{\begin{array}{ccc}
\textrm{3rd term on}\\
\textrm{(\ref{4.3.52}) RHS}
\end{array}}
= -\frac{1}{4\pi}\sum_{\ell=1}^{\infty}\sum_{k=0}^{\infty}\sum_{i=1}^{\ell}\frac{2}{\ell !}\frac{\parens{\ell+i}!}{2^i\parens{i-1}!\parens{\ell-i}!}\frac{\parens{2\ell+1}!!}{\parens{2k}!!\parens{2\ell+2k+1}!!}\frac{\x^{2k}\hat{x}^{L}}{c^{2k+\ell-i+1}} \nonumber\\  
\times\fpa\int_{\mathbb{R}^3}\xpb\frac{\hat{x}'_{L}}{\xp^{\ell+i+2}}\partial_t^{2k+\ell-i+1}\mathcal{M}\parens{h^{\mu\nu}}(u',\mathbf{x'})\ud^3\mathbf{x'}.
\end{eqnarray}
Further, by means of
(\ref{a.1.2})
and
(\ref{a.2.3}),
one can rewrite the fourth term as
\begin{eqnarray}\label{4.3.54}
\fl 
\bracks{\begin{array}{ccc}
\textrm{4th term on}\\
\textrm{(\ref{4.3.52}) RHS}
\end{array}}
=-\frac{1}{4\pi}\sum_{\ell=1}^{\infty}\sum_{k=0}^{\infty}\frac{2}{\parens{\ell-1} !}\frac{\parens{2\ell+1}!!}{\parens{2k}!!\parens{2\ell+2k+1}!!}\frac{\x^{2k}\hat{x}^{L}}{c^{2k+\ell+1}} \nonumber\\ 
\times\fpa\int_{\mathbb{R}^3} \xpb\frac{\hat{x}'_{L}}{\xp^{\ell+2}}\partial_t^{2k+\ell+1}\mathcal{M}\parens{h^{\mu\nu}}(u',\mathbf{x'})\ud^3\mathbf{x'}\nonumber \\  
 -\frac{1}{4\pi}\sum_{\ell=1}^{\infty}\sum_{k=0}^{\infty}\sum_{i=1}^{\ell}\frac{2}{\parens{\ell-1} !}\frac{\parens{\ell+i}!}{2^i i!\parens{\ell-i}!}\frac{\parens{2\ell+1}!!}{\parens{2k}!!\parens{2\ell+2k+1}!!}\frac{\x^{2k}\hat{x}^{L}}{c^{2k+\ell-i+1}} \nonumber\\  
\times\fpa\int_{\mathbb{R}^3} \xpb\frac{\hat{x}'_{L}}{\xp^{\ell+i+2}}\partial_t^{2k+\ell-i+1}\mathcal{M}\parens{h^{\mu\nu}}(u',\mathbf{x'})\ud^3\mathbf{x'}\nonumber \\   
+\frac{1}{4\pi}\sum_{k=0}^{\infty}\frac{2}{\parens{2k}!!\parens{2k+1}!!}\frac{\x^{2k}}{c^{2k+1}}\nonumber \\ 
\times\fpa\int_{\mathbb{R}^3} \xpb\xp^{-2}\partial_t^{2k+1}\mathcal{M}\parens{h^{\mu\nu}}(u',\mathbf{x'})\ud^3\mathbf{x'}\nonumber \\ 
+\frac{1}{4\pi}\sum_{\ell=1}^{\infty}\sum_{k=0}^{\infty}\frac{2\parens{\ell+1}}{\ell !}\frac{\parens{2\ell+1}!!}{\parens{2k}!!\parens{2\ell+2k+1}!!}\frac{\x^{2k}\hat{x}^{L}}{c^{2k+\ell+1}} \nonumber\\  
\times\fpa\int_{\mathbb{R}^3} \xpb\frac{\hat{x}'_{L}}{\xp^{\ell+2}}\partial_t^{2k+\ell+1}\mathcal{M}\parens{h^{\mu\nu}}(u',\mathbf{x'})\ud^3\mathbf{x'}\nonumber \\ 
+\frac{1}{4\pi}\sum_{\ell=1}^{\infty}\sum_{k=0}^{\infty}\sum_{i=1}^{\ell}\frac{2}{\ell !}\frac{\parens{\ell+i+1}!}{2^i i!\parens{\ell-i}!}\frac{\parens{2\ell+1}!!}{\parens{2k}!!\parens{2\ell+2k+1}!!}\frac{\x^{2k}\hat{x}^{L}}{c^{2k+\ell-i+1}} \nonumber\\  
\times\fpa\int_{\mathbb{R}^3} \xpb\frac{\hat{x}'_{L}}{\xp^{\ell+i+2}}\partial_t^{2k+\ell-i+1}\mathcal{M}\parens{h^{\mu\nu}}(u',\mathbf{x'})\ud^3\mathbf{x'}.
\end{eqnarray}
The fifth term on the RHS can also be rewritten as follows:
\begin{eqnarray}\label{4.3.55}
\fl
\bracks{\begin{array}{ccc}
\textrm{5th term on}\\
\textrm{(\ref{4.3.52}) RHS}
\end{array}}
= -\frac{1}{4\pi}\sum_{k=0}^{\infty}\frac{2}{\parens{2k}!!\parens{2k+1}!!}\frac{\x^{2k}}{c^{2k+1}}\nonumber \\ 
\times\fpa\int_{\mathbb{R}^3} \xpb\xp^{-2}\partial_t^{2k+1}\mathcal{M}\parens{h^{\mu\nu}}(u',\mathbf{x'})\ud^3\mathbf{x'}\nonumber \\ 
 -\frac{1}{4\pi}\sum_{\ell=1}^{\infty}\sum_{k=0}^{\infty}\frac{2}{\ell !}\frac{\parens{2\ell+1}!!}{\parens{2k}!!\parens{2\ell+2k+1}!!}\frac{\x^{2k}\hat{x}^{L}}{c^{2k+\ell+1}} \nonumber\\  
 \times\fpa\int_{\mathbb{R}^3} \xpb\frac{\hat{x}'_{L}}{\xp^{\ell+2}}\partial_t^{2k+\ell+1}\mathcal{M}\parens{h^{\mu\nu}}(u',\mathbf{x'})\ud^3\mathbf{x'}\nonumber \\ 
 -\frac{1}{4\pi}\sum_{\ell=1}^{\infty}\sum_{k=0}^{\infty}\sum_{i=1}^{\ell}\frac{2}{\ell !}\frac{\parens{\ell+i}!}{2^i i!\parens{\ell-i}!}\frac{\parens{2\ell+1}!!}{\parens{2k}!!\parens{2\ell+2k+1}!!}\frac{\x^{2k}\hat{x}^{L}}{c^{2k+\ell-i+1}} \nonumber\\  
\times\fpa\int_{\mathbb{R}^3} \xpb\frac{\hat{x}'_{L}}{\xp^{\ell+i+2}}\partial_t^{2k+\ell-i+1}\mathcal{M}\parens{h^{\mu\nu}}(u',\mathbf{x'})\ud^3\mathbf{x'}.
\end{eqnarray}
Substituting
(\ref{4.3.53})-(\ref{4.3.55})
into
(\ref{4.3.52}),
one finds that the sum of the last three terms on the RHS is zero. Hence, we have
\begin{eqnarray}\label{4.3.56}
\fl \bar{h}^{\mu\nu}\finsarg \hspace{-10mm} &\stareq \frac{16\pi G}{c^4}\fpa\gretint{\bar{\tau}^{\mu\nu}\retarg}\nonumber \\  
\hspace{-10mm}& -\frac{1}{4\pi}\sum_{k=0}^{\infty}\sum_{\ell=0}^{\infty}\sum_{i=0}^{\ell}\frac{1}{\ell !}\frac{\parens{\ell+i}!}{2^i i!\parens{\ell-i}!}\frac{\parens{2\ell+1}!!}{\parens{2k}!!\parens{2\ell+2k+1}!!}\frac{\x^{2k}\hat{x}^{L}}{c^{2k+\ell-i}} \nonumber\\  
\hspace{-10mm}&\times\fpa\int_{\mathbb{R}^3}\xpb\frac{\hat{x}'_{L}}{\xp^{\ell+i+1}}\partial_t^{2k+\ell-i}\Lambda^{\mu\nu}\parens{\mathcal{M}\parens{h}}(u',\mathbf{x'})
\ud^3\mathbf{x'}.
\end{eqnarray}
As it was seen, the role of the vanishing expression added to the RHS of
(\ref{4.3.37})
is the elimination of the surface integrals over the surface
$\xp=\mathcal{R}$,
which were brought about after using Gauss' and Green's theorems. The first term on the RHS of
(\ref{4.3.56})
is obviously a particular solution to
(\ref{1.2.31})
and the second term a solution to the homogeneous d'Alembertian equation (to see this, we just need to apply d'Alembertian operator to the second term). Therefore,
$\bar{h}^{\mu\nu}\finsarg$
given in
(\ref{4.3.56})
is indeed a solution to
(\ref{1.2.31}),
and it is legitimate provided that it fulfills the harmonic gauge condition.


\section{Harmonic gauge condition}\label{sec:3}

By applying
$\partial_\mu$
to both sides of
(\ref{4.3.56}),
we find
\begin{eqnarray}\label{4.3.57}
\fl\partial_\mu\bar{h}^{\mu\nu}\finsarg \hspace{-6mm}&\stareq \frac{16\pi G}{c^4}\Bigg\{\partial_\mu\fpa\Bigg[\gnearint{\bar{\tau}^{\mu\nu}\retarg}\Bigg]\nonumber \\  
\hspace{-6mm}&+\partial_\mu\fpa\Bigg[\gfarint{\bar{\tau}^{\mu\nu}\retarg}\Bigg]\Bigg\}\nonumber \\  
\hspace{-6mm}& -\frac{1}{4\pi}\sum_{k=0}^{\infty}\sum_{\ell=0}^{\infty}\sum_{i=0}^{\ell}\frac{1}{\ell !}\frac{\parens{\ell+i}!}{2^i i!\parens{\ell-i}!}\frac{\parens{2\ell+1}!!}{\parens{2k}!!\parens{2\ell+2k+1}!!}\frac{\x^{2k}\hat{x}^{L}}{c^{2k+\ell-i}} \nonumber\\  
\hspace{-6mm}&\times\Bigg[\partial_{0}\fpa\int_{\xp<\mathcal{R}}\xpb\frac{\hat{x}'_{L}}{\xp^{\ell+i+1}}\partial_t^{2k+\ell-i}\Lambda^{0\nu}\parens{\mathcal{M}\parens{h}}(u',\mathbf{x'})\ud^3\mathbf{x'}\nonumber \\  
\hspace{-6mm}&+\partial_{0}\fpa\int_{\mathcal{R}<\xp}\xpb\frac{\hat{x}'_{L}}{\xp^{\ell+i+1}}\partial_t^{2k+\ell-i}\Lambda^{0\nu}\parens{\mathcal{M}\parens{h}}(u',\mathbf{x'})\ud^3\mathbf{x'}\Bigg]\nonumber \\  
\hspace{-6mm}& -\frac{1}{4\pi}\sum_{k=0}^{\infty}\sum_{\ell=0}^{\infty}\sum_{i=0}^{\ell}\frac{1}{\ell !}\frac{\parens{\ell+i}!}{2^i i!\parens{\ell-i}!}\frac{\parens{2\ell+1}!!}{\parens{2k}!!\parens{2\ell+2k+1}!!}\frac{\partial_j\parens{\x^{2k}x^{L}}}{c^{2k+\ell-i}} \nonumber\\  
\hspace{-6mm}&\times\fpa\int_{\mathbb{R}^3}\xpb\frac{\hat{x}'_{L}}{\xp^{\ell+i+1}}\partial_t^{2k+\ell-i}\Lambda^{j\nu}\parens{\mathcal{M}\parens{h}}(u',\mathbf{x'})
\ud^3\mathbf{x'},
\end{eqnarray}
where to write the last term on the RHS, we have also used
(\ref{a.2.3}).
We begin by examining the first term on the RHS of the above equation. In the original domain of analyticity of
$\int_{D_i}\txpb \tbfrac{\bar{\tau}^{\mu\nu}\retarg}{|\mathbf{x}-\mathbf{x'}|}\ud^3\mathbf{x'}$,
where
$D_i$
is as in
(\ref{D}),
one can, by virtue of the Gauss' theorem and the conservation equation, write
\begin{eqnarray}\label{4.3.58}
\partial_\mu\A\Bigg[-\frac{1}{4\pi}\int_{D_i}\xpb \frac{\bar{\tau}^{\mu\nu}\retarg}{|\mathbf{x}-\mathbf{x'}|}\ud^3\mathbf{x'}\Bigg]\nonumber \\ 
=B\Bigg[-\frac{1}{4\pi}\int_{D_i}\xpb \frac{\xp^{-1}n'^j\bar{\tau}^{j\nu}\retarg}{|\mathbf{x}-\mathbf{x'}|}\ud^3\mathbf{x'}\Bigg]\nonumber\\
+\frac{1}{4\pi}\int_{\partial D_i}\xpb \frac{\bar{\tau}^{j\nu}\retarg}{|\mathbf{x}-\mathbf{x'}|}\ud\sigma'_{\parens{i}j}.
\end{eqnarray}
Now, by reasoning analogous to what led us from
(\ref{4.3.45})
to
(\ref{4.3.51}),
we get
\begin{eqnarray}\label{4.3.64}
\partial_\mu\fpa\Bigg[-\frac{1}{4\pi}\int_{D_i}\xpb \frac{\bar{\tau}^{\mu\nu}\retarg}{|\mathbf{x}-\mathbf{x'}|}\ud^3\mathbf{x'}\Bigg]\nonumber \\ 
=\resa\Bigg[-\frac{1}{4\pi}\int_{D_i} \xpb \frac{\xp^{-1}n'^j\bar{\tau}^{j\nu}\retarg}{|\mathbf{x}-\mathbf{x'}|}\ud^3\mathbf{x'}\Bigg]\nonumber \\ 
+\frac{1}{4\pi}\int_{\partial D_i} \frac{\bar{\tau}^{j\nu}\retarg}{|\mathbf{x}-\mathbf{x'}|}\ud\sigma'_{\parens{i}j},
\end{eqnarray}
and hence, taking into account that
$\vec{\ud\sigma'_{\parens{1}}}= - \vec{\ud\sigma'_{\parens{2}}}$
at each point on the surface
$\xp=\mathcal{R}$,
the first term on the RHS of
(\ref{4.3.57})
takes the form
\begin{eqnarray}\label{4.3.65}
\fl
\bracks{\begin{array}{ccc}
\textrm{1st term on}\\
\textrm{(\ref{4.3.57}) RHS}
\end{array}}
=\frac{16\pi G}{c^4}\resa\Bigg[-\frac{1}{4\pi}\int_{\xp<\mathcal{R}} \xpb \frac{\xp^{-1}n'^j\bar{\tau}^{j\nu}\retarg}{|\mathbf{x}-\mathbf{x'}|}\ud^3\mathbf{x'}\Bigg]\nonumber\\
+\frac{16\pi G}{c^4}\resa\Bigg[-\frac{1}{4\pi}\int_{\mathcal{R}<\xp} \xpb \frac{\xp^{-1}n'^j\bar{\tau}^{j\nu}\retarg}{|\mathbf{x}-\mathbf{x'}|}\ud^3\mathbf{x'}\Bigg] \nonumber\\  
 = \frac{16\pi G}{c^4}\resa\Bigg[-\frac{1}{4\pi}\int_{\mathcal{R}<\xp} \xpb \frac{\xp^{-1}n'^j\bar{\tau}^{j\nu}\retarg}{|\mathbf{x}-\mathbf{x'}|}\ud^3\mathbf{x'}\Bigg]\nonumber \\ 
 = \frac{16\pi G}{c^4}\resa\Bigg[-\frac{1}{4\pi}\int_{\mathcal{R}<\xp} \xpb \frac{\xp^{-1}n'^j\Lambda^{j\nu}\parens{\bar{h}}\retarg}{|\mathbf{x}-\mathbf{x'}|}\ud^3\mathbf{x'}\Bigg],\nonumber\\
\end{eqnarray}
where
$\Lambda^{\mu\nu}\parens{\bar{h}}$
is nothing but
$\bar{\Lambda}^{\mu\nu}$,
 and to write the second and third equalities we have respectively used the facts that the integral appearing in the first line is, because of the smoothness of
$\bar{\tau}^{\mu\nu}\finsarg$ in the near zone, analytic at
$B=0$,
and
$\parens{-\bar{g}} \bar{T}^{\mu\nu}$
vanishes outside the near zone. Noting that outside the near zone we have
$\Lambda^{\mu\nu}\parens{\bar{h}}=\Lambda^{\mu\nu}\parens{\mathcal{M}\parens{\bar{h}}}$,
due to the structure of
$\Lambda^{\mu\nu}\parens{\mathcal{M}\parens{\bar{h}}}\finsarg$
being similar to the structure of
$\mathcal{M}\parens{\bar{h}}\finsarg$
in this region, in a manner analogous to the one employed to obtain
(\ref{4.3.31}),
we can rewrite
(\ref{4.3.65})
as
\begin{eqnarray}\label{4.3.66}
\fl
\bracks{\begin{array}{ccc}
\textrm{1st term on}\\
\textrm{(\ref{4.3.57}) RHS}
\end{array}}
=\frac{1}{4\pi}\sum_{k=0}^{\infty}\sum_{\ell=0}^{\infty}\sum_{i=0}^{\ell}\frac{1}{\ell !}\frac{\parens{\ell+i}!}{2^i i!\parens{\ell-i}!}\frac{\parens{2\ell+1}!!}{\parens{2k}!!\parens{2\ell+2k+1}!!}\frac{\x^{2k}\hat{x}^{L}}{c^{2k+\ell-i}} \nonumber\\  
\times\resa\int_{\xp<\mathcal{R}}\xpb\frac{\hat{x}'_{L}}{\xp^{\ell+i+1}}\xp^{-1}n'^j \nonumber\\  
\times\partial_t^{2k+\ell-i}\Lambda^{j\nu}\parens{\mathcal{M}\parens{\bar{h}}}(u',\mathbf{x'})\ud^3\mathbf{x'}.
\end{eqnarray}
Replacing
$\mathcal{M}\parens{\bar{h}^{\mu\nu}}$
by
$\overline{\mathcal{M}\parens{h^{\mu\nu}}}$
with the use of the matching equation, by arguments similar to the ones following
(\ref{4.3.36})-(\ref{4.3.38}),
we reach
\begin{eqnarray}\label{4.3.67}
\fl
\bracks{\begin{array}{ccc}
\textrm{1st term on}\\
\textrm{(\ref{4.3.57}) RHS}
\end{array}}
\stareq\frac{1}{4\pi}\sum_{k=0}^{\infty}\sum_{\ell=0}^{\infty}\sum_{i=0}^{\ell}\frac{1}{\ell !}\frac{\parens{\ell+i}!}{2^i i!\parens{\ell-i}!}\frac{\parens{2\ell+1}!!}{\parens{2k}!!\parens{2\ell+2k+1}!!}\frac{\x^{2k}\hat{x}^{L}}{c^{2k+\ell-i}} \nonumber\\  
\times\resa\int_{\mathbb{R}^3}\xpb\frac{\hat{x}'_{L}}{\xp^{\ell+i+1}}\xp^{-1}n'^j\nonumber \\  
\times\partial_t^{2k+\ell-i}\Lambda^{j\nu}\parens{\mathcal{M}\parens{h}}(u',\mathbf{x'})\ud^3\mathbf{x'}.
\end{eqnarray}
Now we examine the second term on the RHS of
(\ref{4.3.57}).
In the original domain of analyticity of
$\int_{D_i}\txpb\tbfrac{\hat{x}'_{L}}{\xp^{\ell+i+1}}\partial_t^{2k+\ell-i}\Lambda^{0\nu}\parens{\mathcal{M}\parens{h}}(u',\mathbf{x'})\ud^3\mathbf{x'}$,
where
$D_i$
is as defined before, we have
\begin{eqnarray}\label{4.3.68}
\partial_0\A\int_{D_i}\xpb\frac{\hat{x}'_{L}}{\xp^{\ell+i+1}}\partial_t^{2k+\ell-i}\Lambda^{0\nu}\parens{\mathcal{M}\parens{h}}(u',\mathbf{x'})\ud^3\mathbf{x'}\nonumber\\ 
= -\partial_t^{2k+\ell-i}\int_{D_i}\xpb\frac{\hat{x}'_{L}}{\xp^{\ell+i+1}}\parens{\partial'_j\Lambda^{j\nu}\parens{\mathcal{M}\parens{h}}(u',\mathbf{x'})}_{u'}\ud^3\mathbf{x'},
\end{eqnarray}
where we have also used the conservation equation. Considering
\begin{eqnarray}\label{4.3.69}
\partial'_j\Lambda^{j\nu}\parens{\mathcal{M}\parens{h}}(u',\mathbf{x'}) &= - \frac{1}{c}\xp^{-1}x'_j \partial_{t}\Lambda^{j\nu}\parens{\mathcal{M}\parens{h}}(u',\mathbf{x'})\nonumber \\
&+\parens{\partial'_j\Lambda^{j\nu}\parens{\mathcal{M}\parens{h}}(u',\mathbf{x'})}_{u'},
\end{eqnarray}
by using the Gauss' theorem, we get
\begin{eqnarray}\label{4.3.70}
\partial_0\A\int_{D_i}\xpb\frac{\hat{x}'_{L}}{\xp^{\ell+i+1}}\partial_t^{2k+\ell-i}\Lambda^{0\nu}\parens{\mathcal{M}\parens{h}}(u',\mathbf{x'})\ud^3\mathbf{x'}\nonumber\\ 
= B\int_{D_i}\xpb\frac{\hat{x}'_{L}}{\xp^{\ell+i+1}}\xp^{-1} n'^j \partial_t^{2k+\ell-i}\Lambda^{j\nu}\parens{\mathcal{M}\parens{h}}(u',\mathbf{x'})\ud^3\mathbf{x'}\nonumber\\
+\int_{D_i}\xpb\Bigg[\partial'_j\left(\frac{\hat{x}'_{L}}{\xp^{\ell+i+1}}\right)\Bigg]\partial_t^{2k+\ell-i}\Lambda^{j\nu}\parens{\mathcal{M}\parens{h}}(u',\mathbf{x'})\ud^3\mathbf{x'}\nonumber\\
 -\frac{1}{c} \int_{D_i}\xpb\frac{x'_j\hat{x}'_L}{\xp^{\ell+i+2}}\partial_t^{2k+\ell-i+1}\Lambda^{j\nu}\parens{\mathcal{M}\parens{h}}(u',\mathbf{x'})\ud^3\mathbf{x'}\nonumber\\
-\int_{\partial D_i}\xpb\frac{\hat{x}'_{L}}{\xp^{\ell+i+1}}\partial_t^{2k+\ell-i}\Lambda^{j\nu}\parens{\mathcal{M}\parens{h}}(u',\mathbf{x'})\ud\sigma'_{\parens{i}j}.
\end{eqnarray}
Proceeding with the arguments analogous to those employed to derive
(\ref{4.3.51}),
we reach
\begin{eqnarray}\label{4.3.76}
\partial_0\fpa\int_{D_i}\xpb\frac{\hat{x}'_{L}}{\xp^{\ell+i+1}}\partial_t^{2k+\ell-i}\Lambda^{0\nu}\parens{\mathcal{M}\parens{h}}(u',\mathbf{x'})\ud^3\mathbf{x'}\nonumber \\ 
= \resa\int_{D_i}\xpb\frac{\hat{x}'_{L}}{\xp^{\ell+i+1}}\xp^{-1} n'^j \partial_t^{2k+\ell-i}\Lambda^{j\nu}\parens{\mathcal{M}\parens{h}}(u',\mathbf{x'})\ud^3\mathbf{x'}\nonumber\\
+\fpa\int_{D_i}\xpb\Bigg[\partial'_j\left(\frac{\hat{x}'_{L}}{\xp^{\ell+i+1}}\right)\Bigg]\partial_t^{2k+\ell-i}\Lambda^{j\nu}\parens{\mathcal{M}\parens{h}}(u',\mathbf{x'})\ud^3\mathbf{x'}\nonumber\\
-\frac{1}{c} \fpa\int_{D_i}\xpb\frac{x'_j\hat{x}'_L}{\xp^{\ell+i+2}}\partial_t^{2k+\ell-i+1}\Lambda^{j\nu}\parens{\mathcal{M}\parens{h}}(u',\mathbf{x'})\ud^3\mathbf{x'}\nonumber\\
-\int_{\partial D_i}\frac{\hat{x}'_{L}}{\xp^{\ell+i+1}}\partial_t^{2k+\ell-i}\Lambda^{j\nu}\parens{\mathcal{M}\parens{h}}(u',\mathbf{x'})\ud\sigma'_{\parens{i}j}.
\end{eqnarray}
This, together with the fact that 
$\vec{\ud\sigma'_{\parens{1}}}$
is the opposite of
$\vec{\ud\sigma'_{\parens{2}}}$
anywhere on the surface
$\xp=\mathcal{R}$,
enable us to rewrite the second term on the RHS of
(\ref{4.3.57})
as
\begin{eqnarray}\label{4.3.77}
\fl
\bracks{\begin{array}{ccc}
\textrm{2nd term on}\\
\textrm{(\ref{4.3.57}) RHS}
\end{array}}
\stareq -\frac{1}{4\pi}\sum_{k=0}^{\infty}\sum_{\ell=0}^{\infty}\sum_{i=0}^{\ell}\frac{1}{\ell !}\frac{\parens{\ell+i}!}{2^i i!\parens{\ell-i}!}\frac{\parens{2\ell+1}!!}{\parens{2k}!!\parens{2\ell+2k+1}!!}\frac{\x^{2k}\hat{x}^{L}}{c^{2k+\ell-i}} \nonumber\\  
\times\resa\int_{\mathbb{R}^3}\xpb\frac{\hat{x}'_{L}}{\xp^{\ell+i+1}}\xp^{-1} n'^j \partial_t^{2k+\ell-i}\Lambda^{j\nu}\parens{\mathcal{M}\parens{h}}(u',\mathbf{x'})\ud^3\mathbf{x'}\nonumber \\ 
-\frac{1}{4\pi}\sum_{k=0}^{\infty}\sum_{\ell=0}^{\infty}\sum_{i=0}^{\ell}\frac{1}{\ell !}\frac{\parens{\ell+i}!}{2^i i!\parens{\ell-i}!}\frac{\parens{2\ell+1}!!}{\parens{2k}!!\parens{2\ell+2k+1}!!}\frac{\x^{2k}\hat{x}^{L}}{c^{2k+\ell-i}} \nonumber\\  
\times\fpa\int_{\mathbb{R}^3}\xpb\Bigg[\partial'_j\left(\frac{x'_{L}}{\xp^{\ell+i+1}}\right)\Bigg]\partial_t^{2k+\ell-i}\Lambda^{j\nu}\parens{\mathcal{M}\parens{h}}(u',\mathbf{x'})\ud^3\mathbf{x'}\nonumber \\ 
+\frac{1}{4\pi}\sum_{k=0}^{\infty}\sum_{\ell=0}^{\infty}\sum_{i=0}^{\ell}\frac{1}{\ell !}\frac{\parens{\ell+i}!}{2^i i!\parens{\ell-i}!}\frac{\parens{2\ell+1}!!}{\parens{2k}!!\parens{2\ell+2k+1}!!}\frac{\x^{2k}\hat{x}^{L}}{c^{2k+\ell-i+1}} \nonumber\\  
\times\fpa\int_{\mathbb{R}^3}\xpb\frac{x'_{jL}}{\xp^{\ell+i+2}}\partial_t^{2k+\ell-i+1}\Lambda^{j\nu}\parens{\mathcal{M}\parens{h}}(u',\mathbf{x'})\ud^3\mathbf{x'}.
\end{eqnarray}
where to write the second and third terms, we have also used
(\ref{a.2.3}).
By means of
(\ref{a.1.2}),
(\ref{a.2.3})
and
(\ref{a.2.10})
one can rewrite the second term on the RHS of
(\ref{4.3.77})
as
\begin{eqnarray}\label{4.3.78}
\fl
\bracks{\begin{array}{ccc}
\textrm{2nd term on}\\
\textrm{(\ref{4.3.77}) RHS}
\end{array}}
= -\frac{1}{4\pi}\sum_{k=0}^{\infty}\sum_{\ell=0}^{\infty}\frac{\parens{2\ell+1}!}{2^\ell \parens{\ell !}^2}\frac{\parens{2\ell+1}!!}{\parens{2k}!!\parens{2\ell+2k+3}!!}\frac{\x^{2k}\hat{x}^{jL}}{c^{2k+1}} \nonumber\\  
\times\fpa\int_{\mathbb{R}^3}\xpb\frac{\hat{x}'_{L}}{\xp^{2\ell+2}}\partial_t^{2k+1}\Lambda^{j\nu}\parens{\mathcal{M}\parens{h}}(u',\mathbf{x'})\ud^3\mathbf{x'}\nonumber \\  
 +\frac{1}{4\pi}\sum_{k=0}^{\infty}\sum_{\ell=0}^{\infty}\frac{\parens{2\ell+1}!}{2^\ell \parens{\ell !}^2}\frac{\parens{2\ell+1}!!}{\parens{2k}!!\parens{2\ell+2k+1}!!}\frac{\x^{2k}\hat{x}^{L}}{c^{2k}} \nonumber\\  
\times\fpa\int_{\mathbb{R}^3}\xpb\frac{\hat{x}'_{jL}}{\xp^{2\ell+3}}\partial_t^{2k}\Lambda^{j\nu}\parens{\mathcal{M}\parens{h}}(u',\mathbf{x'})\ud^3\mathbf{x'}\nonumber \\  
 -\frac{1}{4\pi}\sum_{k=0}^{\infty}\sum_{\ell=0}^{\infty}\sum_{i=1}^{\ell}\frac{2}{\ell !}\frac{\parens{\ell+i}!}{2^i \parens{i-1}!\parens{\ell-i+1}!}\frac{\parens{2\ell+1}!!}{\parens{2k}!!\parens{2\ell+2k+3}!!}\frac{\x^{2k}\hat{x}^{jL}}{c^{2k+\ell-i+2}} \nonumber\\  
\times\fpa\int_{\mathbb{R}^3}\xpb\frac{\hat{x}'_{L}}{\xp^{\ell+i+1}}\partial_t^{2k+\ell-i+2}\Lambda^{j\nu}\parens{\mathcal{M}\parens{h}}(u',\mathbf{x'})\ud^3\mathbf{x'}\nonumber \\  
+\frac{1}{4\pi}\sum_{k=0}^{\infty}\sum_{\ell=0}^{\infty}\sum_{i=1}^{\ell}\frac{2}{\ell !}\frac{\parens{\ell+i}!}{2^i \parens{i-1}!\parens{\ell-i+1}!}\frac{\parens{2\ell+1}!!}{\parens{2k}!!\parens{2\ell+2k+1}!!}\frac{\x^{2k}\hat{x}^{L}}{c^{2k+\ell-i+1}} \nonumber\\  
\times\fpa\int_{\mathbb{R}^3}\xpb\frac{\hat{x}'_{jL}}{\xp^{\ell+i+2}}\partial_t^{2k+\ell-i+1}\Lambda^{j\nu}\parens{\mathcal{M}\parens{h}}(u',\mathbf{x'})\ud^3\mathbf{x'}.
\end{eqnarray}
Further, with the use of
(\ref{a.2.10}),
the third term on the RHS of
(\ref{4.3.77})
takes the form
\begin{eqnarray}\label{4.3.79}
\fl
\bracks{\begin{array}{ccc}
\textrm{3rd term on}\\
\textrm{(\ref{4.3.77}) RHS}
\end{array}}
= \frac{1}{4\pi}\sum_{k=0}^{\infty}\sum_{\ell=0}^{\infty}\frac{1}{\ell !}\frac{\parens{2\ell+1}!!}{\parens{2k}!!\parens{2\ell+2k+1}!!}\frac{\x^{2k}\hat{x}^{L}}{c^{2k+\ell+1}} \nonumber\\  
\times\fpa\int_{\mathbb{R}^3}\xpb\frac{\hat{x}'_{jL}}{\xp^{\ell+2}}\partial_t^{2k+\ell+1}\Lambda^{j\nu}\parens{\mathcal{M}\parens{h}}(u',\mathbf{x'})\ud^3\mathbf{x'}\nonumber \\  
 +\frac{1}{4\pi}\sum_{k=0}^{\infty}\sum_{\ell=0}^{\infty}\frac{1}{\ell !}\frac{\parens{2\ell+1}!!}{\parens{2k}!!\parens{2\ell+2k+3}!!}\frac{\x^{2k}\hat{x}^{jL}}{c^{2k+\ell+2}} \nonumber\\  
\times\fpa\int_{\mathbb{R}^3}\xpb\frac{\hat{x}'_{L}}{\xp^{\ell+1}}\partial_t^{2k+\ell+2}\Lambda^{j\nu}\parens{\mathcal{M}\parens{h}}(u',\mathbf{x'})\ud^3\mathbf{x'}\nonumber \\
 +\frac{1}{4\pi}\sum_{k=0}^{\infty}\sum_{\ell=0}^{\infty}\frac{\parens{2\ell+1}!}{2^\ell \parens{\ell !}^2}\frac{\parens{2\ell+1}!!}{\parens{2k}!!\parens{2\ell+2k+3}!!}\frac{\x^{2k}\hat{x}^{jL}}{c^{2k+1}} \nonumber\\  
\times\fpa\int_{\mathbb{R}^3}\xpb\frac{\hat{x}'_{L}}{\xp^{2\ell+2}}\partial_t^{2k+1}\Lambda^{j\nu}\parens{\mathcal{M}\parens{h}}(u',\mathbf{x'})\ud^3\mathbf{x'}\nonumber \\
+\frac{1}{4\pi}\sum_{k=0}^{\infty}\sum_{\ell=0}^{\infty}\sum_{i=1}^{\ell}\frac{1}{\ell !}\frac{\parens{\ell+i}!}{2^i i!\parens{\ell-i}!}\frac{\parens{2\ell+1}!!}{\parens{2k}!!\parens{2\ell+2k+1}!!}\frac{\x^{2k}\hat{x}^{L}}{c^{2k+\ell-i+1}} \nonumber\\  
\times\fpa\int_{\mathbb{R}^3}\xpb\frac{\hat{x}'_{jL}}{\xp^{\ell+i+2}}\partial_t^{2k+\ell-i+1}\Lambda^{j\nu}\parens{\mathcal{M}\parens{h}}(u',\mathbf{x'})\ud^3\mathbf{x'}\nonumber\\  
+\frac{1}{4\pi}\sum_{k=0}^{\infty}\sum_{\ell=0}^{\infty}\sum_{i=1}^{\ell}\frac{1}{\ell !}\frac{\parens{\ell+i+1}!}{2^i i!\parens{\ell-i+1}!}\frac{\parens{2\ell+1}!!}{\parens{2k}!!\parens{2\ell+2k+3}!!}\frac{\x^{2k}\hat{x}^{jL}}{c^{2k+\ell-i+2}} \nonumber\\  
\times\fpa\int_{\mathbb{R}^3}\xpb\frac{\hat{x}'_{L}}{\xp^{\ell+i+1}}\partial_t^{2k+\ell-i+2}\Lambda^{j\nu}\parens{\mathcal{M}\parens{h}}(u',\mathbf{x'})\ud^3\mathbf{x'}.
\end{eqnarray}
Finally, we examine the third term on the RHS of
(\ref{4.3.57}).
Using
(\ref{a.1.2}),
(\ref{a.2.3})
and the formula for
$\hat{x}'^L x_{jL}$
obtained through making the change
$\mathbf{x}\leftrightarrow\mathbf{x}'$
in 
(\ref{a.2.10}),
one can reach
\begin{eqnarray}\label{4.3.80}
\fl
\bracks{\begin{array}{ccc}
\textrm{3rd term on}\\
\textrm{(\ref{4.3.57}) RHS}
\end{array}}
=  -\frac{1}{4\pi}\sum_{k=0}^{\infty}\sum_{\ell=0}^{\infty}\frac{1}{\ell !}\frac{\parens{2\ell+1}!!}{\parens{2k}!!\parens{2\ell+2k+3}!!}\frac{\x^{2k}\hat{x}^{jL}}{c^{2k+\ell+2}} \nonumber\\  
\times\fpa\int_{\mathbb{R}^3}\xpb\frac{\hat{x}'_{L}}{\xp^{\ell+1}}\partial_t^{2k+\ell+2}\Lambda^{j\nu}\parens{\mathcal{M}\parens{h}}(u',\mathbf{x'})
\ud^3\mathbf{x'}\nonumber \\  
 -\frac{1}{4\pi}\sum_{k=0}^{\infty}\sum_{\ell=0}^{\infty}\frac{1}{\ell !}\frac{\parens{2\ell+1}!!}{\parens{2k}!!\parens{2\ell+2k+1}!!}\frac{\x^{2k}\hat{x}^{L}}{c^{2k+\ell+1}} \nonumber\\  
\times\fpa\int_{\mathbb{R}^3}\xpb\frac{\hat{x}'_{jL}}{\xp^{\ell+2}}\partial_t^{2k+\ell+1}\Lambda^{j\nu}\parens{\mathcal{M}\parens{h}}(u',\mathbf{x'})
\ud^3\mathbf{x'}\nonumber \\
 -\frac{1}{4\pi}\sum_{k=0}^{\infty}\sum_{\ell=0}^{\infty}\frac{\parens{2\ell+1}!}{2^\ell \parens{\ell !}^2}\frac{\parens{2\ell+1}!!}{\parens{2k}!!\parens{2\ell+2k+1}!!}\frac{\x^{2k}\hat{x}^{L}}{c^{2k}} \nonumber\\  
\times\fpa\int_{\mathbb{R}^3}\xpb\frac{\hat{x}'_{jL}}{\xp^{2\ell+3}}\partial_t^{2k}\Lambda^{j\nu}\parens{\mathcal{M}\parens{h}}(u',\mathbf{x'})
\ud^3\mathbf{x'}\nonumber \\
-\frac{1}{4\pi}\sum_{k=0}^{\infty}\sum_{\ell=0}^{\infty}\sum_{i=1}^{\ell}\frac{1}{\ell !}\frac{\parens{\ell+i}!}{2^i i!\parens{\ell-i}!}\frac{\parens{2\ell+1}!!}{\parens{2k}!!\parens{2\ell+2k+3}!!}\frac{\x^{2k}\hat{x}^{jL}}{c^{2k+\ell-i+2}} \nonumber\\  
\times\fpa\int_{\mathbb{R}^3}\xpb\frac{\hat{x}'_{L}}{\xp^{\ell+i+1}}\partial_t^{2k+\ell-i+2}\Lambda^{j\nu}\parens{\mathcal{M}\parens{h}}(u',\mathbf{x'})
\ud^3\mathbf{x'}\nonumber \\  
 -\frac{1}{4\pi}\sum_{k=0}^{\infty}\sum_{\ell=0}^{\infty}\sum_{i=1}^{\ell}\frac{1}{\ell !}\frac{\parens{\ell+i+1}!}{2^i i!\parens{\ell-i+1}!}\frac{\parens{2\ell+1}!!}{\parens{2k}!!\parens{2\ell+2k+1}!!}\frac{\x^{2k}\hat{x}^{L}}{c^{2k+\ell-i+1}} \nonumber\\  
\times\fpa\int_{\mathbb{R}^3}\xpb\frac{\hat{x}'_{jL}}{\xp^{\ell+i+2}}\partial_t^{2k+\ell-i+1}\Lambda^{j\nu}\parens{\mathcal{M}\parens{h}}(u',\mathbf{x'})
\ud^3\mathbf{x'}.
\end{eqnarray}
Now, substituting
(\ref{4.3.78}) and (\ref{4.3.79})
into
(\ref{4.3.77}),
and then substituting
(\ref{4.3.67}), (\ref{4.3.77}) and (\ref{4.3.80})
into
(\ref{4.3.57}),
it is clear that
$\partial_\mu \bar{h}^{\mu\nu}\finsarg=0$,
or in other words,
$\bar{h}^{\mu\nu}\finsarg$
given by
(\ref{4.3.56})
meets the harmonic gauge condition. Before fixing the unknowns
$\hat{B}^{\mu\nu}_{L}\targ$,
it is also worth noting that, despite the appearance of
$\mathcal{R}$
and
$r_0$,
$\bar{h}^{\mu\nu}\finsarg$
can be shown to be independent of these two constants. While the proof of the independence from
$\mathcal{R}$
is straightforward, the
$r_0$-independence
proof is more technical and we devote appendix B to that.


\section{Determination of the moments $\hat{B}^{\mu\nu}_{L}\targ$}\label{sec:4}

To solve the PN-to-PM matching problem completely, we must determine the moments
$\hat{B}^{\mu\nu}_{L}\targ$.
Let us first rewrite
(\ref{4.3.56})
as
\begin{eqnarray}\label{4.4.2}
\fl\bar{h}^{\mu\nu}\finsarg \hspace{-10mm}& \stareq \frac{16\pi G}{c^4}\fpa\gretint{\bar{\tau}^{\mu\nu}\retarg}\nonumber \\  
\hspace{-10mm} &+\sum_{k=0}^{\infty}\sum_{\ell=0}^{\infty}\frac{1}{c^{2k}}\frac{\parens{2\ell+1}!!}{\parens{2k}!!\parens{2\ell+2k+1}!!}\x^{2k}\hat{x}^{L}\partial^{2k}_t\Bigg\{-\frac{1}{4\pi\ell !} \fpa\int_{\mathbb{R}^3}\xpb\hat{x}'_{L}\nonumber\\  
\hspace{-10mm} &\times\sum_{i=0}^{\ell}\frac{\parens{\ell+i}!}{2^i i!\parens{\ell-i}!}\frac{1}{\xp^{\ell+i+1}}\frac{\partial_t^{\ell-i}\Lambda^{\mu\nu}\parens{\mathcal{M}\parens{h}}(u',\mathbf{x'})}{c^{\ell-i}}\ud^3\mathbf{x'}\Bigg\}.
\end{eqnarray}
Comparing the above equation with
(\ref{1.2.28}),
it can be seen that the second term on the RHS has been already written in the appropriate form. However, one needs to rewrite the first term. Assuming that the (1-variable) Taylor expansion of
$\bar{\tau}^{\mu\nu}\finsarg$
about any arbitrary point
$t$
has an infinite radius of convergence, we can write
\begin{eqnarray}\label{4.4.3}
\frac{16\pi G}{c^4}\fpa\gretint{\bar{\tau}^{\mu\nu}\retarg}\nonumber \\ 
= \frac{16\pi G}{c^4}\sum_{k=0}^{\infty}\frac{1}{c^{2k}}\partial^{2k}_t\fpa\Bigg[-\frac{1}{4\pi}\int_{\mathbb{R}^3} \xpb \frac{{|\mathbf{x}-\mathbf{x'}|}^{2k-1}}{\parens{2k}!}\bar{\tau}^{\mu\nu}(t,\mathbf{x'})\ud^3\mathbf{x'}\Bigg]\nonumber \\  
+\frac{4G}{c^4}\fpa\int_{\xp<\x} \xpb \sum_{k=0}^{\infty}\frac{{|\mathbf{x}-\mathbf{x'}|}^{2k}}{\parens{2k+1}!}\frac{\partial^{2k+1}_t\bar{\tau}^{\mu\nu}(t,\mathbf{x'})}{c^{2k+1}}\ud^3\mathbf{x'}\nonumber \\  
+\frac{4G}{c^4}\fpa\int_{\x<\xp} \xpb \sum_{k=0}^{\infty}\frac{{|\mathbf{x}-\mathbf{x'}|}^{2k}}{\parens{2k+1}!}\frac{\partial^{2k+1}_t\bar{\tau}^{\mu\nu}(t,\mathbf{x'})}{c^{2k+1}}\ud^3\mathbf{x'}.
\end{eqnarray}
In the above equation, the first term on the RHS is in the appropriate form, however, the other two terms have to be rewritten. In order to do so, we rewrite the
$k$
sum by means of the Taylor expansion for functions of three variables. We have
\begin{eqnarray}\label{4.4.4}
\fl\sum_{k=0}^{\infty}\frac{{|\mathbf{x}-\mathbf{x'}|}^{2k}}{\parens{2k+1}!}\frac{\partial^{2k+1}_t}{c^{2k+1}}=\cases{\sum_{k=0}^{\infty}\sum_{j=0}^{\infty}\frac{\parens{-1}^j}{j !\parens{2k+1}!}x'^J\partial_J\x^{2k}\frac{\partial^{2k+1}_t}{c^{2k+1}}, & $\xp<\x$,\\
\sum_{k=0}^{\infty}\sum_{j=0}^{\infty}\frac{\parens{-1}^j}{j !\parens{2k+1}!}x^J\partial'_J\xp^{2k}\frac{\partial^{2k+1}_t}{c^{2k+1}}, & $\x<\xp$,\\}
\end{eqnarray}
where the expressions appearing on the RHS have been obtained through the expansions about
$\mathbf{x}$
and
$\mathbf{x}'$
respectively. Since both of these expressions are of the same general form, it is enough to examine only one of them and then just make the change
$\mathbf{x}\leftrightarrow\mathbf{x}'$
to obtain the result for the other one. With the use of
(\ref{a.2.5})-(\ref{a.2.8}),
we find
\begin{eqnarray}\label{4.4.9}
\fl\sum_{k=0}^{\infty}\sum_{j=0}^{\infty}\frac{\parens{-1}^j}{j !\parens{2k+1}!}x'^J\partial_J\x^{2k}\frac{\partial^{2k+1}_t}{c^{2k+1}} & =\sum_{\ell=0}^{\infty}\sum_{m=0}^{\infty}\sum_{k=0}^{\infty}\frac{\parens{-1}^\ell}{\ell !}\parens{2\ell+1}!!\frac{\xp^{2m}\hat{x}'^{L}}{\parens{2m}!!\parens{2\ell+2m+1}!!}\nonumber \\ 
&\times\frac{\x^{2k}\hat{x}_L}{\parens{2k}!!\parens{2\ell+2k+1}!!}\frac{\partial^{2k+2\ell+2m+1}_t}{c^{2k+2\ell+2m+1}}.
\end{eqnarray}
As stated above, by making the change
$\mathbf{x}\leftrightarrow\mathbf{x}'$,
the expression for
$\x < \xp$
can also be rewritten as
\begin{eqnarray}\label{4.4.10}
\fl\sum_{k=0}^{\infty}\sum_{j=0}^{\infty}\frac{\parens{-1}^j}{j !\parens{2k+1}!}x^J\partial'_J\xp^{2k}\frac{\partial^{2k+1}_t}{c^{2k+1}} & = \sum_{\ell=0}^{\infty}\sum_{m=0}^{\infty}\sum_{k=0}^{\infty}\frac{\parens{-1}^\ell}{\ell !}\parens{2\ell+1}!!\frac{\x^{2m}\hat{x}^{L}}{\parens{2m}!!\parens{2\ell+2m+1}!!}\nonumber \\ 
&\times\frac{\xp^{2k}\hat{x}'_L}{\parens{2k}!!\parens{2\ell+2k+1}!!}\frac{\partial^{2k+2\ell+2m+1}_t}{c^{2k+2\ell+2m+1}}.
\end{eqnarray}
Changing the index of summation
$m$
to
$k$
in the above equation, and vice versa, one finds out that the RHS of
(\ref{4.4.9})
is the same as that of
(\ref{4.4.10}).
Therefore, regardless of whether
$\xp$
is greater or less than
$\x$,
we have
\begin{eqnarray}\label{4.4.11}
\fl\sum_{k=0}^{\infty}\frac{{|\mathbf{x}-\mathbf{x'}|}^{2k}}{\parens{2k+1}!}\frac{\partial^{2k+1}_t}{c^{2k+1}}&= \sum_{k=0}^{\infty}\sum_{\ell=0}^{\infty}\frac{1}{c^{2k}}\frac{\parens{2\ell+1}!!}{\parens{2k}!!\parens{2\ell+2k+1}!!}\x^{2k}\hat{x}^L\nonumber \\ 
&\times\partial^{2k}_t\Bigg[\frac{\parens{-1}^\ell}{\ell !}\sum_{m=0}^{\infty}\frac{\xp^{2m}\hat{x}'_{L}}{\parens{2m}!!\parens{2\ell+2m+1}!!}\frac{\partial^{2\ell+2m+1}_t}{c^{2\ell+2m+1}}\Bigg].
\end{eqnarray}
Using the above equation,
(\ref{4.4.3})
takes the form
\begin{eqnarray}\label{4.4.12}
\fl\frac{16\pi G}{c^4}\fpa\gretint{\bar{\tau}^{\mu\nu}\retarg}\nonumber \\ 
\fl=\frac{16\pi G}{c^4}\sum_{k=0}^{\infty}\frac{1}{c^{2k}}\partial^{2k}_t\fpa\Bigg[-\frac{1}{4\pi}\int_{\mathbb{R}^3} \xpb \frac{{|\mathbf{x}-\mathbf{x'}|}^{2k-1}}{\parens{2k}!}\bar{\tau}^{\mu\nu}(t,\mathbf{x'})\ud^3\mathbf{x'}\Bigg]\nonumber \\  
\fl+\sum_{k=0}^{\infty}\sum_{\ell=0}^{\infty}\frac{1}{c^{2k}}\frac{\parens{2\ell+1}!!}{\parens{2k}!!\parens{2\ell+2k+1}!!}\x^{2k}\hat{x}^L\partial^{2k}_t\Bigg\{ \frac{4G}{c^4}\frac{\parens{-1}^\ell}{\ell !}\fpa\int_{\mathbb{R}^3} \xpb\nonumber \\  
\fl \times \sum_{m=0}^{\infty}\frac{\xp^{2m}\hat{x}'_{L}}{\parens{2m}!!\parens{2\ell+2m+1}!!}\frac{\partial^{2\ell+2m+1}_t \bar{\tau}^{\mu\nu}(t,\mathbf{x'})}{c^{2\ell+2m+1}}\ud^3\mathbf{x'}\Bigg\},
\end{eqnarray}
and hence,
$\bar{h}^{\mu\nu}\finsarg$
can be rewritten as
\begin{eqnarray}\label{4.4.13}
\fl\bar{h}^{\mu\nu}\finsarg &\stareq \frac{16\pi G}{c^4}\sum_{k=0}^{\infty}\frac{1}{c^{2k}}\partial^{2k}_t\fpa\Bigg[-\frac{1}{4\pi}\int_{\mathbb{R}^3} \xpb \frac{{|\mathbf{x}-\mathbf{x'}|}^{2k-1}}{\parens{2k}!}\bar{\tau}^{\mu\nu}(t,\mathbf{x'})\ud^3\mathbf{x'}\Bigg]\nonumber \\  
\hspace{-10mm} &+\sum_{k=0}^{\infty}\sum_{\ell=0}^{\infty}\frac{1}{c^{2k}}\frac{\parens{2\ell+1}!!}{\parens{2k}!!\parens{2\ell+2k+1}!!}\x^{2k}\hat{x}^L\partial^{2k}_t\Bigg\{\frac{4G}{c^4}\frac{\parens{-1}^\ell}{\ell !}\fpa\int_{\mathbb{R}^3} \xpb\nonumber \\  
\hspace{-10mm} &\times \sum_{m=0}^{\infty}\frac{\xp^{2m}\hat{x}'_{L}}{\parens{2m}!!\parens{2\ell+2m+1}!!}\frac{\partial^{2\ell+2m+1}_t \bar{\tau}^{\mu\nu}(t,\mathbf{x'})}{c^{2\ell+2m+1}}\ud^3\mathbf{x'}\Bigg\}\nonumber \\  
\hspace{-10mm} &+\sum_{k=0}^{\infty}\sum_{\ell=0}^{\infty}\frac{1}{c^{2k}}\frac{\parens{2\ell+1}!!}{\parens{2k}!!\parens{2\ell+2k+1}!!}\x^{2k}\hat{x}^{L}\partial^{2k}_t\Bigg\{-\frac{1}{4\pi\ell !} \fpa\int_{\mathbb{R}^3}\xpb\hat{x}'_{L}\nonumber\\  
\hspace{-10mm} &\times\sum_{i=0}^{\ell}\frac{\parens{\ell+i}!}{2^i i!\parens{\ell-i}!}\frac{1}{\xp^{\ell+i+1}}\frac{\partial_t^{\ell-i}\Lambda^{\mu\nu}\parens{\mathcal{M}\parens{h}}(u',\mathbf{x'})}{c^{\ell-i}}\ud^3\mathbf{x'}\Bigg\}.
\end{eqnarray}
Comparing
(\ref{1.2.28}) and (\ref{4.4.13}),
we get
\begin{eqnarray}\label{4.4.14}
\fl\sum_{k=0}^{\infty}\sum_{\ell=0}^{\infty}\frac{1}{c^{2k}}\frac{\parens{2\ell+1}!!}{\parens{2k}!!\parens{2\ell+2k+1}!!}\x^{2k}\hat{x}^{L}\partial^{2k}_t\hat{B}^{\mu\nu}_{L}\targ\nonumber \\ 
\fl= \sum_{k=0}^{\infty}\sum_{\ell=0}^{\infty}\frac{1}{c^{2k}}\frac{\parens{2\ell+1}!!}{\parens{2k}!!\parens{2\ell+2k+1}!!}\x^{2k}\hat{x}^L\partial^{2k}_t\Bigg\{\frac{4G}{c^4}\frac{\parens{-1}^\ell}{\ell !}\fpa\int_{\mathbb{R}^3} \xpb\nonumber \\  
\fl \times \sum_{m=0}^{\infty}\frac{\xp^{2m}\hat{x}'_{L}}{\parens{2m}!!\parens{2\ell+2m+1}!!}\frac{\partial^{2\ell+2m+1}_t \bar{\tau}^{\mu\nu}(t,\mathbf{x'})}{c^{2\ell+2m+1}}\ud^3\mathbf{x'}\Bigg\}\nonumber \\  
\fl +\sum_{k=0}^{\infty}\sum_{\ell=0}^{\infty}\frac{1}{c^{2k}}\frac{\parens{2\ell+1}!!}{\parens{2k}!!\parens{2\ell+2k+1}!!}\x^{2k}\hat{x}^{L}\partial^{2k}_t\Bigg\{-\frac{1}{4\pi\ell !} \fpa\int_{\mathbb{R}^3}\xpb\hat{x}'_{L}\nonumber\\  
\fl \times\sum_{i=0}^{\ell}\frac{\parens{\ell+i}!}{2^i i!\parens{\ell-i}!}\frac{1}{\xp^{\ell+i+1}}\frac{\partial_t^{\ell-i}\Lambda^{\mu\nu}\parens{\mathcal{M}\parens{h}}(u',\mathbf{x'})}{c^{\ell-i}}\ud^3\mathbf{x'}\Bigg\}.
\end{eqnarray}
We demand
(\ref{4.4.14})
hold for all
$\mathbf{x}$.
Thus, for any
$k$
and
$\ell$
we must have
\begin{eqnarray}\label{4.4.15}
\partial^{2k}_t\hat{B}^{\mu\nu}_{L}\targ &= \partial^{2k}_t\Bigg\{\frac{4G}{c^4}\frac{\parens{-1}^\ell}{\ell !}\fpa\int_{\mathbb{R}^3} \xpb\nonumber \\  
&\times \sum_{m=0}^{\infty}\frac{\xp^{2m}\hat{x}'_{L}}{\parens{2m}!!\parens{2\ell+2m+1}!!}\frac{\partial^{2\ell+2m+1}_t \bar{\tau}^{\mu\nu}(t,\mathbf{x'})}{c^{2\ell+2m+1}}\ud^3\mathbf{x'}\Bigg\}\nonumber \\  
&+\partial^{2k}_t\Bigg\{-\frac{1}{4\pi\ell !} \fpa\int_{\mathbb{R}^3}\xpb\hat{x}'_{L}\nonumber\\  
&\times\sum_{i=0}^{\ell}\frac{\parens{\ell+i}!}{2^i i!\parens{\ell-i}!}\frac{1}{\xp^{\ell+i+1}}\frac{\partial_t^{\ell-i}\Lambda^{\mu\nu}\parens{\mathcal{M}\parens{h}}(u',\mathbf{x'})}{c^{\ell-i}}\ud^3\mathbf{x'}\Bigg\}.
\end{eqnarray}
Taking
$k=0$,
we find
\begin{eqnarray}\label{4.4.16}
\fl\hat{B}^{\mu\nu}_{L}\targ \nonumber\\
\fl= \frac{4G}{c^4}\frac{\parens{-1}^\ell}{\ell !}\fpa\int_{\mathbb{R}^3} \xpb \sum_{m=0}^{\infty}\frac{\xp^{2m}\hat{x}'_{L}}{\parens{2m}!!\parens{2\ell+2m+1}!!}\frac{\partial^{2\ell+2m+1}_t \bar{\tau}^{\mu\nu}(t,\mathbf{x'})}{c^{2\ell+2m+1}}\ud^3\mathbf{x'}\nonumber\\  
\fl -\frac{1}{4\pi\ell !} \fpa\int_{\mathbb{R}^3}\xpb\hat{x}'_{L}\sum_{i=0}^{\ell}\frac{\parens{\ell+i}!}{2^i i!\parens{\ell-i}!}\frac{1}{\xp^{\ell+i+1}}\frac{\partial_t^{\ell-i}\Lambda^{\mu\nu}\parens{\mathcal{M}\parens{h}}(u',\mathbf{x'})}{c^{\ell-i}}\ud^3\mathbf{x'}.
\end{eqnarray}
The first term on the RHS of the above equation is past-zero since
$\bar{\tau}^{\mu\nu}(t,\mathbf{x'})$
is past-stationary, and hence, due to
$2\ell+2m+1\ge 1$,
$\partial^{2\ell+2m+1}_t \bar{\tau}^{\mu\nu}(t,\mathbf{x'})$
past-zero. Writing
$\Lambda^{\mu\nu}\parens{\mathcal{M}\parens{h}}(u',\mathbf{x'})$
as
$\Lambda^{\mu\nu}_{\mathrm{AS}}\parens{\mathcal{M}\parens{h}}\xparg+\Lambda^{\mu\nu}_{\mathrm{PZ}}\parens{\mathcal{M}\parens{h}}(u',\mathbf{x'})$,
one can easily deduce that, since
$\partial_t^{\ell-i}\Lambda^{\mu\nu}_{\mathrm{AS}}\parens{\mathcal{M}\parens{h}}\xarg=0$
if
$i<\ell$,
and if
$i=\ell$,
we have
\begin{eqnarray}
\A\int_{\xp<\mathcal{R}}\xpb\frac{\hat{x}'_{L}}{\xp^{\ell+i+1}}\Lambda^{\mu\nu}_{\mathrm{AS}}\parens{\mathcal{M}\parens{h}}\xparg\ud^3\mathbf{x'}\nonumber \\ 
= -\A\int_{\mathcal{R}<\xp}\xpb\frac{\hat{x}'_{L}}{\xp^{\ell+i+1}}\Lambda^{\mu\nu}_{\mathrm{AS}}\parens{\mathcal{M}\parens{h}}\xparg\ud^3\mathbf{x'},
\end{eqnarray}
which is itself due to the particular structure of
$\Lambda^{\mu\nu}_{\mathrm{AS}}\parens{\mathcal{M}\parens{h}}\xarg$,
the always-stationary constituting part of the second term is actually zero. Furthermore, it is evident that the remaining constituting part of the second term is past-zero. In light of these considerations,
$\hat{B}^{\mu\nu}_{L}\targ$
is past-zero and can be rewritten as
\begin{eqnarray}\label{4.4.17}
\fl\hat{B}^{\mu\nu}_{L}\targ \nonumber\\
\fl= \frac{4G}{c^4}\frac{\parens{-1}^\ell}{\ell !}\fpa\int_{\mathbb{R}^3} \xpb \sum_{m=0}^{\infty}\frac{\xp^{2m}\hat{x}'_{L}}{\parens{2m}!!\parens{2\ell+2m+1}!!}\frac{\partial^{2\ell+2m+1}_t \bar{\tau}^{\mu\nu}(t,\mathbf{x'})}{c^{2\ell+2m+1}}\ud^3\mathbf{x'}\nonumber\\  
\fl -\frac{1}{4\pi\ell !} \fpa\int_{\mathbb{R}^3}\xpb\hat{x}'_{L}\sum_{i=0}^{\ell}\frac{\parens{\ell+i}!}{2^i i!\parens{\ell-i}!}\frac{1}{\xp^{\ell+i+1}}\frac{\partial_t^{\ell-i}\Lambda^{\mu\nu}_{\mathrm{PZ}}\parens{\mathcal{M}\parens{h}}(u',\mathbf{x'})}{c^{\ell-i}}\ud^3\mathbf{x'}.
\end{eqnarray}


\section{Closed form of $\bar{h}^{\mu\nu}\finsarg$}\label{sec:5}

To better compare the result given at the end of section
\ref{sec:2}
with that of
\cite{PB2002},
we next obtain the closed form of the sum of the last two terms on the RHS of its equivalence, that is
(\ref{4.4.13}).
To do so, we first need to find the closed form of
$\hat{B}^{\mu\nu}_{L}\targ$.
The first term on the RHS of 
(\ref{4.4.17})
can be written as
\begin{eqnarray}\label{4.4.18}
\fl
\bracks{\begin{array}{ccc}
\textrm{1st term on}\\
\textrm{(\ref{4.4.17}) RHS}
\end{array}}
=\frac{1}{c^{2\ell+1}\parens{2\ell+1}!!}\partial^{2\ell+1}_t \Bigg\{ \frac{4G}{c^4}\frac{\parens{-1}^\ell}{\ell !}\fpa\int_{\mathbb{R}^3} \xpb \hat{x}'_{L}\nonumber\\
\times\Bigg[ \sum_{m=0}^{\infty}\frac{\parens{2\ell+1}!!}{2^m m!\parens{2\ell+2m+1}!!}\parens{\frac{\xp}{c}}^{2m}\partial^{2m}_t \bar{\tau}^{\mu\nu}(t,\mathbf{x'})\Bigg]\ud^3\mathbf{x'}\Bigg\}.
\end{eqnarray}
Since we have assumed that the Taylor expansion of
$\bar{\tau}^{\mu\nu}\finsarg$
has an infinite radius of convergence about any arbitrary point
$t$
(this assumption first was stated above
(\ref{4.4.3})),
it is apparent that we can write
\begin{eqnarray}\label{4.4.19}
\fl \sum_{m=0}^{\infty}\frac{\parens{2\ell+1}!!}{2^m m!\parens{2\ell+2m+1}!!}\parens{\frac{\xp}{c}}^{2m}\partial^{2m}_t \bar{\tau}^{\mu\nu}(t,\mathbf{x'}) &= \int_{-1}^{1} f(z)\bar{\tau}^{\mu\nu}(t+z\frac{\xp}{c},\mathbf{x'})\ud z,\qquad
\end{eqnarray}
provided that
$ f(z)$
is an even function of
$z$
fulfilling
\begin{equation}\label{fz}
\int_{-1}^{1} f(z) z^{2m}\ud z=\frac{\parens{2\ell+1}!!\parens{2m}!}{2^m m!\parens{2\ell+2m+1}!!}.
\end{equation}
We have
\begin{eqnarray}\label{4.4.20}
\fl\frac{\parens{2m}!}{2^m m!\parens{2\ell+2m+1}!!} &= \frac{\parens{2m}!\parens{2\ell+2m+2}!!}{2^m m!\parens{2\ell+2m+2}!} = \frac{2^{\ell+m+1}\parens{2m}!\parens{\ell+m+1}!}{2^m m!\parens{2\ell+2m+2}!} \nonumber\\
&= 2^{\ell+1}\frac{\parens{2m}!}{m!}\frac{\parens{\ell+m+1}!}{\parens{2\ell+2m+2}!}.
\end{eqnarray}
Taking into account
(\ref{a.3.2})
and
(\ref{a.3.3}),
(\ref{4.4.20})
takes the form
\begin{eqnarray}\label{4.4.23}
\fl\frac{\parens{2m}!}{2^m m!\parens{2\ell+2m+1}!!} &= 2^{\ell+1}\parens{\frac{4^m \Gamma (m+\frac{1}{2})}{\sqrt{\pi}}}\parens{\frac{\sqrt{\pi}}{4^{\ell+m+1}\Gamma(\ell+m+\frac{3}{2})}}\nonumber\\
&= \frac{1}{2^{\ell+1}}\frac{\Gamma (m+\frac{1}{2})}{\Gamma(\ell+m+\frac{3}{2})}=\frac{1}{2^{\ell+1}\ell !}\frac{\Gamma (\ell+1)\Gamma (m+\frac{1}{2})}{\Gamma(\ell+m+\frac{3}{2})},
\end{eqnarray}
and by using
(\ref{a.3.8}),
one reaches
\begin{eqnarray}\label{4.4.25}
\fl\frac{\parens{2m}!}{2^m m!\parens{2\ell+2m+1}!!} &= \frac{1}{2^{\ell+1}\ell !}\int_{0}^{1} t^{m-\frac{1}{2}} \parens{1-t}^{\ell} \ud t = \frac{2}{2^{\ell+1}\ell !}\int_{0}^{1} \parens{z^2}^{m-\frac{1}{2}} \parens{1-z^2}^{\ell} z\ud z \nonumber\\
&= \frac{2}{2^{\ell+1}\ell !}\int_{0}^{1} z^{2m} \parens{1-z^2}^{\ell} \ud z = \frac{1}{2^{\ell+1}\ell !}\int_{-1}^{1} z^{2m} \parens{1-z^2}^{\ell} \ud z.
\end{eqnarray}
Combining
(\ref{fz}) and (\ref{4.4.25})
and then comparing the LHS and the RHS of the resultant equation,
$f(z)$
can be found as
\begin{eqnarray}\label{4.4.26}
f(z) &= \delta_\ell (z) = \frac{\parens{2\ell+1}!!}{2^{\ell+1}\ell !}\parens{1-z^2}^\ell,
\end{eqnarray}
and hence,
(\ref{4.4.18})
becomes
\begin{eqnarray}\label{4.4.27}
\fl
\bracks{\begin{array}{ccc}
\textrm{1st term on}\\
\textrm{(\ref{4.4.17}) RHS}
\end{array}}
=\frac{1}{c^{2\ell+1}\parens{2\ell+1}!!}\partial^{2\ell+1}_t\Bigg\{ \frac{4G}{c^4}\frac{\parens{-1}^\ell}{\ell !}\fpa\int_{\mathbb{R}^3} \xpb \hat{x}'_{L}\nonumber\\
\times\int_{-1}^{1}\delta_\ell (z)\bar{\tau}^{\mu\nu}(t+z\frac{\xp}{c},\mathbf{x'})\ud z \ud^3\mathbf{x'}\Bigg\}\nonumber\\
= \frac{1}{c^{2\ell+1}\parens{2\ell+1}!!}\partial^{2\ell+1}_t \Bigg\{ \frac{4G}{c^4}\frac{\parens{-1}^\ell}{\ell !}\fpa\int_{\mathbb{R}^3} \xpb \hat{x}'_{L}\nonumber\\
\times\int_{-1}^{1}\delta_\ell (z)\bar{\tau}^{\mu\nu}(t-z\frac{\xp}{c},\mathbf{x'})\ud z \ud^3\mathbf{x'}\Bigg\},
\end{eqnarray}
where to obtain the final form we changed the integration variable from
$z$
to
$-z$
and then used the fact that
$\delta_\ell (z)$
is an even function of
$z$.
(\ref{4.4.27})
is called the {\it closed form} of 
(\ref{4.4.18}).
Before examining the second term on the RHS of
(\ref{4.4.17}),
first note that
\begin{eqnarray}\label{4.4.28}
\frac{\partial}{\partial z}\bigg[P_\ell^{\parens{n}}(z)\partial^m_t w(t-z\frac{\xp}{c},\mathbf{x'})\bigg] &= P_\ell^{\parens{n+1}}(z)\partial^m_t w(t-z\frac{\xp}{c},\mathbf{x'}) \nonumber\\
 &-\frac{\xp}{c} P_\ell^{\parens{n}}(z)\partial^{m+1}_t w(t-z\frac{\xp}{c},\mathbf{x'}),
\end{eqnarray}
where
$w\finsarg$
is a general function and
$P_\ell(z)$
denotes the
$\ell$th-degree
Legendre polynomial. Integrating both sides of
(\ref{4.4.28})
with respect to
$z$,
we reach
\begin{eqnarray}\label{4.4.29}
\fl\int_{a}^{b} P_\ell^{\parens{n}}(z)\partial^{m+1}_t w(t-z\frac{\xp}{c},\mathbf{x'}) \ud z &= -\frac{c}{\xp} P_\ell^{\parens{n}}(z)\partial^m_t w(t-z\frac{\xp}{c},\mathbf{x'}){\bigg|}_{z=a}^{z=b}\nonumber \\  
&+\frac{c}{\xp}\int_{a}^{b} P_\ell^{\parens{n+1}}(z)\partial^m_t w(t-z\frac{\xp}{c},\mathbf{x'})\ud z.
\end{eqnarray}
Since the LHS of the above equation and the integral appearing in the last term on its RHS are of the same general form, one can write
\begin{eqnarray}\label{4.4.30}
\fl\int_{a}^{b} P_\ell^{\parens{n}}(z)\partial^{m+1}_t w(t-z\frac{\xp}{c},\mathbf{x'}) \ud z &= -\frac{c}{\xp} P_\ell^{\parens{n}}(z)\partial^m_t w(t-z\frac{\xp}{c},\mathbf{x'}){\bigg|}_{z=a}^{z=b}\nonumber \\  
& -\parens{\frac{c}{\xp}}^2 P_\ell^{\parens{n+1}}(z)\partial^{m-1}_t w(t-z\frac{\xp}{c},\mathbf{x'}){\bigg|}_{z=a}^{z=b}\nonumber\\
&+\parens{\frac{c}{\xp}}^2\int_{a}^{b} P_\ell^{\parens{n+2}}(z)\partial^{m-1}_t w(t-z\frac{\xp}{c},\mathbf{x'})\ud z.\nonumber\\
\end{eqnarray}
By repeating this process of substitution into the last term on the RHS,
$k$
times in all,
we get
\begin{eqnarray}\label{4.4.31}
\fl\int_{a}^{b} P_\ell^{\parens{n}}(z)\partial^{m+1}_t w(t-z\frac{\xp}{c},\mathbf{x'}) \ud z &= -\sum_{i=0}^{k-1}\parens{\frac{c}{\xp}}^{i+1} P_\ell^{\parens{n+i}}(z)\partial^{m-i}_t w(t-z\frac{\xp}{c},\mathbf{x'}){\bigg|}_{z=a}^{z=b}\nonumber \\  
&+\parens{\frac{c}{\xp}}^{k}\int_{a}^{b} P_\ell^{\parens{n+k}}(z)\partial^{m+1-k}_t w(t-z\frac{\xp}{c},\mathbf{x'})\ud z,\nonumber\\
\end{eqnarray}
which, by taking
$m=\ell$,
$n=0$
and
$k=\ell+1$
and noting that
$P_\ell^{\parens{j}}(z)$
is identically zero if
$j > \ell$
(due to
$P_\ell(z)$
being a polynomial of degree
$\ell$),
takes the form
\begin{eqnarray}\label{4.4.32}
\fl\int_{a}^{b} P_\ell(z)\partial^{\ell+1}_t w(t-z\frac{\xp}{c},\mathbf{x'}) \ud z &=-\sum_{i=0}^{\ell}\parens{\frac{c}{\xp}}^{i+1} P_\ell^{\parens{i}}(z)\partial^{\ell-i}_t w(t-z\frac{\xp}{c},\mathbf{x'}){\bigg|}_{z=a}^{z=b}.
\end{eqnarray}
Furthermore, providing that
$w\finsarg$
is a past-zero function, after choosing
$a=1$
and
$b\to\infty$,
we find
\begin{eqnarray}\label{4.4.33}
\fl\int_{1}^{\infty} P_\ell(z)\partial^{\ell+1}_t w(t-z\frac{\xp}{c},\mathbf{x'}) \ud z &=\sum_{i=0}^{\ell}\parens{\frac{c}{\xp}}^{i+1} P_\ell^{\parens{i}}(1)\partial^{\ell-i}_t w(t-\frac{\xp}{c},\mathbf{x'}),
\end{eqnarray}
which by virtue of
(\ref{a.4.4})
becomes
\begin{eqnarray}\label{4.4.35}
\fl\int_{1}^{\infty} P_\ell(z)\partial^{\ell+1}_t w(t-z\frac{\xp}{c},\mathbf{x'}) \ud z &= c^{\ell+1}\xp^\ell \sum_{i=0}^{\ell}\frac{\parens{\ell+i}!}{2^i i!\parens{\ell-i}!}\frac{1}{\xp^{\ell+i+1}}\frac{\partial^{\ell-i}_t w(t-\frac{\xp}{c},\mathbf{x'})}{c^{\ell-i}},\nonumber\\
\end{eqnarray}
or equivalently,
\begin{eqnarray}\label{4.4.36}
\sum_{i=0}^{\ell}\frac{\parens{\ell+i}!}{2^i i!\parens{\ell-i}!}\frac{1}{\xp^{\ell+i+1}}\frac{\partial^{\ell-i}_t w(t-\frac{\xp}{c},\mathbf{x'})}{c^{\ell-i}} \nonumber \\ 
= \frac{1}{c^{\ell+1}}\partial^{\ell+1}_t \Bigg[\frac{1}{\xp^\ell} \int_{1}^{\infty} P_\ell(z) w(t-z\frac{\xp}{c},\mathbf{x'}) \ud z\Bigg]\nonumber\\
=\frac{1}{c^{\ell+1}}\partial^{\ell+1}_t \Bigg[\frac{\parens{-1}^\ell}{2^\ell \ell !\xp^\ell} \int_{1}^{\infty} \bigg[\frac{\ud^\ell}{\ud z^\ell}\parens{1-z^2}^\ell\bigg] w(t-z\frac{\xp}{c},\mathbf{x'}) \ud z\Bigg],
\end{eqnarray}
where to obtain the last equality we have used the Rodrigues' formula
(\ref{a.4.2}).
Integrating by parts
$\ell$
times and noting that we have assumed
$w\finsarg$
is a past-zero function and the
$n$th
derivative of
$\parens{z^2 - 1}^\ell$
vanishes at
$z=1$
for
$n < \ell$,
the integral in the last line of
(\ref{4.4.36})
can be expressed as
\begin{eqnarray}\label{4.4.38}
\fl\int_{1}^{\infty} \bigg[\frac{\ud^\ell}{\ud z^\ell}\parens{1-z^2}^\ell\bigg] w(t-z\frac{\xp}{c},\mathbf{x'}) \ud z &= \parens{\frac{\xp}{c}}^\ell \int_{1}^{\infty} \parens{1-z^2}^\ell \partial^\ell_t w(t-z\frac{\xp}{c},\mathbf{x'}) \ud z.\nonumber\\
\end{eqnarray}
Using the above equation,
(\ref{4.4.36})
takes the form
\begin{eqnarray}\label{4.4.39}
\sum_{i=0}^{\ell}\frac{\parens{\ell+i}!}{2^i i!\parens{\ell-i}!}\frac{1}{\xp^{\ell+i+1}}\frac{\partial^{\ell-i}_t w(t-\frac{\xp}{c},\mathbf{x'})}{c^{\ell-i}} \nonumber \\ 
= \frac{1}{c^{2\ell+1}}\partial^{2\ell+1}_t \int_{1}^{\infty}\frac{\parens{1-z^2}^\ell}{2^\ell \ell !} w(t-z\frac{\xp}{c},\mathbf{x'}) \ud z\nonumber\\
= \frac{1}{c^{2\ell+1}\parens{2\ell+1}!!}\partial^{2\ell+1}_t \Bigg\{\parens{-1}^\ell\int_{1}^{\infty}2\delta_\ell (z) w(t-z\frac{\xp}{c},\mathbf{x'}) \ud z \Bigg\},
\end{eqnarray}
and by means of
(\ref{4.4.39}),
one finds that the closed form of the second term on the RHS of
(\ref{4.4.17})
reads
\begin{eqnarray}\label{4.4.40}
\fl
\bracks{\begin{array}{ccc}
\textrm{2nd term on}\\
\textrm{(\ref{4.4.17}) RHS}
\end{array}}
=\frac{1}{c^{2\ell+1}\parens{2\ell+1}!!}\partial^{2\ell+1}_t \Bigg\{\frac{1}{4\pi}\frac{\parens{-1}^\ell}{\ell !}\fpa\int_{\mathbb{R}^3}\xpb\hat{x}'_{L}\nonumber\\
 \times \int_{1}^{\infty} \gamma_\ell (z) \Lambda^{\mu\nu}_{\mathrm{PZ}}\parens{\mathcal{M}\parens{h}}(t-z\frac{\xp}{c},\mathbf{x'})\ud z \ud^3\mathbf{x'}\Bigg\},\nonumber\\
\end{eqnarray}
where
$\gamma_\ell (z)=-2\delta_\ell (z)$.
However, in order to use the closed form
(\ref{4.4.40}) in computations, we must rewrite it in terms of
$\Lambda^{\mu\nu}\parens{\mathcal{M}\parens{h}}=\Lambda^{\mu\nu}_{\mathrm{AS}}\parens{\mathcal{M}\parens{h}}+\Lambda^{\mu\nu}_{\mathrm{PZ}}\parens{\mathcal{M}\parens{h}}$.
Such a closed form is obtainable. In order to derive it, first note that, although
$\int_{1}^{\infty} \gamma_\ell (z) \ud z$
is divergent and hence meaningless,
$\ellfpa \int_{1}^{\infty} \gamma_j (z) \ud z$,
where
$j\in \mathbb{C}$
and
\begin{equation}\label{lambdaj}
\gamma_j (z)=\frac{\parens{-1}^{j+1}}{2^{2j}}\frac{\Gamma (2j+2)}{{\left[\Gamma (j+1)\right]}^2}\parens{z^2 -1}^j,
\end{equation}
in which
$\Gamma$
denotes the gamma function defined in appendix A,
is well-defined, and by being well-defined we mean that
$\int_{1}^{\infty} \gamma_j (z) \ud z$
can be analytically continued down to some neighborhood of any point with
$j=\ell\in\mathbb{N}$.
To see this, first we need to show that there exists a region in the complex plane where
$\int_{1}^{\infty} \gamma_j (z) \ud z$
is defined, and further, is analytic (if instead of
(\ref{lambdaj}),
we had considered the former expression for
$\gamma_j (z)$,
i.e.,
$\parens{-1}^{j+1}\tbfrac{\parens{2j+1}!!}{2^j j !}\parens{z^2 -1}^j$,
which is identical to the newly defined
$\gamma_j (z)$
for
$j\in\mathbb{N}$,
$\int_{1}^{\infty} \gamma_j (z) \ud z$
would have been defined nowhere in the complex plane due to the divergence for
$j\in\mathbb{N}$
and
$\parens{2j+1}!!$
and
$j !$
being ill-defined for
$j\in\mathbb{C}-\mathbb{N}$).
Making the change
$t=\tfrac{1}{z^2}$,
we reach
\begin{eqnarray}\label{4.4.41}
\int_{1}^{\infty} \gamma_j (z) \ud z & =  \frac{\parens{-1}^{j+1}}{2^{2j+1}}\frac{\Gamma (2j+2)}{{\left[\Gamma (j+1)\right]}^2}\int_{0}^{1}\parens{1 - t}^j t^{-j-\frac{3}{2}}\ud t \nonumber \\ 
& =  \frac{\parens{-1}^{j+1}}{2^{2j+1}}\frac{\Gamma (2j+2)}{{\left[\Gamma (j+1)\right]}^2}\int_{0}^{1} t^{\parens{-j-\frac{1}{2}}-1}\parens{1 - t}^{\parens{j+1}-1}\ud t.
\end{eqnarray}
Taking
$j$
to be in the strip
$-1<\re(j)<-\frac{1}{2}$,
the integral appearing on the RHS of the last equality in the above equation is nothing but
$B (-j-\frac{1}{2},j+1)$
($B$
denotes the beta function defined in appendix A).
With the use of
(\ref{a.3.8}),
(\ref{4.4.41})
can be rewritten as
\begin{eqnarray}\label{4.4.42}
\int_{1}^{\infty} \gamma_j (z) \ud z & =  \frac{\parens{-1}^{j+1}}{2^{2j+1}}\frac{\Gamma (2j+2)}{{\left[\Gamma (j+1)\right]}^2}\frac{\Gamma (-j-\frac{1}{2})\Gamma (j+1)}{\Gamma(\frac{1}{2})} \nonumber\\
&= \frac{\parens{-1}^{j+1}}{2^{2j+1}\sqrt{\pi}}\frac{\Gamma (2j+2)}{\Gamma (j+1)}\Gamma (-j-\frac{1}{2}).
\end{eqnarray}
Since the gamma function is everywhere-analytic and nowhere-zero in its original domain of definition, from
(\ref{4.4.42})
it can be deduced that
$\int_{1}^{\infty} \gamma_j (z) \ud z$
is analytic in the aforementioned vertical strip. By means of
(\ref{a.3.6}),
one can rewrite
(\ref{4.4.42})
as
\begin{eqnarray}\label{4.4.46}
\fl\int_{1}^{\infty} \gamma_j (z) \ud z \hspace{-2mm}& = \frac{\parens{-1}^{j+1}}{2^{2j+1}\sqrt{\pi}}\bracks{\frac{\Gamma (2j+2+2n)}{\prod_{i=0}^{2n-1}\parens{2j+2+i}}}\bracks{\frac{\Gamma (j+1+n)}{\prod_{m=0}^{n-1}\parens{j+1+m}}}^{-1}\nonumber\\
\hspace{-2mm}& \times\bracks{\frac{\Gamma (-j-\frac{1}{2}+n)}{\prod_{k=0}^{n-1}\parens{-j-\frac{1}{2}+k}}}\nonumber\\
\hspace{-2mm}& = \frac{\parens{-1}^{j+1}}{2^{2j+1}\sqrt{\pi}}\bracks{\frac{\Gamma (2j+2+2n)}{\left[\prod_{i_{1}=0}^{n-1}\parens{2j+3+2 i_{1}}\right] \left[\prod_{i_{2}=0}^{n-1}\parens{2j+2+2 i_{2}}\right]}}\nonumber\\
\hspace{-2mm}& \times\bracks{\frac{2^n\Gamma (j+1+n)}{\prod_{m=0}^{n-1}\parens{2j+2+2m}}}^{-1}\bracks{\frac{2^n\Gamma (-j-\frac{1}{2}+n)}{\parens{-1}^n\prod_{k=0}^{n-1}\parens{2j+1-2k}}}\nonumber\\
\hspace{-2mm}& = \frac{\parens{-1}^{j+1+n}}{2^{2j+1}\sqrt{\pi}\prod_{m=-n}^{n-1}\parens{2j+3+2m}}\frac{\Gamma (2j+2+2n)}{\Gamma (j+1+n)}\Gamma (-j-\frac{1}{2}+n).
\end{eqnarray}
By virtue of the identity theorem, the equality between the analytic continuations of
$\int_{1}^{\infty} \gamma_j (z) \ud z$
and the RHS of the last equality in
(\ref{4.4.46})
must hold wherever they are both defined. Therefore, we have
\begin{eqnarray}\label{4.4.47}
\fl\A\int_{1}^{\infty} \gamma_j (z) \ud z = \frac{\parens{-1}^{j+1+n}}{2^{2j+1}\sqrt{\pi}\prod_{m=-n}^{n-1}\parens{2j+3+2m}}\frac{\A\Gamma (2j+2+2n)}{\A\Gamma (j+1+n)}\A\Gamma (-j-\frac{1}{2}+n),\nonumber\\
\end{eqnarray}
which in the strip
$-n-1 < \re(j) <-\frac{1}{2}+n $
reads
\begin{eqnarray}\label{4.4.48}
\fl\A\int_{1}^{\infty} \gamma_j (z) \ud z  =  \frac{\parens{-1}^{j+1+n}}{2^{2j+1}\sqrt{\pi}\prod_{m=-n}^{n-1}\parens{2j+3+2m}}\frac{\Gamma (2j+2+2n)}{\Gamma (j+1+n)}\Gamma (-j-\frac{1}{2}+n).\quad
\end{eqnarray}
As it can be seen, in this strip, which is
$n$
units wider from each side in comparison with the original striplike domain of analyticity of
$\int_{1}^{\infty} \gamma_j (z) \ud z$,
$\A\int_{1}^{\infty} \gamma_j (z) \ud z$
has
$2n$
simple poles at the points with
$j$
half-integer (namely at
$j=-m-\frac{3}{2}$
with
$-n\le m\le n-1$)
and hence is defined in some neighborhood of any point with
$j$
integer. Therefore, taking
$n=\ell+1$
(which is also compatible with the implicit requirement
$n\ge 1$),
$\ellfpa \int_{1}^{\infty} \gamma_j (z) \ud z$
is well-defined as claimed earlier, and moreover, since
$\A\int_{1}^{\infty} \gamma_j (z) \ud z$
is analytic at
$j=\ell$,
we find
\begin{eqnarray}\label{4.4.49}
\fl\ellfpa\int_{1}^{\infty} \gamma_j (z) \ud z & =  \A\int_{1}^{\infty} \gamma_j (z) \ud z{\bigg|}_{j=\ell} \nonumber\\
&=\frac{\parens{-1}^{2\ell+2}}{2^{2\ell+1}\sqrt{\pi}\prod_{m=-\ell-1}^{\ell}\parens{2\ell+3+2m}}\frac{\Gamma (4\ell+4)}{\Gamma (2\ell+2)}\Gamma (\frac{1}{2})\nonumber\\
&= \frac{1}{2^{2\ell+1}\parens{4\ell+3}!!}\frac{\parens{4\ell+3}!}{\parens{2\ell+1}!}=\frac{\parens{4\ell+2}!!}{2^{2\ell+1}\parens{2\ell+1}!}=\frac{\parens{4\ell+2}!!}{\parens{4\ell+2}!!}=1.
\end{eqnarray}
Similarly, although
$\fpa\int_{\mathbb{R}^3}\txpb\hat{x}'_{L}\int_{1}^{\infty} \gamma_\ell (z) \Lambda^{\mu\nu}_{\mathrm{AS}}\parens{\mathcal{M}\parens{h}}\xparg\ud z \ud^3\mathbf{x'}$
is not defined, it is now obvious
$\fpa\int_{\mathbb{R}^3}\txpb\hat{x}'_{L}\ellfpa\int_{1}^{\infty} \gamma_j (z) \Lambda^{\mu\nu}_{\mathrm{AS}}\parens{\mathcal{M}\parens{h}}\xparg\ud z \ud^3\mathbf{x'}$
is well-defined, and in fact
\begin{eqnarray}\label{4.4.50}
\fpa\int_{\mathbb{R}^3}\xpb\hat{x}'_{L}\ellfpa\int_{1}^{\infty} \gamma_j (z) \Lambda^{\mu\nu}_{\mathrm{AS}}\parens{\mathcal{M}\parens{h}}\xparg\ud z \ud^3\mathbf{x'} \nonumber\\ 
= \fpa\int_{\mathbb{R}^3}\xpb\hat{x}'_{L}\Lambda^{\mu\nu}_{\mathrm{AS}}\parens{\mathcal{M}\parens{h}}\xparg\ellfpa\int_{1}^{\infty} \gamma_j (z) \ud z \ud^3\mathbf{x'}\nonumber \\ 
=\fpa\int_{\mathbb{R}^3}\xpb\hat{x}'_{L}\Lambda^{\mu\nu}_{\mathrm{AS}}\parens{\mathcal{M}\parens{h}}\xparg\ud^3\mathbf{x'}\nonumber\\
= \fpa\int_{\xp<\mathcal{R}}\xpb\hat{x}'_{L}\Lambda^{\mu\nu}_{\mathrm{AS}}\parens{\mathcal{M}\parens{h}}\xparg\ud^3\mathbf{x'}\nonumber\\
+\fpa\int_{\mathcal{R}<\xp}\xpb\hat{x}'_{L}\Lambda^{\mu\nu}_{\mathrm{AS}}\parens{\mathcal{M}\parens{h}}\xparg\ud^3\mathbf{x'}\nonumber\\
= \fpa\int_{\xp<\mathcal{R}}\xpb\hat{x}'_{L}\Lambda^{\mu\nu}_{\mathrm{AS}}\parens{\mathcal{M}\parens{h}}\xparg\ud^3\mathbf{x'}\nonumber\\
 -\fpa\int_{\xp<\mathcal{R}}\xpb\hat{x}'_{L}\Lambda^{\mu\nu}_{\mathrm{AS}}\parens{\mathcal{M}\parens{h}}\xparg\ud^3\mathbf{x'}=0,
\end{eqnarray}
where we could write the fourth equality due to the particular structure of
$\Lambda^{\mu\nu}_{\mathrm{AS}}\parens{\mathcal{M}\parens{h}}\xarg$.
Further, it can be proven that
$\int_{1}^{\infty} \gamma_j (z) \Lambda^{\mu\nu}_{\mathrm{PZ}}\parens{\mathcal{M}\parens{h}}(t-z\tfrac{\xp}{c},\mathbf{x'})\ud z$
is analytic in the half-plane
$\re(j)>-1$
owing to
$\Lambda^{\mu\nu}_{\mathrm{PZ}}\parens{\mathcal{M}\parens{h}}\finsarg$
being a past-zero function. Hence, we have
\begin{eqnarray}\label{4.4.51}
\ellfpa\int_{1}^{\infty} \gamma_j (z) \Lambda^{\mu\nu}_{\mathrm{PZ}}\parens{\mathcal{M}\parens{h}}(t-z\frac{\xp}{c},\mathbf{x'})\ud z \nonumber\\
= \mathop{\mathrm{FP}}\limits_{j=\ell} \int_{1}^{\infty} \gamma_j (z) \Lambda^{\mu\nu}_{\mathrm{PZ}}\parens{\mathcal{M}\parens{h}}(t-z\frac{\xp}{c},\mathbf{x'})\ud z\nonumber\\
 = \int_{1}^{\infty} \gamma_\ell (z) \Lambda^{\mu\nu}_{\mathrm{PZ}}\parens{\mathcal{M}\parens{h}}(t-z\frac{\xp}{c},\mathbf{x'})\ud z.
\end{eqnarray}
With the use of
(\ref{4.4.50}) and (\ref{4.4.51}), we get
\begin{eqnarray}\label{4.4.52}
\fl\fpa\int_{\mathbb{R}^3}\xpb\hat{x}'_{L}\int_{1}^{\infty} \gamma_\ell (z) \Lambda^{\mu\nu}_{\mathrm{PZ}}\parens{\mathcal{M}\parens{h}}(t-z\frac{\xp}{c},\mathbf{x'})\ud z \ud^3\mathbf{x'} \nonumber \\ 
\fl= \fpa\int_{\mathbb{R}^3}\xpb\hat{x}'_{L}\ellfpa\int_{1}^{\infty} \gamma_j (z) \Lambda^{\mu\nu}\parens{\mathcal{M}\parens{h}}(t-z\frac{\xp}{c},\mathbf{x'})\ud z \ud^3\mathbf{x'}.
\end{eqnarray}
Thus, the closed form
(\ref{4.4.40}) takes the form
\begin{eqnarray}\label{4.4.53}
\fl
\bracks{\begin{array}{ccc}
\textrm{2nd term on}\\
\textrm{(\ref{4.4.17}) RHS}
\end{array}}
=\frac{1}{c^{2\ell+1}\parens{2\ell+1}!!}\partial^{2\ell+1}_t \Bigg\{\frac{1}{4\pi}\frac{\parens{-1}^\ell}{\ell !}\fpa\int_{\mathbb{R}^3}\xpb\hat{x}'_{L}\nonumber\\
\times \ellfpa\int_{1}^{\infty} \gamma_j (z) \Lambda^{\mu\nu}\parens{\mathcal{M}\parens{h}}(t-z\frac{\xp}{c},\mathbf{x'})\ud z \ud^3\mathbf{x'}\Bigg\}.
\end{eqnarray}
Having obtained the closed forms
(\ref{4.4.27}) and (\ref{4.4.53}),
one finds the closed form of
$\hat{B}^{\mu\nu}_{L}\targ$
as
\begin{eqnarray}\label{4.4.54}
\hat{B}^{\mu\nu}_{L}\targ &= \frac{1}{c^{2\ell+1}\parens{2\ell+1}!!}\partial^{2\ell+1}_t \Bigg\{ \frac{4G}{c^4}\frac{\parens{-1}^\ell}{\ell !}\fpa\int_{\mathbb{R}^3} \xpb \hat{x}'_{L}\nonumber \\
&\times\int_{-1}^{1}\delta_\ell (z)\bar{\tau}^{\mu\nu}(t-z\frac{\xp}{c},\mathbf{x'})\ud z \ud^3\mathbf{x'}\nonumber \\
&+\frac{1}{4\pi}\frac{\parens{-1}^\ell}{\ell !}\fpa\int_{\mathbb{R}^3}\xpb\hat{x}'_{L}\nonumber \\
&\times \ellfpa\int_{1}^{\infty} \gamma_j (z) \Lambda^{\mu\nu}\parens{\mathcal{M}\parens{h}}(t-z\frac{\xp}{c},\mathbf{x'})\ud z \ud^3\mathbf{x'}\Bigg\}.
\end{eqnarray}
Now note that we can write
\begin{eqnarray}
\sum_{k=0}^{\infty}\sum_{\ell=0}^{\infty}\frac{1}{c^{2k}}\frac{\parens{2\ell+1}!!}{\parens{2k}!!\parens{2\ell+2k+1}!!}\x^{2k}\hat{x}^{L}\partial^{2k}_t\hat{B}^{\mu\nu}_{L}\targ\nonumber\\
=\sum_{\ell=0}^{\infty}\hat{\partial}_L\parens{\frac{\hat{U}_L^{\mu\nu}\uarg + \hat{V}_L^{\mu\nu}\varg}{\x}},
\end{eqnarray}
which is due to the LHS being a solution to the homogeneous d'Alembertian equation whose most general solution is given as on the RHS
\cite{BD1986}.
In order to determine
$\hat{U}_L^{\mu\nu}(t-\tfrac{\x}{c})$
and
$\hat{V}_L^{\mu\nu}(t+\tfrac{\x}{c})$
corresponding to the LHS, the following two properties must be taken into account:
\begin{enumerate}
\item Retardation effects are small in the near zone.
\item The solution is smooth in the near zone.
\end{enumerate}
The property (i) allows us to use the Taylor expansion. Using it as well as
(\ref{a.2.7}) and (\ref{a.2.8}),
we get
\begin{eqnarray}\label{4.4.55}
\fl\sum_{\ell=0}^{\infty}\hat{\partial}_L\parens{\frac{\hat{U}_L^{\mu\nu}\uarg + \hat{V}_L^{\mu\nu}\varg}{\x}} \nonumber \\ 
\fl= \sum_{k=0}^{\infty}\frac{\bigg[ {}^{\parens{2k}}\hat{U}^{\mu\nu}\targ +  {}^{\parens{2k}}\hat{V}^{\mu\nu}\targ\bigg]}{c^{2k} \parens{2k}!} \x^{2k-1} \nonumber\\
\fl +\sum_{k=0}^{\infty}\sum_{\ell=1}^{\infty}\frac{\bigg[ {}^{\parens{2k}}\hat{U}_L^{\mu\nu}\targ +  {}^{\parens{2k}}\hat{V}_L^{\mu\nu}\targ\bigg]}{c^{2k} \parens{2k}!}\parens{2k-1}\parens{2k-3}\cdots\parens{2k-2\ell+1}\hat{n}_L \x^{2k-\ell-1} \nonumber\\
\fl - \sum_{k=0}^{\infty}\sum_{\ell=0}^{\infty}\frac{\bigg[{}^{\parens{2k+2\ell+1}}\hat{U}_L^{\mu\nu}\targ - {}^{\parens{2k+2\ell+1}}\hat{V}_L^{\mu\nu}\targ\bigg]}{c^{2k+2\ell+1} \parens{2k+2\ell+1}!!}\frac{\x^{2k}\hat{x}_L}{\parens{2k}!!}.
\end{eqnarray}
Obviously, the property (ii) requires that each of the first two terms on the RHS of the above equation vanishes. Therefore, we must have
$\hat{V}_L^{\mu\nu}\targ =-\hat{U}_L^{\mu\nu}\targ $,
and hence,
\begin{equation}\label{4.4.56}
\hat{B}^{\mu\nu}_L\targ=\frac{1}{c^{2\ell+1}\parens{2\ell+1}!!}\partial_t^{2\ell+1}\left[-2\hat{U}_L^{\mu\nu}\targ\right].
\end{equation}
Comparing the above equation with
(\ref{4.4.54}),
one finds
\begin{equation}\label{4.4.57}
\hat{U}^{\mu\nu}_L\targ = -\frac{4G}{c^4}\frac{\parens{-1}^\ell}{\ell !}\frac{\hat{A}^{\mu\nu}_L\targ}{2}+\hat{C}^{\mu\nu}_L\targ,
\end{equation}
where
$\hat{C}^{\mu\nu}_L\targ$
is a constant STF tensor, and
\begin{equation}\label{4.4.58}
\fl\hat{A}^{\mu\nu}_L\targ = \hat{F}^{\mu\nu}_L\targ+\hat{R}^{\mu\nu}_L\targ,
\end{equation}
\begin{equation}\label{4.4.59}
\fl\hat{F}^{\mu\nu}_L\targ = \fpa\int_{\mathbb{R}^3} \xpb \hat{x}'_{L}\int_{-1}^{1}\delta_\ell (z)\bar{\tau}^{\mu\nu}(t-z\frac{\xp}{c},\mathbf{x'})\ud z \ud^3\mathbf{x'},
\end{equation}
\begin{equation}\label{4.4.60}
\fl\hat{R}^{\mu\nu}_L\targ = \fpa\int_{\mathbb{R}^3}\xpb\hat{x}'_{L} \ellfpa\int_{1}^{\infty} \gamma_j (z) \mathcal{M}\parens{\tau^{\mu\nu}}(t-z\frac{\xp}{c},\mathbf{x'})\ud z \ud^3\mathbf{x'},
\end{equation}
where in
(\ref{4.4.60}),
by
$\mathcal{M}\parens{\tau^{\mu\nu}}$
we mean
$\tbfrac{c^4}{16\pi G} \Lambda^{\mu\nu}\parens{\mathcal{M}\parens{h}}$.
All in all,
$\bar{h}^{\mu\nu}\finsarg$
given by
(\ref{4.3.56})
can be written as
\begin{eqnarray}\label{4.4.61}
\fl\bar{h}^{\mu\nu}\finsarg \hspace{-10mm} &\stareq \frac{16\pi G}{c^4}\sum_{k=0}^{\infty}\frac{1}{c^{2k}}\partial^{2k}_t\fpa\Bigg[-\frac{1}{4\pi}\int_{\mathbb{R}^3}\xpb\frac{{|\mathbf{x}-\mathbf{x'}|}^{2k-1}}{\parens{2k}!}\bar{\tau}^{\mu\nu}(t,\mathbf{x'})\ud^3\mathbf{x'}\Bigg]\nonumber \\  
\hspace{-10mm} &-\frac{4G}{c^4}\sum_{\ell=0}^{\infty}\frac{\parens{-1}^\ell}{\ell !}\hat{\partial}_L \parens{\frac{\hat{A}_L^{\mu\nu}\uarg - \hat{A}_L^{\mu\nu}\varg}{2\x}},
\end{eqnarray}
which is in full agreement with the result derived in
\cite{PB2002}.
As a bonus, since the second term on the RHS of
(\ref{4.4.17}),
whose closed form is given by
(\ref{4.4.53}),
corresponds only to the second term on the RHS of
(\ref{4.4.2}),
and vice versa, we can also write
\begin{eqnarray}\label{4.4.62}
\bar{h}^{\mu\nu}\finsarg&\stareq \frac{16\pi G}{c^4}\fpa\gretint{\bar{\tau}^{\mu\nu}\retarg}\nonumber \\  
& -\frac{4G}{c^4}\sum_{\ell=0}^{\infty}\frac{\parens{-1}^\ell}{\ell !}\hat{\partial}_L \parens{\frac{\hat{R}_L^{\mu\nu}\uarg - \hat{R}_L^{\mu\nu}\varg}{2\x}},
\end{eqnarray}
which is nothing but the result stated in
\cite{BFN2005}.


\section{Summary}\label{sec:6}

Starting with a trivial equality, by virtue of a specific process of analytic continuation, we derived an expression for the post-Newtonian approximation of
$h^{\mu\nu}\finsarg$.
We then demonstrated that this approximate solution to the Einstein field equation satisfies the harmonic gauge condition. Finally, we obtained the closed form of the stated solution, mainly by means of the properties of the gamma and beta functions, and in this way, verified the compatibility of the result of this paper with that of the 2002 paper by Poujade and Blanchet.


\ack

The author would like to express his profound gratitude toward Nahid Ahmadi for helpful discussions.


\appendix


\section{A collection of useful definitions, lemmas and formulae}

\subsection*{A.1. Basic formulae}

\begin{eqnarray}\label{a.1.1}
\Delta \parens{\frac{F\uarg}{\x}} = \frac{1}{c^2}\frac{F^{\left(2\right)}\uarg}{|\mathbf{x}|}.
\end{eqnarray}
\begin{eqnarray}\label{a.1.2}
\partial_k x^L=\cases{0, & $\ell=0$,\\
\delta^{k\{ i_1}x^{ i_2... i_{\ell}\}}, & $\ell \ge 1$.\\}
\end{eqnarray}
\begin{eqnarray}\label{a.1.3}
\frac{\partial}{\partial B}\xb=\xb\ln{\frac{\x}{r_0}}.
\end{eqnarray}
\begin{eqnarray}
f(t+a)=\sum_{j=0}^{\infty}\frac{1}{j!}a^j f^{\parens{j}}\targ. \hspace{13mm} \textrm{({\it 1-variable Taylor expansion})}
\end{eqnarray}
\begin{eqnarray}
f(\mathbf{x}+\mathbf{a})=\sum_{j=0}^{\infty}\frac{1}{j!}a^J\partial_Jf\xarg. \hspace{10mm} \textrm{({\it3-variable Taylor expansion})}
\end{eqnarray}


\subsection*{A.2. Symmetric-trace free Cartesian tensors}

\begin{defa1}
A Cartesian tensor is called
\textit{symmetric-trace-free}
if its any single contraction vanishes. We denote (the components of) a symmetric-trace-free Cartesian tensor of rank
$q$
by
$\hat{T}_Q$
or
$T_{<Q>}$.
\end{defa1}
\begin{defa2}
The symmetric-trace-free part of a general tensor
$T_Q$
is defined as
\begin{equation}\label{a.2.1}
\hat{T}_Q=\sum_{m=0}^{\left[\frac{q}{2}\right]}a_{qm}\delta_{(i_1 i_2}\cdots\delta_{i_{2m-1}i_{2m}}T_{(i_{2m+1}...i_q))a_1a_1...a_ma_m},
\end{equation}
where
\begin{equation}\label{a.2.2}
a_{qm}={\parens{-1}}^m \frac{q!}{\parens{q-2m}!}\frac{\parens{2q-2m-1}!!}{\parens{2q-1}!!\parens{2m}!!}.
\end{equation}
\end{defa2}
\noindent Two general Cartesian tensors
$F_L$
and
$G_L$
satisfy the following formula:
\begin{eqnarray}\label{a.2.3}
\hat{F}^L\hat{G}_L=\hat{F}^LG_L.
\end{eqnarray}
Further, two important lemmas containing symmetric-trace-free tensors (whose proofs are given in
\cite{THESIS})
are as follows.
\begin{lemma1}
When
$\x > 0$,
the integral
$\int_{|\mathbf{x'}|<\mathcal{R}} \tbfrac{\hat{n}'^Q \xp^a \parens{\ln{\xp}}^p g\sinsarg}{|\mathbf{x}-\mathbf{x'}|}\ud^3\mathbf{x'}$
with
$0 < \mathcal{R} <\infty$,
$q \in \mathbb{N}$,
$a \in \mathbb{R}$,
$p  \in \mathbb{R}^{\ge 0}$
and
$g\finsarg$
a bounded function in
$\mathbb{R}^4$
converges if
$a > -q-3$.
\end{lemma1}
\begin{lemma2}
When
$\x < \infty$,
the integral
$\int_{\mathcal{R}<|\mathbf{x'}|} \tbfrac{\hat{n}'^Q \xp^a \parens{\ln{\xp}}^p g\sinsarg}{|\mathbf{x}-\mathbf{x'}|}\ud^3\mathbf{x'}$
with
$0 < \mathcal{R} <\infty$,
$q \in \mathbb{N}$,
$a \in \mathbb{R}$,
$p  \in \mathbb{R}^{\ge 0}$
and
$g\finsarg$
a bounded function in
$\mathbb{R}^4$
converges if
$a < q-2$.
\end{lemma2}
\noindent Other useful formulae including symmetric-trace-free Cartesian tensors are
\begin{eqnarray}\label{a.2.4}
n_L=\sum_{k=0}^{\left[\frac{\ell}{2}\right]}\frac{\parens{2\ell-4k+1}!!}{\parens{2\ell-2k+1}!!}\delta_{\{i_1 i_2}\cdots\delta_{i_{2k-1}i_{2k}}\hat{n}_{i_{2k+1}...i_\ell\}},
\end{eqnarray}
\begin{equation}\label{a.2.5}
\partial_L=\sum_{k=0}^{\left[\frac{\ell}{2}\right]}\frac{\parens{2\ell-4k+1}!!}{\parens{2\ell-2k+1}!!}\delta_{\{i_1 i_2}\cdots\delta_{i_{2k-1}i_{2k}}\hat{\partial}_{i_{2k+1}...i_\ell\}}\Delta^k,
\end{equation}
\begin{equation}\label{a.2.6}
\Delta \left[\hat{n}^Q \x^{a+2}\right]=\parens{a-q+2}\parens{a+q+3}\hat{n}^Q \x^a,
\end{equation}
\begin{equation}\label{a.2.7}
\hat{\partial}_L \x^\lambda=\lambda\parens{\lambda-2}\cdots\parens{\lambda-2\ell+2}\hat{n}_L\x^{\lambda-\ell},
\end{equation}
\begin{equation}\label{a.2.8}
\hat{\partial}_L \x^{2j}=0 \qquad  \textrm{if}\;\; j=0,1,2,...,\ell-1,
\end{equation}
\begin{eqnarray}\label{a.2.9}
\hat{\partial}_L\parens{\frac{F(t-\frac{\x}{c})}{\x}} = {\parens{-1}}^\ell\hat{n}^L\sum_{j=0}^{\ell}\frac{\parens{\ell+j}!}{2^j j!\parens{\ell-j}!}\frac{F^{\parens{\ell-j}}(t-\frac{\x}{c})}{c^{\ell-j}\x^{j+1}},
\end{eqnarray}
and last but not least,
\begin{eqnarray}\label{a.2.10}
\hat{x}^L x'_{jL}&=\hat{x}^L\sum_{k=0}^{\left[\frac{\ell+1}{2}\right]}\frac{\parens{2\parens{\ell+1}-4k+1}!!}{\parens{2\parens{\ell+1}-2k+1}!!}\delta_{\{i_1 i_2}\cdots\delta_{i_{2k-1}i_{2k}}\hat{x}'_{i_{2k+1}...i_\ell j\}}\xp^{2k}\nonumber \\  
&=\hat{x}^L\Bigg[\frac{\parens{2\ell+3}!!}{\parens{2\ell+3}!!}\hat{x}'_{jL}+\frac{\parens{2\ell-1}!!}{\parens{2\ell+1}!!}\delta_{j\{ i_1}\hat{x}'_{i_2...i_\ell\}}\xp^2\Bigg]\nonumber \\  
&=\hat{x}^L\hat{x}'_{jL}+\frac{\ell}{2\ell+1}\hat{x}^{jL-1}\hat{x}'_{L-1}\xp^2.
\end{eqnarray}


\subsection*{A.3. Gamma and beta functions}

\begin{defa3}
The gamma function is defined as
\begin{equation}\label{a.3.1}
\Gamma (x)=\int_{0}^{\infty} t^{x-1} e^{-t} \ud t \qquad  \textrm{for}\; \re(x) > 0.
\end{equation}
\end{defa3}
\noindent One can show
\begin{equation}\label{a.3.2}
\Gamma (n)=\parens{n-1}! \hspace{20.45mm}  \textrm{for}\; n \in \mathbb{Z}^{\ge 1},
\end{equation}
\begin{equation}\label{a.3.3}
\Gamma (n+\frac{1}{2})=\frac{\parens{2n}!}{4^n n!}\sqrt{\pi} \hspace{10.75mm}  \textrm{for}\; n \in \mathbb{Z}^{\ge 0}.
\end{equation}
Moreover, by integrating by parts, one obtains
\begin{equation}\label{a.3.4}
\fl\Gamma (x+1)=\int_{0}^{\infty} t^{x} e^{-t} \ud t = -t^x e^{-t}{\bigg|}_{0}^{\infty} + \int_{0}^{\infty} x t^{x-1} e^{-t} \ud t = x \Gamma (x) \quad \textrm{for}\;\re(x) > 0,
\end{equation}
or equivalently,
\begin{equation}\label{a.3.5}
\Gamma (x)=\frac{\Gamma (x+1)}{x}.
\end{equation}
Repeatedly using the above equation,
$n$
times in all, one gets
\begin{equation}\label{a.3.6}
\Gamma (x)=\frac{\Gamma (x+n)}{x\parens{x+1}\cdots\parens{x+n-1}}=\frac{\Gamma (x+n)}{\prod_{k=0}^{n-1}\parens{x+k}}.
\end{equation}
\begin{defa4}
The beta function is defined as
\begin{equation}\label{a.3.7}
B (x,y)=\int_{0}^{1} t^{x-1} \parens{1-t}^{y-1} \ud t \qquad  \textrm{for}\; \re(x) > 0\; \textrm{and}\; \re(y) > 0.
\end{equation}
\end{defa4}
\noindent It can be shown
\begin{equation}\label{a.3.8}
B (x,y)=\frac{\Gamma (x)\Gamma (y)}{\Gamma(x+y)}.
\end{equation}


\subsection*{A.4. Other mathematical relations}

\begin{defkirch}
A general function
$f\finsarg$
fulfills this condition if
\cite{BW1999}
\begin{eqnarray}\label{a.4.1}
\mathop{\mathop{\lim}\limits_{\x\to \infty}}\limits_{t+\frac{\x}{c}=\mathrm{const}} \left[ \frac{\partial}{\partial  \x}\parens{\x f\finsarg} +\frac{1}{c}\frac{\partial}{\partial t}\parens{\x f\finsarg}\right]=0.
\end{eqnarray}
\end{defkirch}
\begin{formurod}
It is a formula for Legendre polynomials given by
\begin{eqnarray}\label{a.4.2}
P_\ell (z)=\frac{1}{2^\ell \ell !}\frac{\ud^\ell}{\ud z^\ell} \parens{z^2 - 1}^\ell.
\end{eqnarray}
\end{formurod}
\noindent By means of the Rodrigues' formula and the Leibniz's formula for the
$n$th
derivative of a product of more than two factors, which reads
\begin{eqnarray}\label{a.4.3}
\parens{f_1 f_2 \cdots f_m}^{\parens{n}} = \mathop{\sum_{k_1,k_2,...,k_m}}\limits_{k_1+k_2+\cdots+k_m=n}\frac{n!}{k_1!k_2!\cdots k_m!}\prod_{j=1}^{m}f_j^{\parens{k_j}},
\end{eqnarray}
we can show
\begin{eqnarray}\label{a.4.4}
 P_\ell^{\parens{i}}(1)=\cases{\frac{\parens{\ell+i}!}{2^i i!\parens{\ell-i}!}, & $ i \le\ell$,\\
0, & $ i>\ell$.\\}
\end{eqnarray}
The expression for
$i>\ell$
comes from the fact that
$P_\ell(z)$
is a
$\ell$th-degree
polynomial . In order to obtain the expression for
$i\le\ell$, 
first we note that
\begin{eqnarray}\label{a.4.5}
 P_\ell^{\parens{i}}(z) = \frac{1}{2^\ell \ell!}\frac{\ud^{\ell+i}}{\ud z^{\ell+i}}\parens{z^2 - 1}^\ell.
\end{eqnarray}
Taking
$n=\ell+i$,
$m=\ell$
and
$f_1=f_2=...=f_m=z^2 - 1$
in the Leibniz's formula, we get
\begin{eqnarray}\label{a.4.6}
 P_\ell^{\parens{i}}(z) = \frac{1}{2^\ell \ell!} \mathop{\sum_{k_1,k_2,...,k_\ell}}\limits_{k_1+k_2+\cdots+k_\ell=\ell+i}\frac{\parens{\ell+i}!}{k_1!k_2!\cdots k_\ell!}\prod_{j=1}^{\ell}\parens{z^2 - 1}^{\parens{k_j}}.
\end{eqnarray}
From the above equation, it is obvious that only the terms with all
$k_j$'s
fulfilling
$1\le k_j\le 2$
contribute to
$ P_\ell^{\parens{i}}(1)$.
This, together with the fact that the number of
$k_j$'s
is
$\ell$, lead us to deduce from the constraint
$\sum_{j=1}^{\ell}k_j=\ell+i$
that in each nonzero term of the sum, exactly
$\parens{\ell-i}$
first derivatives
and
$i$
second derivatives appear. Therefore, for all these terms
$\tfrac{1}{\parens{k_1!k_2!\cdots k_\ell!}}$
equals
$\tfrac{1}{2^i}$.
Further, it is clear that
$\prod_{j=1}^{\ell}\parens{z^2 - 1}^{\parens{k_j}}{\big|}_{z=1}$
in each of these terms is equal to
$2^\ell$,
and hence, each of the nonzero terms contributing to the sum has the value of
$2^{\ell-i}\parens{\ell+i}!$.
This makes it possible to evaluate the sum by multiplying the number of its nonzero terms by
$2^{\ell-i}\parens{\ell+i}!$.
The problem of finding that number is equivalent to the problem of counting the number of ways in which one can put
$i$
indistinguishable objects in
$\ell$
distinguishable boxes
($i\le\ell$).
The answer is
${\ell\choose i}$,
and therefore,
\begin{eqnarray}\label{a.4.7}
 P_\ell^{\parens{i}}(1)  =  \frac{1}{2^\ell \ell!}\left[\frac{2^\ell}{2^i}{\ell\choose i}\parens{\ell+i}!\right]=\frac{\parens{\ell+i}!}{2^i i!\parens{\ell-i}!}.
\end{eqnarray}


\section{Independence of $\bar{h}^{\mu\nu}\finsarg$ from $r_0$}\label{app:b}

To prove the independence from
$r_0$,
we need to show that the RHS of
(\ref{4.3.56})
equals the same expression with
$r_0$
replaced by another arbitrary constant
$r_1$.
Making the substitution
$\txpb=\parens{\tfrac{\xp}{r_1}}^B \parens{\tfrac{r_1}{r_0}}^B$
and using the Taylor expansion of
$\parens{\tfrac{r_1}{r_0}}^B$,
we get
\begin{eqnarray}\label{r_0-Independence-first part}
\fl\bar{h}^{\mu\nu}\finsarg \hspace{-10mm}&\stareq \frac{16\pi G}{c^4}\fpa\retint{\parens{\frac{\xp}{r_1}}^{B}\bar{\tau}^{\mu\nu}\retarg}\nonumber\\  
\hspace{-10mm}&+\frac{16\pi G}{c^4}\fp\Bigg\{\Bigg(\sum_{j=1}^{\infty}\frac{1}{j!}\parens{\ln{\frac{r_1}{r_0}}}^j B^j\Bigg)\cdot\A\retint{\parens{\frac{\xp}{r_1}}^{B}\bar{\tau}^{\mu\nu}\retarg}\Bigg\}\nonumber \\  
\hspace{-10mm}& -\frac{1}{4\pi}\sum_{k=0}^{\infty}\sum_{\ell=0}^{\infty}\sum_{i=0}^{\ell}\frac{1}{\ell !}\frac{\parens{\ell+i}!}{2^i i!\parens{\ell-i}!}\frac{\parens{2\ell+1}!!}{\parens{2k}!!\parens{2\ell+2k+1}!!}\frac{\x^{2k}\hat{x}^{L}}{c^{2k+\ell-i}} \nonumber\\  
\hspace{-10mm}& \times\fpa\int_{\mathbb{R}^3}\parens{\frac{\xp}{r_1}}^{B}\frac{\hat{x}'_{L}}{\xp^{\ell+i+1}}\partial_t^{2k+\ell-i}\Lambda^{\mu\nu}\parens{\mathcal{M}\parens{h}}(u',\mathbf{x'})\ud^3\mathbf{x'}\nonumber\\
\hspace{-10mm}& -\frac{1}{4\pi}\sum_{k=0}^{\infty}\sum_{\ell=0}^{\infty}\sum_{i=0}^{\ell}\frac{1}{\ell !}\frac{\parens{\ell+i}!}{2^i i!\parens{\ell-i}!}\frac{\parens{2\ell+1}!!}{\parens{2k}!!\parens{2\ell+2k+1}!!}\frac{\x^{2k}\hat{x}^{L}}{c^{2k+\ell-i}} \nonumber\\  
\hspace{-10mm}&\times\fp\Bigg\{\Bigg(\sum_{j=1}^{\infty}\frac{1}{j!}\parens{\ln{\frac{r_1}{r_0}}}^j B^j\Bigg)\cdot\A\int_{\mathbb{R}^3}\parens{\frac{\xp}{r_1}}^{B}\frac{\hat{x}'_{L}}{\xp^{\ell+i+1}}\nonumber\\
\hspace{-10mm}&\times\partial_t^{2k+\ell-i}\Lambda^{\mu\nu}\parens{\mathcal{M}\parens{h}}(u',\mathbf{x'})\ud^3\mathbf{x'}\Bigg\},
\end{eqnarray}
which, after using
\begin{eqnarray}
\fl\bar{\tau}^{\mu\nu}\finsarg=\frac{c^4}{16\pi G}\bar{\Lambda}^{\mu\nu}\finsarg=\frac{c^4}{16\pi G}\Lambda^{\mu\nu}\parens{\bar{h}}\finsarg=\frac{c^4}{16\pi G}\Lambda^{\mu\nu}\parens{\mathcal{M}\parens{\bar{h}}}\finsarg,
\end{eqnarray}
whose domain of validity is outside the near zone, takes the form
\begin{eqnarray}\label{r_0-Independence-second part}
\fl\bar{h}^{\mu\nu}\finsarg\nonumber \\ 
\fl\stareq\frac{16\pi G}{c^4}\fpa\retint{\parens{\frac{\xp}{r_1}}^{B}\bar{\tau}^{\mu\nu}\retarg}\nonumber\\
\fl -\frac{1}{4\pi}\sum_{k=0}^{\infty}\sum_{\ell=0}^{\infty}\sum_{i=0}^{\ell}\frac{1}{\ell !}\frac{\parens{\ell+i}!}{2^i i!\parens{\ell-i}!}\frac{\parens{2\ell+1}!!}{\parens{2k}!!\parens{2\ell+2k+1}!!}\frac{\x^{2k}\hat{x}^{L}}{c^{2k+\ell-i}} \nonumber\\  
\fl\times\fpa\int_{\mathbb{R}^3}\parens{\frac{\xp}{r_1}}^{B}\frac{\hat{x}'_{L}}{\xp^{\ell+i+1}}\partial_t^{2k+\ell-i}\Lambda^{\mu\nu}\parens{\mathcal{M}\parens{h}}(u',\mathbf{x'})\ud^3\mathbf{x'}\nonumber\\  
\fl+\frac{16\pi G}{c^4}\fp\Bigg\{\Bigg(\sum_{j=1}^{\infty}\frac{1}{j!}\parens{\ln{\frac{r_1}{r_0}}}^j B^j\Bigg)\cdot\A\Bigg[-\frac{1}{4\pi}\int_{\xp<\mathcal{R}}\parens{\frac{\xp}{r_1}}^{B}\frac{\bar{\tau}^{\mu\nu}\retarg}{|\mathbf{x}-\mathbf{x'}|}\ud^3\mathbf{x'}\Bigg]\Bigg\}\nonumber \\  
\fl+\fp\Bigg\{\Bigg(\sum_{j=1}^{\infty}\frac{1}{j!}\parens{\ln{\frac{r_1}{r_0}}}^j B^j\Bigg)\cdot\A\Bigg[-\frac{1}{4\pi}\int_{\mathcal{R}<\xp}\parens{\frac{\xp}{r_1}}^{B}\frac{\Lambda^{\mu\nu}\parens{\mathcal{M}\parens{\bar{h}}}\retarg}{|\mathbf{x}-\mathbf{x'}|}\ud^3\mathbf{x'}\Bigg]\Bigg\}\nonumber \\  
\fl -\frac{1}{4\pi}\sum_{k=0}^{\infty}\sum_{\ell=0}^{\infty}\sum_{i=0}^{\ell}\frac{1}{\ell !}\frac{\parens{\ell+i}!}{2^i i!\parens{\ell-i}!}\frac{\parens{2\ell+1}!!}{\parens{2k}!!\parens{2\ell+2k+1}!!}\frac{\x^{2k}\hat{x}^{L}}{c^{2k+\ell-i}} \nonumber\\  
\fl\times\fp\Bigg\{\Bigg(\sum_{j=1}^{\infty}\frac{1}{j!}\parens{\ln{\frac{r_1}{r_0}}}^j B^j\Bigg)\cdot\A\int_{\mathbb{R}^3}\parens{\frac{\xp}{r_1}}^{B}\frac{\hat{x}'_{L}}{\xp^{\ell+i+1}}\nonumber\\
\fl\times\partial_t^{2k+\ell-i}\Lambda^{\mu\nu}\parens{\mathcal{M}\parens{h}}(u',\mathbf{x'})\ud^3\mathbf{x'}\Bigg\},
\end{eqnarray}
The third term on the RHS of the above equation is zero due to the near-zone integral appearing in it being analytic at
$B=0$
and having the coefficients
$B^k$
with
$k\ge 1$.
Furthermore, by proceeding as we did to obtain
(\ref{4.3.67}),
the fourth term on the RHS can be rewritten as
\begin{eqnarray}\label{r_0-Independence-4th term}
\fl
\bracks{\begin{array}{ccc}
\textrm{4th term on}\\
\textrm{(\ref{r_0-Independence-second part}) RHS}
\end{array}}
=\frac{1}{4\pi}\sum_{k=0}^{\infty}\sum_{\ell=0}^{\infty}\sum_{i=0}^{\ell}\frac{1}{\ell !}\frac{\parens{\ell+i}!}{2^i i!\parens{\ell-i}!}\frac{\parens{2\ell+1}!!}{\parens{2k}!!\parens{2\ell+2k+1}!!}\frac{\x^{2k}\hat{x}^{L}}{c^{2k+\ell-i}} \nonumber\\  
\times\fp\Bigg\{\Bigg(\sum_{j=1}^{\infty}\frac{1}{j!}\parens{\ln{\frac{r_1}{r_0}}}^j B^j\Bigg)\cdot\A\int_{\mathbb{R}^3}\parens{\frac{\xp}{r_1}}^{B}\frac{\hat{x}'_{L}}{\xp^{\ell+i+1}}\nonumber\\
\times\partial_t^{2k+\ell-i}\Lambda^{\mu\nu}\parens{\mathcal{M}\parens{h}}(u',\mathbf{x'})\ud^3\mathbf{x'}\Bigg\}.
\end{eqnarray}
Thus, we have
\begin{eqnarray}\label{r_0-Independence-third part}
\fl\bar{h}^{\mu\nu}\finsarg \hspace{-10mm} &\stareq \frac{16\pi G}{c^4}\fpa\retint{\parens{\frac{\xp}{r_1}}^{B}\bar{\tau}^{\mu\nu}\retarg}\nonumber\\
\hspace{-10mm} &-\frac{1}{4\pi}\sum_{k=0}^{\infty}\sum_{\ell=0}^{\infty}\sum_{i=0}^{\ell}\frac{1}{\ell !}\frac{\parens{\ell+i}!}{2^i i!\parens{\ell-i}!}\frac{\parens{2\ell+1}!!}{\parens{2k}!!\parens{2\ell+2k+1}!!}\frac{\x^{2k}\hat{x}^{L}}{c^{2k+\ell-i}} \nonumber\\  
\hspace{-10mm} &\times\fpa\int_{\mathbb{R}^3}\parens{\frac{\xp}{r_1}}^{B}\frac{\hat{x}'_{L}}{\xp^{\ell+i+1}}\partial_t^{2k+\ell-i}\Lambda^{\mu\nu}\parens{\mathcal{M}\parens{h}}(u',\mathbf{x'})\ud^3\mathbf{x'},
\end{eqnarray}
which means, as stated earlier,
$\bar{h}^{\mu\nu}\finsarg$
is independent of
$r_0$.



\end{document}